\pdfoutput=1
\documentclass[12pt,a4paper]{article}

\usepackage{ifthen} 
\newboolean{pdflatex}
\setboolean{pdflatex}{true} 

\newboolean{articletitles}
\setboolean{articletitles}{true} 

\newboolean{uprightparticles}
\setboolean{uprightparticles}{false} 

\def\paperauthors{LHCb collaboration} 
\def\paperasciititle{Observation of CP violation in two-body B(s)-meson decays to charged pions and kaons}
\def\papertitle{Observation of \CP violation\\in two-body \Bds-meson decays\\to charged pions and kaons}

\def\paperkeywords{{High Energy Physics}, {LHCb}, {CP violation}, {B physics}} 
\def\papercopyright{\the\year\ CERN for the benefit of the LHCb collaboration} 
\def\paperlicence{CC BY 4.0 licence}
\def\paperlicenceurl{https://creativecommons.org/licenses/by/4.0/}

\usepackage[top=1in, bottom=1.25in, left=1in, right=1in]{geometry}

\usepackage{microtype}
\usepackage{lineno}  
\usepackage{xspace} 
\usepackage{caption} 

\usepackage{graphicx}  
\usepackage{color}
\usepackage{colortbl}
\graphicspath{{./figs/}} 

\usepackage{amsmath} 
\usepackage{amssymb}
\usepackage{amsfonts}
\usepackage{upgreek} 

\newcommand*\patchAmsMathEnvironmentForLineno[1]{%
\expandafter\let\csname old#1\expandafter\endcsname\csname #1\endcsname
\expandafter\let\csname oldend#1\expandafter\endcsname\csname
end#1\endcsname
 \renewenvironment{#1}%
   {\linenomath\csname old#1\endcsname}%
   {\csname oldend#1\endcsname\endlinenomath}%
}
\newcommand*\patchBothAmsMathEnvironmentsForLineno[1]{%
  \patchAmsMathEnvironmentForLineno{#1}%
  \patchAmsMathEnvironmentForLineno{#1*}%
}
\AtBeginDocument{%
\patchBothAmsMathEnvironmentsForLineno{equation}%
\patchBothAmsMathEnvironmentsForLineno{align}%
\patchBothAmsMathEnvironmentsForLineno{flalign}%
\patchBothAmsMathEnvironmentsForLineno{alignat}%
\patchBothAmsMathEnvironmentsForLineno{gather}%
\patchBothAmsMathEnvironmentsForLineno{multline}%
\patchBothAmsMathEnvironmentsForLineno{eqnarray}%
}

\usepackage{hyperxmp}

\usepackage[pdftex,
            pdfauthor={\paperauthors},
            pdftitle={\paperasciititle},
            pdfkeywords={\paperkeywords},
            pdfcopyright={Copyright (C) \papercopyright},
            pdflicenseurl={\paperlicenceurl}]{hyperref}

\usepackage[colorinlistoftodos,textsize=scriptsize]{todonotes}

\usepackage[bottom,flushmargin,hang,multiple]{footmisc}

\usepackage[all]{hypcap} 

\usepackage{xspace} 
\usepackage{upgreek}

\def\lhcb   {\mbox{LHCb}\xspace}

\def\babar  {\mbox{BaBar}\xspace}
\def\belle  {\mbox{Belle}\xspace}

\def\cdf    {\mbox{CDF}\xspace}

\def\lhc    {\mbox{LHC}\xspace}

\def\MagUp {\mbox{\em Mag\kern -0.05em Up}\xspace}

\ifthenelse{\boolean{uprightparticles}}%
{

 \def\Pmu         {\ensuremath{\upmu}\xspace}                 
 \def\Pnu         {\ensuremath{\upnu}\xspace}                 
                  
 \def\Ppi         {\ensuremath{\uppi}\xspace}

 \def\Ppsi        {\ensuremath{\uppsi}\xspace}

 \def\PDelta      {\ensuremath{\Delta}\xspace}                 
 \def\PXi         {\ensuremath{\Xi}\xspace}                 
 \def\PLambda     {\ensuremath{\Lambda}\xspace}                 
 \def\PSigma      {\ensuremath{\Sigma}\xspace}                 
 \def\POmega      {\ensuremath{\Omega}\xspace}                 
 \def\PUpsilon    {\ensuremath{\Upsilon}\xspace}

 \def\PB      {\ensuremath{\mathrm{B}}\xspace}                 
                  
 \def\PD      {\ensuremath{\mathrm{D}}\xspace}

 \def\PJ      {\ensuremath{\mathrm{J}}\xspace}                 
 \def\PK      {\ensuremath{\mathrm{K}}\xspace}

 \def\PX      {\ensuremath{\mathrm{X}}\xspace}

 \def\Pb      {\ensuremath{\mathrm{b}}\xspace}                 
 \def\Pc      {\ensuremath{\mathrm{c}}\xspace}                 
 \def\Pd      {\ensuremath{\mathrm{d}}\xspace}                 
                  
 \def\Pf      {\ensuremath{\mathrm{f}}\xspace}

 \def\Pi      {\ensuremath{\mathrm{i}}\xspace}

 \def\Pp      {\ensuremath{\mathrm{p}}\xspace}

 \def\Ps      {\ensuremath{\mathrm{s}}\xspace}

 \def\thebaroffset{0.0em}
}
{

 \def\Pmu         {\ensuremath{\mu}\xspace}                 
 \def\Pnu         {\ensuremath{\nu}\xspace}                 
                  
 \def\Ppi         {\ensuremath{\pi}\xspace}

 \def\Ppsi        {\ensuremath{\psi}\xspace}                 
                  
 \mathchardef\PDelta="7101
 \mathchardef\PXi="7104
 \mathchardef\PLambda="7103
 \mathchardef\PSigma="7106
 \mathchardef\POmega="710A
 \mathchardef\PUpsilon="7107
                  
 \def\PB      {\ensuremath{B}\xspace}                 
                  
 \def\PD      {\ensuremath{D}\xspace}

 \def\PJ      {\ensuremath{J}\xspace}                 
 \def\PK      {\ensuremath{K}\xspace}

 \def\PX      {\ensuremath{X}\xspace}

 \def\Pb      {\ensuremath{b}\xspace}                 
 \def\Pc      {\ensuremath{c}\xspace}                 
 \def\Pd      {\ensuremath{d}\xspace}                 
                  
 \def\Pf      {\ensuremath{f}\xspace}

 \def\Pi      {\ensuremath{i}\xspace}

 \def\Pp      {\ensuremath{p}\xspace}

 \def\Ps      {\ensuremath{s}\xspace}

 \def\thebaroffset{0.18em}
}
\newcommand{\offsetoverline}[2][\thebaroffset]{\kern #1\overline{\kern -#1 #2}}%

\makeatletter
\ifcase \@ptsize \relax
  \newcommand{\miniscule}{\@setfontsize\miniscule{4}{5}}
\or
  \newcommand{\miniscule}{\@setfontsize\miniscule{5}{6}}
\or
  \newcommand{\miniscule}{\@setfontsize\miniscule{5}{6}}
\fi
\makeatother

\DeclareRobustCommand{\optbar}[1]{\shortstack{{\miniscule (\rule[.5ex]{1.25em}{.18mm})}
  \\ [-.7ex] $#1$}}

\def\mup        {{\ensuremath{\Pmu^+}}\xspace}
\def\mun        {{\ensuremath{\Pmu^-}}\xspace} 

\def\ellp       {{\ensuremath{\ell^+}}\xspace}

\def\neu        {{\ensuremath{\Pnu}}\xspace}

\def\dquark    {{\ensuremath{\Pd}}\xspace}
\def\dquarkbar {{\ensuremath{\overline \dquark}}\xspace}

\def\squark    {{\ensuremath{\Ps}}\xspace}
\def\squarkbar {{\ensuremath{\overline \squark}}\xspace}

\def\cquark    {{\ensuremath{\Pc}}\xspace}

\def\bquark    {{\ensuremath{\Pb}}\xspace}
\def\bquarkbar {{\ensuremath{\overline \bquark}}\xspace}
\def\bbbar     {{\ensuremath{\bquark\bquarkbar}}\xspace}

\def\pion   {{\ensuremath{\Ppi}}\xspace}

\def\pip    {{\ensuremath{\pion^+}}\xspace}
\def\pim    {{\ensuremath{\pion^-}}\xspace}

\def\pimp   {{\ensuremath{\pion^\mp}}\xspace}

\def\kaon    {{\ensuremath{\PK}}\xspace}
\def\Kbar    {{\ensuremath{\offsetoverline{\PK}}}\xspace}

\def\KorKbar {\kern \thebaroffset\optbar{\kern -\thebaroffset \PK}{}\xspace}
\def\Kz      {{\ensuremath{\kaon^0}}\xspace}
\def\Kzb     {{\ensuremath{\Kbar{}^0}}\xspace}
\def\Kp      {{\ensuremath{\kaon^+}}\xspace}
\def\Km      {{\ensuremath{\kaon^-}}\xspace}
\def\Kpm     {{\ensuremath{\kaon^\pm}}\xspace}

\def\Dbar    {{\ensuremath{\offsetoverline{\PD}}}\xspace}
\def\D       {{\ensuremath{\PD}}\xspace}

\def\DorDbar {\kern \thebaroffset\optbar{\kern -\thebaroffset \PD}\xspace}
\def\Dz      {{\ensuremath{\D^0}}\xspace}
\def\Dzb     {{\ensuremath{\Dbar{}^0}}\xspace}
\def\Dp      {{\ensuremath{\D^+}}\xspace}
\def\Dm      {{\ensuremath{\D^-}}\xspace}

\def\DpDm    {\ensuremath{\Dp {\kern -0.16em \Dm}}\xspace}

\def\Dstarp  {{\ensuremath{\D^{*+}}}\xspace}

\def\Dsm     {{\ensuremath{\D^-_\squark}}\xspace}

\def\B       {{\ensuremath{\PB}}\xspace}
\def\Bbar    {{\ensuremath{\offsetoverline{\PB}}}\xspace}
\def\Bb      {{\ensuremath{\Bbar}}\xspace}
\def\BorBbar {\kern \thebaroffset\optbar{\kern -\thebaroffset \PB}\xspace}
\def\Bz      {{\ensuremath{\B^0}}\xspace}

\def\Bd      {{\ensuremath{\B^0}}\xspace}
\def\Bdb     {{\ensuremath{\Bbar{}^0}}\xspace}
\def\BdorBdbar {\kern \thebaroffset\optbar{\kern -\thebaroffset \Bd}\xspace}
\def\Bu      {{\ensuremath{\B^+}}\xspace}

\def\Bp      {{\ensuremath{\Bu}}\xspace}

\def\Bs      {{\ensuremath{\B^0_\squark}}\xspace}
\def\Bsb     {{\ensuremath{\Bbar{}^0_\squark}}\xspace}
\def\BsorBsbar {\kern \thebaroffset\optbar{\kern -\thebaroffset \Bs}\xspace}
\def\Bc      {{\ensuremath{\B_\cquark^+}}\xspace}

\def\Bds     {{\ensuremath{\B_{(\squark)}^0}}\xspace}
\def\Bdsb    {{\ensuremath{\Bbar{}_{(\squark)}^0}}\xspace}

\def\jpsi     {{\ensuremath{{\PJ\mskip -3mu/\mskip -2mu\Ppsi}}}\xspace}

\def\Y#1S{\ensuremath{\PUpsilon{(#1S)}}\xspace}
\def\OneS  {{\Y1S}}

\def\proton      {{\ensuremath{\Pp}}\xspace}

\def\Lz          {{\ensuremath{\PLambda}}\xspace}

\def\LorLbar     {\kern \thebaroffset\optbar{\kern -\thebaroffset \PLambda}\xspace}

\def\Lb           {{\ensuremath{\Lz^0_\bquark}}\xspace}

\newcommand{\decay}[2]{\ensuremath{#1\!\to #2}\xspace} 

\def\to                 {\ensuremath{\rightarrow}\xspace}

\def\CP                {{\ensuremath{C\!P}}\xspace}

\newcommand{\dms}{{\ensuremath{\Delta m_{\squark}}}\xspace}
\newcommand{\dmd}{{\ensuremath{\Delta m_{\dquark}}}\xspace}
\newcommand{\DG}{{\ensuremath{\Delta\Gamma}}\xspace}
\newcommand{\DGs}{{\ensuremath{\Delta\Gamma_{\squark}}}\xspace}
\newcommand{\DGd}{{\ensuremath{\Delta\Gamma_{\dquark}}}\xspace}
\newcommand{\Gs}{{\ensuremath{\Gamma_{\squark}}}\xspace}
\newcommand{\Gd}{{\ensuremath{\Gamma_{\dquark}}}\xspace}

\newcommand{\SKK}{\ensuremath{{S}_{\kaon\kaon}}\xspace}
\newcommand{\CKK}{\ensuremath{{C}_{\kaon\kaon}}\xspace}
\newcommand{\Spipi}{\ensuremath{{S}_{\pion\pion}}\xspace}
\newcommand{\Cpipi}{\ensuremath{{C}_{\pion\pion}}\xspace}

\newcommand{\ADGKK}{\ensuremath{{\cal A}^{\Delta\Gamma}_{\kaon\kaon}}\xspace}
\newcommand{\ACPBd}{\ensuremath{A_{\CP}^{\Bd}}\xspace}
\newcommand{\ACPBs}{\ensuremath{A_{\CP}^{\Bs}}\xspace}

\def\SSK      {{\ensuremath{{\rm SS}\kaon}}\xspace}

\def\BdTopipi     {\decay{\Bd}{\pip\pim}}
\def\BdToKpi      {\decay{\Bd}{\Kp\pim}}
\def\BsToKK       {\decay{\Bs}{\Kp\Km}}
\def\BsTopiK      {\decay{\Bs}{\Km\pip}}

\def\LbTopK       {\decay{\Lb}{p\Km}\xspace}

\def\JpsiTomumu  {\decay{ \jpsi}{\mup\mun}}
\def\BsToDspi     {\decay{\Bs}{\Dsm\pip\xspace}}
\def\BsTopipi     {\decay{\Bs}{\pip\pim\xspace}}
\def\BdToKK       {\decay{\Bd}{\Kp\Km\xspace}}

\def\LbTopK       {\decay{\Lb}{p\Km}\xspace}

\def\BdToDpi      {\decay{\Bd}{\Dm\pip}}
\def\BsToDspi      {\decay{\Bs}{\Dsm\pip}}

\def\AT#1     {\ensuremath{A_{\mathrm{T}}^{#1}}\xspace}           

\def\C#1      {\ensuremath{\mathcal{C}_{#1}}\xspace}                       
\def\Cp#1     {\ensuremath{\mathcal{C}_{#1}^{'}}\xspace}                    
\def\Ceff#1   {\ensuremath{\mathcal{C}_{#1}^{\mathrm{(eff)}}}\xspace}        
\def\Cpeff#1  {\ensuremath{\mathcal{C}_{#1}^{'\mathrm{(eff)}}}\xspace}       
\def\Ope#1    {\ensuremath{\mathcal{O}_{#1}}\xspace}                       
\def\Opep#1   {\ensuremath{\mathcal{O}_{#1}^{'}}\xspace}                    

\newcommand{\nospaceunit}[1]{\ensuremath{\text{#1}}}       
\newcommand{\aunit}[1]{\ensuremath{\text{\,#1}}}       

\newcommand{\tev}{\aunit{Te\kern -0.1em V}\xspace}
\newcommand{\gev}{\aunit{Ge\kern -0.1em V}\xspace}
\newcommand{\mev}{\aunit{Me\kern -0.1em V}\xspace}
\newcommand{\kev}{\aunit{ke\kern -0.1em V}\xspace}
\newcommand{\ev}{\aunit{e\kern -0.1em V}\xspace}
 
\newcommand{\mevc}{\ensuremath{\aunit{Me\kern -0.1em V\!/}c}\xspace}
\newcommand{\gevc}{\ensuremath{\aunit{Ge\kern -0.1em V\!/}c}\xspace}
\newcommand{\mevcc}{\ensuremath{\aunit{Me\kern -0.1em V\!/}c^2}\xspace}
\newcommand{\gevcc}{\ensuremath{\aunit{Ge\kern -0.1em V\!/}c^2}\xspace}

\def\mum  {\ensuremath{\,\upmu\nospaceunit{m}}\xspace}

\def\fb   {\ensuremath{\aunit{fb}}\xspace}
\def\invfb   {\ensuremath{\fb^{-1}}\xspace}

\def\ps   {\ensuremath{\aunit{ps}}\xspace}
\def\fs   {\aunit{fs}}

\def\invps{\ensuremath{\ps^{-1}}\xspace}

\newcommand{\chisq}{\ensuremath{\chi^2}\xspace}

\newcommand{\chisqip}{\ensuremath{\chi^2_{\text{IP}}}\xspace}

\def\gsim{{~\raise.15em\hbox{$>$}\kern-.85em
          \lower.35em\hbox{$\sim$}~}\xspace}
\def\lsim{{~\raise.15em\hbox{$<$}\kern-.85em
          \lower.35em\hbox{$\sim$}~}\xspace}

\def\sPlot{\mbox{\em sPlot}\xspace}
\def\sFit{\mbox{\em sFit}\xspace}

\def\pt         {\ensuremath{p_{\mathrm{T}}}\xspace}

\def\ptot       {\ensuremath{p}\xspace}

\def\evtgen     {\mbox{\textsc{EvtGen}}\xspace}

\def\geant      {\mbox{\textsc{Geant4}}\xspace}

\def\photos     {\mbox{\textsc{Photos}}\xspace}

\def\pythia     {\mbox{\textsc{Pythia}}\xspace}

\def\tell1  {TELL1\xspace}
\def\ukl1   {UKL1\xspace}

\newcommand{\ie}{\mbox{\itshape i.e.}\xspace}

\usepackage{cite} 
\usepackage{mciteplus}

\usepackage{longtable} 

\begin{document}

\renewcommand{\thefootnote}{\fnsymbol{footnote}}
\setcounter{footnote}{1}


\begin{titlepage}
\pagenumbering{roman}

\vspace*{-1.5cm}
\centerline{\large EUROPEAN ORGANIZATION FOR NUCLEAR RESEARCH (CERN)}
\vspace*{1.5cm}
\noindent
\begin{tabular*}{\linewidth}{lc@{\extracolsep{\fill}}r@{\extracolsep{0pt}}}
\ifthenelse{\boolean{pdflatex}}
{\vspace*{-1.5cm}\mbox{\!\!\!\includegraphics[width=.14\textwidth]{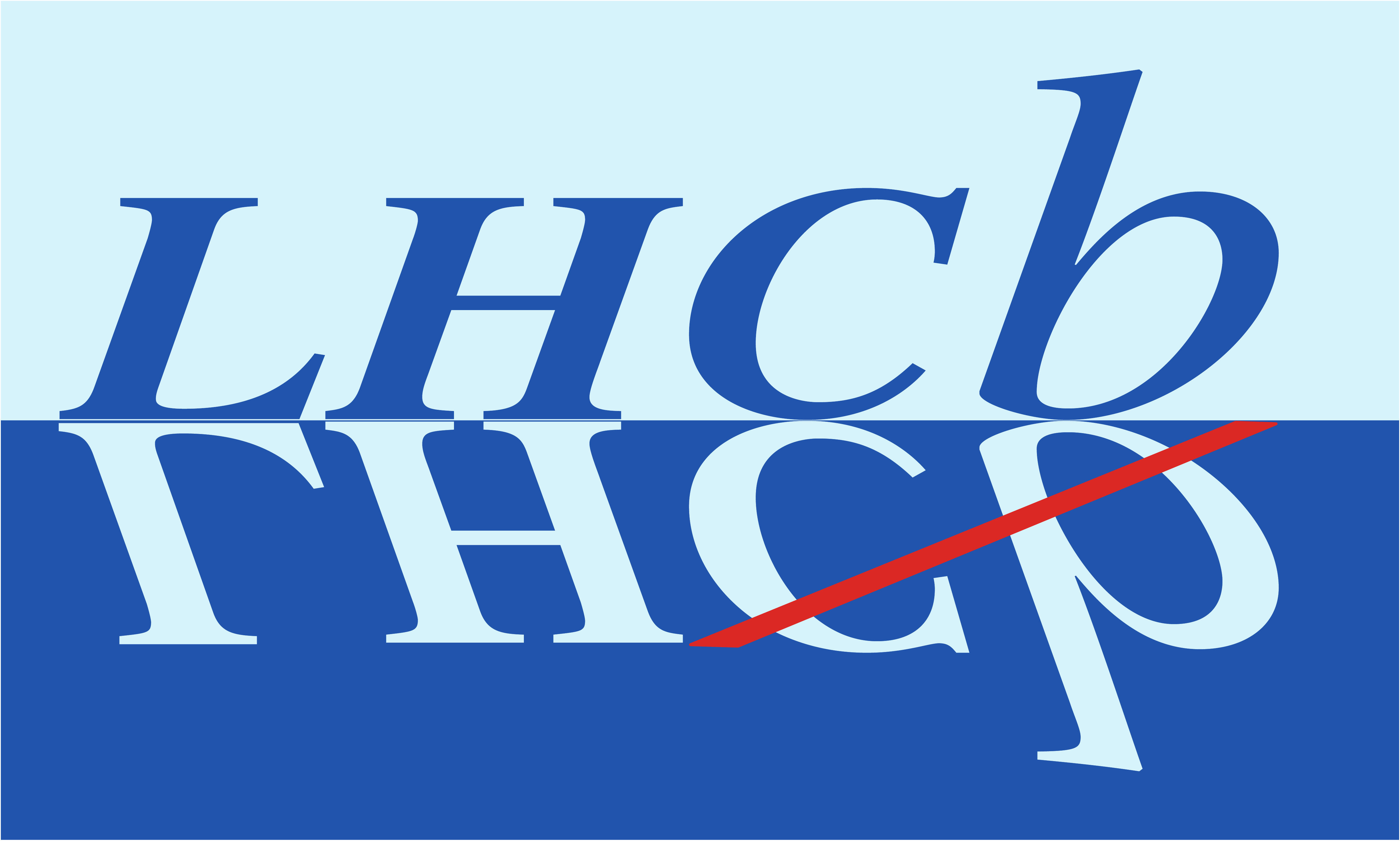}} & &}%
{\vspace*{-1.2cm}\mbox{\!\!\!\includegraphics[width=.12\textwidth]{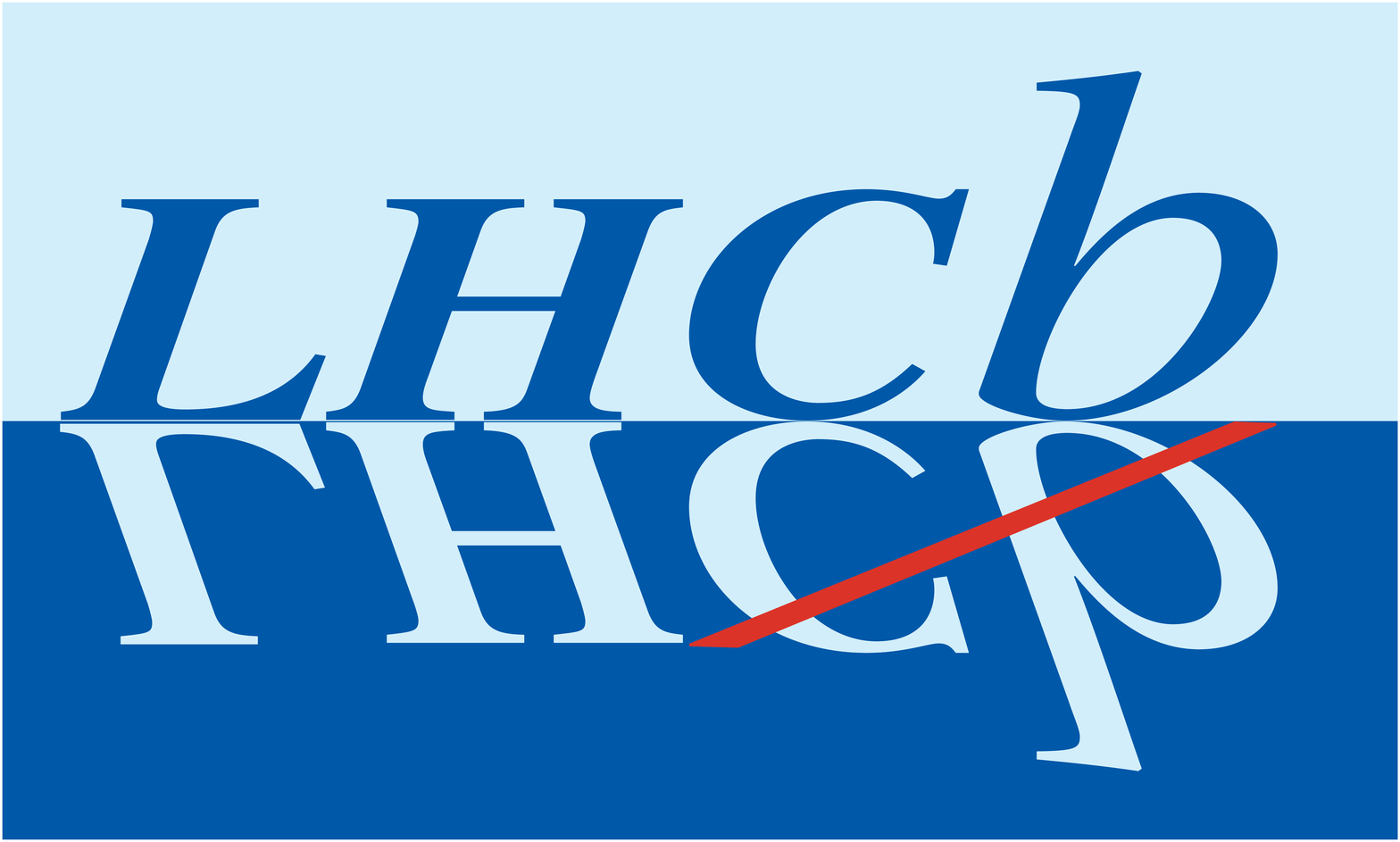}} & &}%
\\
 & & CERN-EP-2020-215 \\  
 & & LHCb-PAPER-2020-029 \\  
 & & March 8, 2021 \\
 & & \\
\end{tabular*}

\vspace*{2.0cm}

{\normalfont\bfseries\boldmath\huge
\begin{center}
  \papertitle 
\end{center}
}

\vspace*{0.2cm}

\begin{center}
\paperauthors\footnote{Authors are listed at the end of this paper.}
\end{center}

\vspace{\fill}

\begin{abstract}
  \noindent
  The time-dependent \CP asymmetries of \BdTopipi and \BsToKK decays are measured using a data sample of \proton\proton collisions corresponding to an integrated luminosity of 1.9~\invfb, collected with the \lhcb detector at a centre-of-mass energy of 13~\tev. The results are 
  \begin{eqnarray*}
    \Cpipi & = & -0.311 \pm 0.045 \pm 0.015, \\
    \Spipi & = & -0.706 \pm 0.042 \pm 0.013, \\
    \CKK   & = & \phantom{-}0.164 \pm 0.034 \pm 0.014, \\
    \SKK   & = & \phantom{-}0.123 \pm 0.034 \pm 0.015, \\
    \ADGKK & = & -0.83\phantom{0} \pm 0.05\phantom{0} \pm 0.09,
  \end{eqnarray*}
  where the first uncertainties are statistical and the second systematic. The same data sample is used to measure the time-integrated \CP asymmetries of \BdToKpi and \BsTopiK decays and the results are 
  \begin{eqnarray*}
    \ACPBd & = & -0.0824 \pm 0.0033 \pm 0.0033, \\
    \ACPBs & = & \phantom{-}0.236\phantom{0} \pm 0.013\phantom{0} \pm 0.011.
  \end{eqnarray*}
  All results are consistent with earlier measurements. A combination of \lhcb measurements provides the first observation of time-dependent \CP violation in \Bs decays.
\end{abstract}

\vspace*{1.5cm}

\begin{center}
  Published in
  JHEP 03 (2021) 075.
\end{center}

\vspace{\fill}

{\footnotesize 
\centerline{\copyright~\papercopyright. \href{\paperlicenceurl}{\paperlicence}.}}
\vspace*{2mm}

\end{titlepage}


\newpage
\setcounter{page}{2}
\mbox{~}
%
%
%
%


\renewcommand{\thefootnote}{\arabic{footnote}}
\setcounter{footnote}{0}

\cleardoublepage


\pagestyle{plain} 
\setcounter{page}{1}
\pagenumbering{arabic}



\section{Introduction}

Charge-parity (\CP) asymmetries of charmless \Bds-meson decays to two-body charged final states are important inputs to the validation of the Cabibbo-Kobayashi-Maskawa (CKM) mechanism~\cite{Cabibbo:1963yz,Kobayashi:1973fv}, which models  \CP violation in charged-current quark transitions. Deviations from Standard Model (SM) predictions may reveal the presence of phenomena not included in the SM, manifested as modifications to the amplitudes of these decays.~\cite{PhysRevLett.75.1703,He1999,Fleischer:1999pa,GRONAU200071,LIPKIN2005126,Fleischer:2007hj,Fleischer:2010ib}. The \CP asymmetry in the \BdTopipi decay is a fundamental input to the isospin analysis of \decay{\B}{\pion\pion} decays that allows the determination of the CKM angle $\alpha$~\cite{PhysRevLett.65.3381,PhysRevD.76.014015,Charles2017}. The analysis can be extended by exploiting the approximate U-spin symmetry~\cite{Gronau:2000zy} that relates the hadronic parameters entering the decay amplitudes of the \BdTopipi and \BsToKK decays.\footnote{Unless stated otherwise, the inclusion of charge-conjugate decay modes is implied throughout this paper.} It has been shown that, by incorporating the \CP asymmetry and branching fraction of the \BsToKK decay into the standard isospin analysis, stringent constraints on the CKM angle $\gamma$ and on the \CP-violating phase $-2\beta_s$ can be set, even when allowing for U-spin breaking effects~\cite{Ciuchini:2012gd,LHCb-PAPER-2014-045}. Furthermore, a substantial reduction of uncertainties on the determination of $-2\beta_s$ can be achieved by combining the \CP asymmetries of the \BdTopipi and \BsToKK decays with information provided by the semileptonic decays \decay{\Bd}{\pim\ellp\neu}
and \decay{\Bs}{\Km\ellp\neu}~\cite{Fleischer:2016jbf,Fleischer:2016ofb}. The \CP asymmetries and branching fractions of the \BdToKpi and \BsTopiK provides the test of the SM, assuming U-spin symmetry, proposed in Ref.~\cite{LIPKIN2005126}. The \CP asymmetry of the \BdToKpi decay is also a key input to the long-standing \decay{\B}{\kaon\pion} puzzle~\cite{Gronau:1998ep,Buras:2003yc,Baek:2007yy}. Strategies have been proposed to combine information from several decays of the \decay{\B}{\pion\pion} and \decay{\B}{\kaon\pion} systems in order to investigate the presence of physics beyond the SM~\cite{Fleischer:2018bld,Crivellin:2019isj,Calibbi:2019lvs}.

This paper presents measurements of time-dependent \CP asymmetries in \mbox{\BdTopipi} and \BsToKK decays and of time-integrated \CP asymmetries in \BdToKpi and \BsTopiK decays. The analysis is based on a data sample of \proton\proton collisions corresponding to an integrated luminosity of 1.9\invfb, collected with the \lhcb detector at a centre-of-mass energy of 13\tev during 2015 and 2016. These results are combined with previous LHCb results, published in Ref.~\cite{LHCb-PAPER-2018-006}, based on a sample corresponding to 3.0\invfb, collected at 7 and 8\tev in the Run~1 data taking.

In decays of \Bds mesons to a final state $f$, where $f$ is a \CP eigenstate ($f=\overline{f}$), \CP violation originates from the interference between the decay and \Bds-\Bdsb mixing. The latter can be modelled by an effective Hamiltonian whose mass eigenstates are linear combinations of the two flavour eigenstates, $p|\Bds\rangle \pm q|\Bdsb\rangle$, where $p$ and $q$ are complex parameters, normalised such that $|p|^2 + |q|^2 = 1$.
The \CP asymmetry as a function of decay time for \decay{\Bds}{\Pf} decays is given by
\begin{equation}\label{eq:acpTDDef}
A_{\CP}(t)=\frac{\Gamma_{\Bdsb \to f}(t)-\Gamma_{\Bds \to f}(t)}{\Gamma_{\Bdsb \to f}(t)+\Gamma_{\Bds \to f}(t)}=\frac{-C_f \cos(\Delta m_{\dquark(\squark)} t) + S_f \sin(\Delta m_{\dquark(\squark)} t)}{\cosh\left(\frac{\Delta\Gamma_{\dquark(\squark)}}{2} t\right) + A^{\Delta\Gamma}_f \sinh\left(\frac{\Delta\Gamma_{\dquark(\squark)}}{2} t\right)},
\end{equation}
where $\Delta m_{\dquark(\squark)}$ and $\Delta\Gamma_{\dquark(\squark)}$ are the mass and width differences of the mass eigenstates of the \Bds system. In accordance with current experimental knowledge, the value of \DGd is assumed to be negligible. The quantities $C_f$, $S_f$ and $A^{\Delta\Gamma}_f$ are defined as
\begin{equation}
	C_{f} \equiv \frac{1-|\lambda_f|^2}{1+|\lambda_f|^2},\qquad
	S_{f} \equiv  \frac{2 {\rm Im} \lambda_f}{1+|\lambda_f|^2},\qquad
	A^{\Delta\Gamma}_f \equiv  - \frac{2 {\rm Re} \lambda_f}{1+|\lambda_f|^2},
\label{eq:adirmix} 
\end{equation}
where $\lambda_f$ is given by
\begin{equation}
\lambda_f \equiv \frac{q}{p}\frac{\overline{A}_f}{A_f}
\end{equation}
and $A_{f}$ ($\overline{A}_{f}$) is the decay amplitude for the \Bds (\Bdsb)~$\to f$ transition.
As current experimental determinations~\cite{HFLAV18,LHCb-PAPER-2016-013,LHCb-PAPER-2014-053} confirm the SM expectation~\cite{UTfit-UT,CKMfitter2015} of negligible \CP violation in the \Bds-\Bdsb mixing (implying $\left|q/p\right| = 1$), a nonzero value of $C_f$ and $S_f$ indicates the presence of \CP violation in the decay and in the interference between mixing and decay, respectively. 
The quantities $C_f$, $S_f$ and $A_f^{\DG}$ are related through 
the unitary condition $\left(C_f\right)^2+\left(S_f\right)^2+\left(A_f^{\DG}\right)^2 = 1$. This constraint is not imposed in this analysis and is instead used as a cross-check of the consistency of the results. Previous determinations of \Cpipi and \Spipi were performed by \babar~\cite{Lees:2012mma}, \belle~\cite{Adachi:2013mae} and \lhcb~\cite{LHCb-PAPER-2018-006} experiments, while only \lhcb has measured \CKK, \SKK and \ADGKK~\cite{LHCb-PAPER-2018-006}.

The time-integrated \CP asymmetry for a \Bds decay to a flavour-specific final state $f$, such as \BdToKpi and \BsTopiK, is defined as
\begin{equation}\label{eq:acpDef}
A_{\CP} = \frac{\left|\overline{A}_{\overline{f}}\right|^2-\left|A_f\right|^2}{\left|\overline{A}_{\overline{f}}\right|^2+\left|A_f\right|^2}.
\end{equation}
Measurements of $A_{\CP}$ for the \BdToKpi decay~(\ACPBd) were carried out by \babar~\cite{Lees:2012mma}, \belle~\cite{Duh:2012ie}, \cdf~\cite{Aaltonen:2014vra} and \lhcb~\cite{LHCb-PAPER-2018-006}, while $A_{\CP}$ for the \BsTopiK decay~(\ACPBs) was measured only by \cdf~\cite{Aaltonen:2014vra} and \lhcb~\cite{LHCb-PAPER-2018-006}.

This paper is organised as follows. The \lhcb detector, its trigger system and the simulation process are briefly introduced in Sec.~\ref{sec:detector}, while the sample selection is described in  Sec.~\ref{sec:selection}.
The \CP asymmetries are determined by means of unbinned maximum-likelihood fits to the invariant-mass and decay-time distributions of \Bds candidates reconstructed in the \pip\pim, $\Kp\!\Km$ and \Kpm\pimp final states.
In order to measure the time-dependent \CP asymmetries, it is necessary to determine the flavour of the \Bds meson at its production. In addition, a precise determination of the \Bds decay time is important, in particular for the \Bs meson, due to its fast oscillation frequency. The flavour-tagging algorithms and their calibration are presented in Sec.~\ref{sec:flavourTagging}, while the determination of the decay-time resolution is discussed in Sec.~\ref{sec:timeResolution}. The models used in the fits are described in Sec.~\ref{sec:fitMethod}. Two measurements of the \CP-violating parameters for the \BdTopipi and \BsToKK decays are performed with different experimental techniques. 
The first method, referred to as the \textit{simultaneous method}, fits all the signal decays simultaneously and uses a fit model similar to that  described in Ref.~\cite{LHCb-PAPER-2018-006}. The second method, referred to as the \textit{per-candidate method}, describes the selection efficiency
as a function of the decay time of the \Bds meson on a per-candidate basis using the {\it swimming} technique~\cite{Bailey:1985zz,Rademacker:2005ay,Gligorov:2018lb,LHCb-PAPER-2011-014}. 
The determination of the detection asymmetry between the \BdToKpi and \BsTopiK decays and their charge-conjugate final states, necessary to measure $A_{\CP}$, is discussed in Sec.~\ref{sec:asymmetrydet}. The results are given in Sec.~\ref{sec:fitResult} and the assessment of systematic uncertainties is presented in Sec.~\ref{sec:systematics}. 
The statistical and systematic uncertainties on the simultaneous method are found to be, in general, smaller than those for the per-candidate method.  The results from the simultaneous method are therefore given as the main results of this paper.
The final results and their combination with previous \lhcb measurements from Ref.~\cite{LHCb-PAPER-2018-006} are presented in Sec.~\ref{sec:finalResults}, while considerations on the combined measurements are reported in Sec.~\ref{sec:conclusions}.

\section{Detector, trigger and simulation}\label{sec:detector}

The \lhcb detector~\cite{LHCb-DP-2008-001,LHCb-DP-2014-002} is a single-arm forward spectrometer covering the \mbox{pseudorapidity} in the range between 2 and 5, designed for the study of particles containing \bquark or \cquark quarks. The detector includes a high-precision tracking system
consisting of a silicon-strip vertex detector surrounding the \proton\proton-interaction region, a large-area silicon-strip detector located upstream of a dipole magnet with a bending power of about $4{\mathrm{\,Tm}}$, and three stations of silicon-strip detectors and straw drift tubes placed downstream of the magnet~\cite{LHCb-DP-2014-001,LHCb-DP-2013-003}. The tracking system provides a measurement of momentum, \ptot, of charged particles with a relative uncertainty that varies from 0.5\% at low momentum to 1.0\% at 200\gevc.
The minimum distance of a track to a primary \proton\proton-collision vertex (PV), the impact parameter (IP), is measured with a resolution of $(15+29/\pt)\mum$,
where \pt is the component of the momentum transverse to the beam, in\,\gevc. Different types of charged hadrons are distinguished using information from two ring-imaging Cherenkov (RICH) detectors~\cite{LHCb-DP-2012-003}. Photons, electrons and hadrons are identified by a calorimeter system consisting of scintillating-pad and preshower detectors, an electromagnetic calorimeter and a hadronic calorimeter. Muons are identified by a system composed of alternating layers of iron and multiwire proportional chambers.
The online event selection is performed by a trigger~\cite{LHCb-DP-2012-004}, which consists of a hardware stage, based on information from the calorimeter and muon systems, followed by a software stage, which applies a full event reconstruction.

At the hardware trigger stage, events are required to have a muon with high \pt, or a hadron, photon or electron with high transverse energy in the calorimeters. For hadrons, the transverse energy threshold is 3.5\gev. The software trigger requires the presence in the event of at least one charged particle with $\pt > 1.6\gevc$ and inconsistent with originating from any PV. The tracks identified at this stage are used by a trigger selection dedicated for two-body \bquark-hadron decays. The selection algorithm imposes requirements on the quality of the reconstructed tracks, their \pt and minimum \chisqip with respect to every PV in the event, where the \chisqip is defined as the difference in the vertex-fit \chisq of a given PV reconstructed with and without the track under consideration. Pairs of oppositely charged tracks must have a small distance of closest approach and a large scalar sum of their \pt in order to be eligible to form a \Bds candidate. Finally, the \Bds candidates are required to pass criteria based on their \pt, \chisqip, flight distance with respect to their associated PV, and angle between the direction of the \Bds candidate momentum and the direction defined by its decay vertex and associated PV. Candidates are associated with the PV that is most consistent with their flight direction.

Simulation is used to study the discrimination between signal and background candidates, and to assess differences between signal and calibration decays. The $pp$ collisions are generated using \pythia~\cite{Sjostrand:2007gs,Sjostrand:2006za} with a specific \lhcb configuration~\cite{LHCb-PROC-2010-056}.  Decays of hadronic particles are described by \evtgen~\cite{Lange:2001uf}, in which final-state radiation is generated using \photos~\cite{Golonka:2005pn}. The interaction of the generated particles with the detector, and its response, are implemented using the \geant toolkit~\cite{Allison:2006ve, *Agostinelli:2002hh} as described in Ref.~\cite{LHCb-PROC-2011-006}.

\section{Selection}\label{sec:selection}

The \Bds candidates selected by the dedicated software trigger are further filtered, requiring that either the decay products or particles from the rest of the event are responsible for the positive decision of the hadronic hardware trigger. Candidates are then classified into mutually exclusive samples of different final states (\pip\pim, $\Kp\!\Km$ and \Kpm\pimp) using particle identification (PID) information. Finally, a boosted decision tree (BDT) algorithm~\cite{Breiman,Roe} is used to separate signal candidates from combinatorial background candidates for each of the final states.

Four types of background contributions are considered: two-body \bquark-hadron decays with misidentified pions, kaons or protons in the final state (cross-feed background); pairs of randomly associated and oppositely charged tracks (combinatorial background); pairs of oppositely charged tracks from partially reconstructed three-body decays of \bquark hadrons (three-body background); \Bds mesons produced in \Bc decays rather than at a PV, whose measured decay time is biased due to the finite lifetime of the \Bc meson. Given the small production rate of \Bc mesons~\cite{LHCb-PAPER-2013-044}, this background contribution is neglected in the analysis and a systematic uncertainty is assessed in Sec.~\ref{sec:systematics}. Since the three-body background candidates give rise to \Bds candidates with invariant-mass values well separated from the mass peak, the candidate selection is customised to reject mainly the cross-feed and combinatorial background candidates, as they affect the invariant-mass region around the \Bd and \Bs nominal masses.

The requirements imposed on the PID variables, used to identify the \pip\pim and \Kp\Km samples, are optimised using pseudoexperiments that take into account the different background contributions. First the PID efficiencies and misidentification probabilities for kaons and pions are determined, for different requirements, using samples of \decay{\Dstarp}{\Dz(\to\Km\pip)\pip} decays\cite{LHCb-PUB-2016-021}  and are used to estimate the cross-feed background yields in each of the final states. The results of the PID calibration and the fitting model described in Sec.~\ref{sec:fitMethod} are used to generate pseudoexperiments that are fitted with the same model. The results of the fits are used to find the configuration of PID requirements giving the best trade-off between the statistical sensitivity to the \CP-violation parameters of the \BdTopipi and \BsToKK decays and the systematic effects due to large contributions of cross-feed background candidates. The PID selection used to identify the \Kpm\pimp samples is, instead, optimised to reduce the amount of the \BdTopipi and \BsToKK cross-feed background yields to approximately 10\% of the \BsTopiK yield.

The BDT algorithm exploits the following properties of the \Bds decay products: the \pt of the two tracks; the \chisqip of each track with respect to their associated PV; the distance of closest approach between the two tracks, and the quality of their common vertex. The BDT classifier also uses properties of the reconstructed \Bds candidate, particularly the \pt, the \chisqip and the \chisq of the flight distance with respect to the associated PV.
Separate BDT algorithms are trained and optimised for the selection of the \BdTopipi and the \BsToKK decays. Simulated events of the two decay modes are used to model the signal candidates, while data from their high-mass sidebands (from 5.6\gevcc to 6.2\gevcc) are used to model the combinatorial background candidates.
The optimal threshold on the response of the BDT algorithm is chosen to maximise $S/\sqrt{S+B}$, where $S$ and $B$ represent the estimated yield of signal and combinatorial background candidates within $\pm 60\mevcc$ (corresponding to about $\pm 3$ times the invariant-mass resolution) around the known \Bds mass. The \Kpm\pimp samples are selected using the BDT classifier optimised for the \BdTopipi decay.\footnote{A BDT classifier optimised for \BdToKpi decays was found to have a comparable performance to that optimised for \BdTopipi decays and applied to the \Kpm\pimp sample.} Multiple candidates are present in less than 0.06\% of the events satisfying the offline selections. Only one candidate is accepted at random from each event.

The optimisation of the selection criteria preferentially rejects short-lived candidates over longer lived ones. This introduces a distorted decay-time efficiency that must be corrected for. The selection criteria present in the analysis that produce this efficiency are the requirements on the \chisqip of all particles, the \chisq of the \Bds flight distance, the direction defined by its decay vertex and associated PV, and the outputs of the BDT algorithm. In addition, there are also decay-time biasing selection criteria due to the geometry of the detector. These are the limit on the radial flight distance of the \Bds, which is required to avoid secondary interactions with the vertex detector material, and the minimal number of the vertex-detector sensors required to have track hits, which is imposed by the software triggers. The bias introduced by the radial flight distance is only present in the per-candidate method.

\section{Flavour tagging}\label{sec:flavourTagging}

Tagging of the initial flavour of the \Bds meson plays a crucial role in measuring the time-dependent \CP asymmetries of decays to \CP
eigenstates, since the sensitivity to the $C_{f}$ and $S_{f}$ coefficients, defined in Eq.~\eqref{eq:acpTDDef}, is related to the tagging performance.
The flavour of the \Bds candidates is inferred by two classes of the flavour-tagging algorithms called opposite-side (OS) and 
same-side (SS) taggers. The OS taggers~\cite{LHCb-PAPER-2011-027} exploit the fact that in \proton\proton collisions beauty quarks 
are almost exclusively produced in \bbbar pairs. Thus the flavour of the decaying signal \Bds meson can be determined by 
looking at the decay products of the other \bquark hadron in the event, for example, the charge of the lepton originating from semileptonic decays, the charge of the kaon from 
the $\bquark\to\cquark\to\squark$ transition, or the charge of a charm hadron.
An additional OS tagger is based on the inclusive reconstruction of the opposite \bquark-hadron decay vertex by computing the \pt-weighted average of the charges of all tracks associated to that vertex. The SS taggers are based on the identification of the particles produced in the hadronisation of the 
signal beauty quarks. In contrast to the OS taggers, which to a very good approximation act equally on \Bd and \Bs mesons, SS taggers 
are specific to the light quark of the \Bds meson under study. Additional \dquarkbar (\dquark) or \squarkbar (\squark) 
quarks produced in association with a \Bd (\Bdb) or a \Bs (\Bsb) meson, respectively, can form charged pions and protons, 
in the down-quark case, or charged kaons, in the strange-quark case. The so-called 
SS\pion and SS\proton taggers~\cite{LHCb-PAPER-2016-039} are used to determine the initial flavour of \Bd mesons, while 
the SS\kaon tagger~\cite{LHCb-PAPER-2015-056} is used for \Bs mesons.

For each tagger, the probability of misidentifying the flavour of the \Bds meson at production, the mistag probability, $\eta$, 
is estimated by means of a multivariate classifier, and is defined in the range $0 \leq \eta \leq 0.5$. 
The flavour-tagging performance of each tagger can be quantified by means of the tagging power, defined as 
\begin{align}
\varepsilon_{\rm eff} = \frac{1}{N} \sum\limits_{i}^{}{|\xi_i|\,(1-2\eta_i)^2},
\end{align}
where $\xi_i$ and $\eta_i$ are the tagging decision and the probability of misidentifying the flavour of the $i$-th out of $N$ \Bds candidates,  respectively. The tagging decision $\xi_i$ takes the value of $+1$ when the candidate is tagged as \Bds, $-1$ when the candidate is tagged as \Bdsb, and zero for untagged candidates. 
Multivariate algorithms are used to determine the values of $\eta$ for the OS and SS taggers, denoted as $\eta_{\rm OS}$ and $\eta_{\rm SS}$. These are trained using specific 
\B-meson decay channels and selections. The differences between the training samples and the selected signal \Bds candidates 
can lead to an imperfect determination of the mistag probability. Hence, a more accurate estimate, denoted as $\omega$ 
hereafter, is obtained by means of a calibration procedure that takes into account the specific kinematics of selected 
signal \Bds mesons. 
The relation between $\eta$ and $\omega$ is calibrated using \decay{\Bu}{\Dzb\pip}, \BsToDspi and \BdToDpi decays for the OS, 
SS\kaon, and SS\pion and SS\proton taggers, respectively.
The flavour for the \Bp meson is tagged by the charge of the pion in the final state. For the \Bz and \Bs modes, which decay into flavour-specific final states, the amplitude of the tagged time-dependent asymmetry is proportional to $1-2\omega$.
When the response of more than one OS tagger is available per candidate, the different decisions and associated calibrated
mistag probabilities are combined into a unique decision $\xi_{\rm OS}$ and a single $\eta_{\rm OS}$. 
A similar combination is also performed between the SS\pion and SS\proton taggers to create a combined same-side tagger, SS$_{c}$, where a combined tagging decision 
$\xi_{\rm SSc}$ and mistag probability $\eta_{\rm SSc}$ is evaluated, as discussed in App.~\ref{sec:ssdCombination}.

In the simultaneous method, the OS and SSc combinations are recalibrated in the final fit, discussed in Sec.~\ref{sec:fitMethod}, using the \BdToKpi decays in order to correct for possible correlations between the individual algorithms not taken into account in the combination procedure.
For the SS\kaon case, since the small yield of \BsTopiK decays is insufficient for a reliable recalibration, the 
original calibration is kept and a systematic uncertainty is assigned. In the per-candidate method, the OS and SS combinations are further combined into a unique tagging decision and mistag probability using the calibrations determined by the simultaneous method. This combination is again recalibrated with the calibration samples. The description of the implementation of the flavour tagging into the fit models is presented in Sec.~\ref{sec:fitMethod}.

\section{Decay-time resolution}\label{sec:timeResolution}

\begin{figure}[t]
  \begin{center}
    \includegraphics[width=0.7\textwidth]{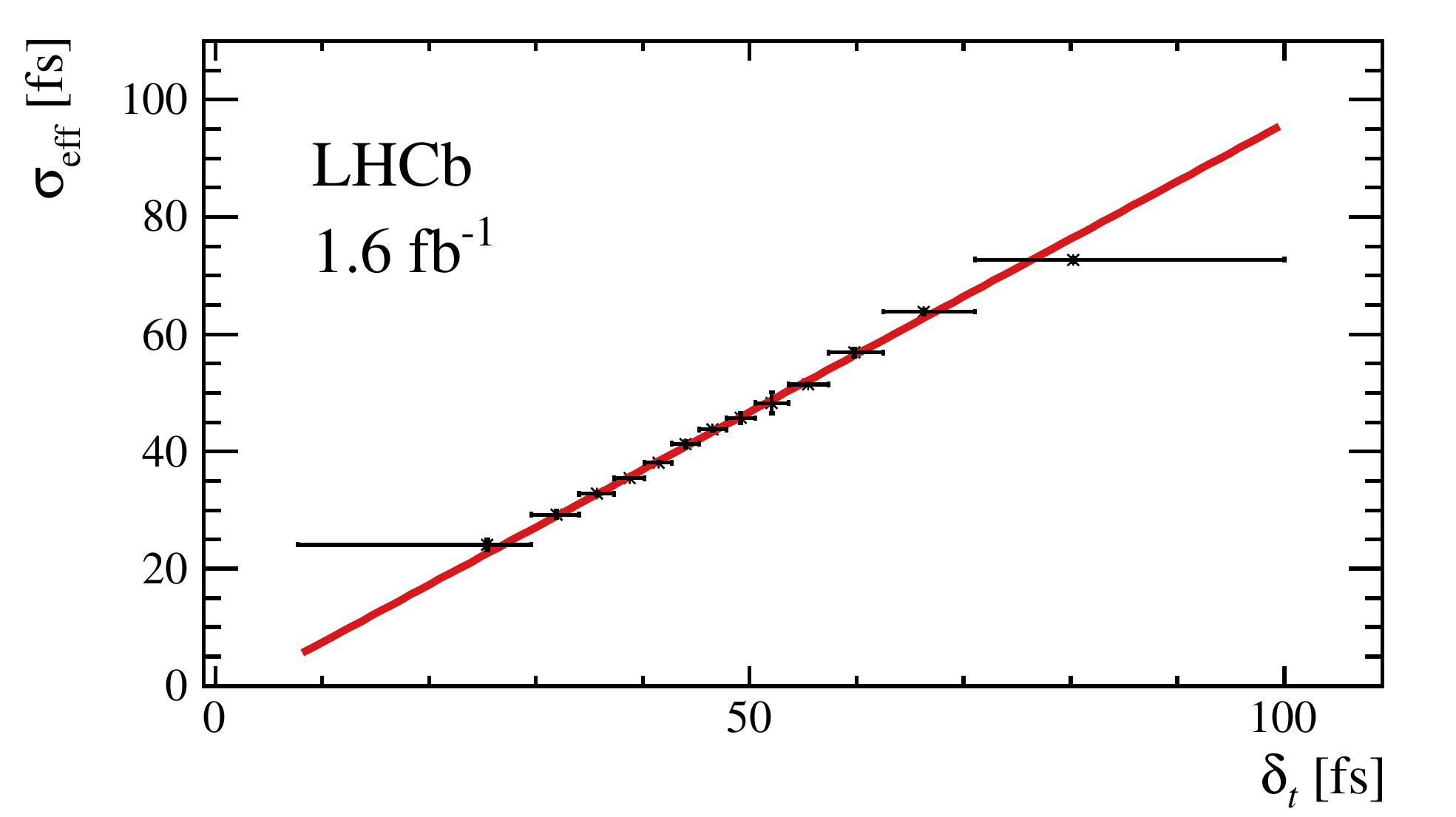}
 \end{center}
  \caption{\small Dependence of the effective decay-time resolution, $\sigma_{\rm eff}$, on the estimated decay-time uncertainty, $\delta_t$, for the background-subtracted data sample of \JpsiTomumu decays. The result of 
  a linear fit is superimposed.}
  \label{fig:resoDependence1PV}
\end{figure}

The decay-time resolution is modelled with a Gaussian function, whose mean and width are calibrated with a sample of \JpsiTomumu decays produced directly in \proton\proton collisions. The background contribution in the \jpsi sample is subtracted using the \sPlot technique~\cite{Pivk:2004ty} with the dimuon invariant mass acting as a discriminating variable. The background-subtracted sample is separated in intervals of decay-time uncertainty, $\delta_t$, which is determined for each candidate from the kinematic fit used to measure the decay time. The decay-time distribution in each bin of $\delta_t$ is fitted with a model comprising three Gaussian functions with shared mean and independent widths. According to Ref.~\cite{LHCb-PAPER-2019-013} the parameters obtained from the fits are combined into an effective resolution, $\sigma_{\rm eff}$, such that a single-Gaussian resolution model with width $\sigma_{\rm eff}$ gives the same dilution effect on the amplitude of the time-dependent asymmetry as the triple-Gaussian model. The value of $\sigma_{\rm eff}$ is calibrated assuming all the signal decays have the same mixing frequency as the \Bs meson. This assumption does not impact the analysis for \Bz mesons, since for them the effect of the decay-time resolution is negligible. Figure~\ref{fig:resoDependence1PV} shows the dependence of $\sigma_{\rm eff}$ on $\delta_t$ and is found to be well modelled with a linear function with an intercept $q_0$ and slope $q_1$. The fit is repeated for different numbers of bins of $\delta_t$, and the obtained mean values of the slope and intercept are found to be 0.94$\,\pm\,$0.02 and 1.64$\,\pm\,$1.09\,fs, respectively. Differences in the decay-time resolution between \JpsiTomumu and two-body \bquark-hadron decays are studied using samples of fully simulated \JpsiTomumu and \BsToKK decays. The calibrated decay-time resolution as a function of $\delta_t$ is
\begin{equation}\label{eq:calibratedResolution}
	\sigma_t(\delta_t) = \sigma_{\rm eff}(\delta_t)\frac{\sigma^{\Kp\Km}_{\rm eff}(\delta_t)}{\sigma^{\mup\mun}_{\rm eff}(\delta_t)},
\end{equation}
where $\sigma^{\Kp\Km}_{\rm eff}(\delta_t)$ and $\sigma^{\mup\mun}_{\rm eff}(\delta_t)$ are the effective resolution widths for the simulated \BsToKK and \JpsiTomumu decays, respectively. 

For the per-candidate method, the calibrated resolution in Eq.~\ref{eq:calibratedResolution} is applied to each candidate in the fit to the \Bs \to \Kp\Km decay-time spectrum.\footnote{A calibrated per-candidate resolution is not required for \BdTopipi decays as the \Bz oscillation is significantly slower than that of the \Bs meson.} For the simultaneous method, the decay-time resolution is not used on a per-candidate basis, but an average model is used instead. The consequence of using the average model is a small loss in the statistical precision for \CKK and \SKK, corresponding to a relative 1\% difference on the final uncertainties, while the effect on the other \CP-violation parameters is negligible. The loss is compensated by a significant simplification of the fit model, as will be discussed in detail in Sec.~\ref{sec:simFitMethod}. To obtain the average resolution, $\sigma_t(\delta_t)$ in Eq.~\eqref{eq:calibratedResolution} is integrated over the distribution of $\delta_t$ from background-subtracted \BsToKK decays, and an averaged resolution of $\hat{\sigma}_t = 42.9 \pm 0.1\fs$ is obtained. A dependence of the resolution on the decaying particle mass is found when repeating the procedure using a sample of  \decay{\OneS}{\mup\mun} decays instead of the \JpsiTomumu sample, which yields $\hat{\sigma}_t = 44.1 \pm 0.1\fs$. The average between the two calibrations, $\hat{\sigma}_t = 43.5\fs$, is used in the fit to data with the simultaneous method, and the difference between them is considered in the determination of the related systematic uncertainty.

In the fit to the \JpsiTomumu data sample, an offset of the mean of the triple-Gaussian model is observed and attributed to a misalignment in the vertex detector. The size of the bias, $\mu_t = -6.5\fs$, is used as mean value in the resolution model in both fit methods.

\section{Fitting methods}\label{sec:fitMethod}

Two independent methods, called simultaneous and per-candidate, are used to determine the \CP-violation parameters in the \BdTopipi and \BsToKK decays, while the simultaneous method also determines the direct \CP-asymmetries in \BdToKpi and \BsTopiK. A comparison of their respective results serves as validation of the measurements. The common aspects of the two methods are described in Sec.~\ref{sec:commonMethod} and~\ref{sec:signalModel}, while the specific details of each one are discussed in Sec.~\ref{sec:simFitMethod} and~\ref{sec:perEventMethod}.

\subsection{Components of the fit models}\label{sec:commonMethod}

For each component, the distributions of the final-state invariant mass, decay time and flavour-tagging assignment with the associated mistag probability are modelled for \Bds candidates. Signal components are \BdToKpi and \BsTopiK decays in the \Kpm\pimp samples, the \BdTopipi decay in the \pip\pim sample, and the \BsToKK decay in the $\Kp\!\Km$ sample. In the \pip\pim and $\Kp\!\Km$ samples, a small contribution from \BsTopipi and \BdToKK decays is present and must be taken into account. Cross-feed, combinatorial and three-body background contributions are described by the model. Apart from \B-meson decays, the only relevant source of cross-feed background is the \LbTopK decay with the proton misidentified as a kaon in the $\Kp\!\Km$ sample. Considering the PID efficiencies, the branching fractions and the relative hadronisation probabilities~\cite{HFLAV18}, the contribution of this background component is expected to be about 2.5\% relative to the \BsToKK decay and is included in the fit. Components describing partially reconstructed three-body \Bds-meson decays and combinatorial background candidates are necessary in all of the three final states. 

\subsection{Decay-time model for two-body \Bds decays}\label{sec:signalModel}

The time-dependent decay rate of a flavour-specific \decay{\B}{f} decay and of its \CP conjugate \decay{\Bb}{\overline{f}}, as for the \BdToKpi and \BsTopiK decays, is given by the probability density function (PDF)
\begin{equation}\label{eq:decayTimeB2KPI}
  \begin{split}
    T_{\rm FS}\left(t,\,\psi,\,\vec{\xi},\,\vec{\eta}\right) = & K_{\rm FS}\left(1-\psi A_{\CP}\right)\left(1-\psi A_{\rm D}\right) \times \\
    & \left\{\left[\left(1\!-\!A_{\rm P}\right)\!\Omega_{\rm sig}(t,\vec{\xi},\vec{\eta})\!+\!\left(1\!+\!A_{\rm P}\right)\!\overline{\Omega}_{\rm sig}(t,\vec{\xi},\vec{\eta})\right]\!H_{+}\left(t\right)\!+\!\right. \\
    & \left. \psi \!\left[\left(1\!-\!A_{\rm P}\right)\!\Omega_{\rm sig}(t,\vec{\xi},\vec{\eta})\!-\!\left(1\!+\!A_{\rm P}\right)\!\overline{\Omega}_{\rm sig}(t,\vec{\xi},\vec{\eta})\right]\!H_{-}\left(t\right) \!\right \}, 
  \end{split}
\end{equation}
where $K_{\rm FS}$ is a normalisation factor and the discrete variable $\psi$ assumes the value $+1$ for the final state $f$ and $-1$ for the final state $\overline{f}$. The functions $H_{\pm}$, $\Omega_{\rm sig}$ and $\overline{\Omega}_{\rm sig}$ are defined below. The direct \CP asymmetry, $A_{\CP}$, is defined in Eq.~\eqref{eq:acpDef}, while the final-state detection asymmetry, $A_{\rm D}$, and the \Bds-meson production asymmetry, $A_{\rm P}$, are defined as
\begin{equation}
  A_{\rm D} = \frac{\varepsilon_{\rm tot}\left(\overline{f}\right)-\varepsilon_{\rm tot}\left(f\right)}{\varepsilon_{\rm tot}\left(\overline{f}\right)+\varepsilon_{\rm tot}\left(f\right)},\qquad
  A_{\rm P} = \frac{\sigma_{\Bdsb}-\sigma_{\Bds}}{\sigma_{\Bdsb}+\sigma_{\Bds}},\label{eq:nuisanceAsymDef}
\end{equation}
where $\varepsilon_{\rm tot}$ is the time-integrated efficiency in reconstructing and selecting the final state $f$ or $\overline{f}$, and $\sigma_{\Bds}$ ($\sigma_{\Bdsb}$) is the production cross-section of the given \Bds (\Bdsb) meson. The asymmetry $A_{\rm P}$ arises because production rates of \Bds and \Bdsb mesons are not identical in \proton\proton collisions. It is measured to be of the order of one percent at \lhc energies~\cite{LHCb-PAPER-2016-062}. 
From the time-dependent fit it is possible to determine simultaneously $A_{\rm P}$ and the sum $A_{\CP}+A_{\rm D}$. The contribution of $A_{\rm D}$ is subtracted {\it a posteriori} as described in Sec.~\ref{sec:asymmetrydet}.

The variable $\vec{\xi}=\left(\xi_{\rm OS},\,\xi_{\rm SS}\right)$ is the pair of flavour-tagging assignments of the OS and SS algorithms used to identify the \Bds-meson flavour at production, and $\vec{\eta}=\left(\eta_{\rm OS},\,\eta_{\rm SS}\right)$ is the pair of associated mistag probabilities defined in Sec.~\ref{sec:flavourTagging}. The functions $\Omega_{\rm sig}(t,\vec{\xi},\vec{\eta})$ and $\overline{\Omega}_{\rm sig}(t,\vec{\xi},\vec{\eta})$ describe how the flavour tagging modifies the time-dependent decay rate. The functions $H_{+}\left(t\right)$ and $H_{-}\left(t\right)$ are defined as
\begin{eqnarray}
  H_{+}\left(t\right) & = & \left[e^{-\Gamma_{\dquark(\squark)} t^{\prime}}\cosh{\left(\frac{\Delta\Gamma_{\dquark(\squark)}}{2}t^{\prime}\right)}\right]\otimes R\left(t-t^{\prime}\right),\label{eq:flavourSpecificHfunctions} \\
  H_{-}\left(t\right) & = & \left[e^{-\Gamma_{\dquark(\squark)} t^{\prime}}\cos{\left(\Delta m_{\dquark(\squark)}t^{\prime}\right)}\right]\otimes R\left(t-t^{\prime}\right),\nonumber
\end{eqnarray}
where $\Gamma_\dquark$ and $\Gamma_\squark$ are the \Bz and \Bs mean decay widths, respectively, $R\left(t-t^{\prime}\right)$ is the decay-time resolution model described in Sec.~\ref{sec:timeResolution} and $\otimes$ denotes the convolution product.

In the case of a decay to a \CP eigenstate $f$, as it is for the \BdTopipi and \BsToKK decays, the decay-time PDF is given by
\begin{equation}
	\begin{split}\label{eq:decayTimeB2HH}
	T_{\CP}\left(t,\,\vec{\xi},\vec{\eta}\right) = K_{\CP} & \left\{ \left[\left(1-A_{\rm P}\right)\Omega_{\rm sig}\left(t,\vec{\xi},\vec{\eta}\right)+\left(1+A_{\rm P}\right)\overline{\Omega}_{\rm sig}\left(t,\vec{\xi},\vec{\eta}\right)\right]I_{+}\left(t\right)+\right. \\
	& \left. ~ \left[\left(1-A_{\rm P}\right)\Omega_{\rm sig}\left(t,\vec{\xi},\vec{\eta}\right)-\left(1+A_{\rm P}\right)\overline{\Omega}_{\rm sig}\left(t,\vec{\xi},\vec{\eta}\right)\right]I_{-}\left(t\right) \right\},
	\end{split}
\end{equation}
where $K_{\CP}$ is a normalisation factor and the functions $I_{+}\left(t\right)$ and $I_{-}\left(t\right)$ are
\begin{eqnarray}
			I_{+}\left(t\right) & = & \left\{e^{-\Gamma_{\dquark(\squark)} t^{\prime}}\left[\cosh{\left(\frac{\Delta\Gamma_{\dquark(\squark)}}{2}t^{\prime}\right)}+A_{f}^{\Delta\Gamma}\sinh{\left(\frac{\Delta\Gamma_{\dquark(\squark)}}{2}t^{\prime}\right)}\right]\right\}\otimes R\left(t-t^{\prime}\right), \\
			I_{-}\left(t\right) & = & \left\{e^{-\Gamma_{\squark(\squark)} t^{\prime}}\left[C_f\cos{\left(\Delta m_{\dquark(\squark)}t^{\prime}\right)}-S_f\sin{\left(\Delta m_{\dquark(\squark)}t^{\prime}\right)}\right]\right\}\otimes R\left(t-t^{\prime}\right). \nonumber
\end{eqnarray}
In this case $f$ is equal to $\overline{f}$, hence the final-state detection asymmetry $A_{\rm D}$ is zero. The parameters $\Delta m_{\dquark(\squark)}$, $\Gamma_{\dquark(\squark)}$, and $\Delta\Gamma_{\dquark(\squark)}$ are fixed in the fit to data to the values reported in Table~\ref{tab:lifetimeParameters}. 
\begin{table}[!htbp]
  \caption{ \small Values of the parameters \dmd, \dms, \Gd~\cite{HFLAV18}, \Gs and \DGs~\cite{LHCb-PAPER-2019-013} used in the two methods. For \Gs and \DGs the correlation factor, $\rho$, between the two quantities is also reported. The decay width difference \DGd is fixed to zero.}
  \begin{center}
    \begin{tabular}{l|cccc}
      \hline
      Parameter        & Value \\
      \hline
      \dmd             & $\phantom{0}0.5065 \pm 0.0019\invps$ \\
      \Gd              & $\phantom{0}0.6579 \pm 0.0017\invps$ \\
      \DGd             & $\hspace{0.2cm}0\invps$ \\
      \dms             & $17.757\phantom{0} \pm 0.021\invps\phantom{0}$ \\
      \Gs              & $\phantom{0}0.6562 \pm 0.0021\invps$ \\
      \DGs             & $\phantom{0}0.082 \pm 0.005\invps$ \\
      $\rho(\Gs,\DGs)$ & \hspace{-0.2cm}$-0.170$ \\
      \hline
    \end{tabular}
  \end{center}
  \label{tab:lifetimeParameters}
\end{table}

\subsection{Simultaneous fit method}\label{sec:simFitMethod}

The simultaneous method relies on a concurrent fit to all the final-state samples (\pip\pim, \Kp\!\Km and \Kpm\pimp), modelling the multidimensional space defined by the final-state invariant mass, \Bds decay time, flavour-tagging decision and associated mistag probability for the signal and background components. The models used in the fit are a modification of those described in Ref.~\cite{LHCb-PAPER-2018-006}.

The model describing the invariant-mass shape of the signal components comprises a sum of two Gaussian functions and a Johnson function~\cite{JohnsonFunc}, while the model for cross-feed background is based on a kernel estimation (KDE) method~\cite{Cranmer:2001aa} and tuned with simulated decays. The normalisation of each cross-feed background component is determined by rescaling the yields of the decay reconstructed with the correct mass hypothesis by the ratio between the misidentification probability and the  PID efficiency for the wrong and correct mass hypotheses.

The decay-time model of the signal components is also used for the cross-feed background components originating from the signal decays reconstructed with the wrong mass hypothesis. This is valid under the assumption that the decay-time calculated under the wrong mass hypothesis is equal to that calculated using the correct hypothesis, and is verified using samples of simulated decays. 
The flavour-tagging assignments and related mistag probabilities for OS and SS taggers enter the time-dependent decay rates of Eqs.~\eqref{eq:decayTimeB2KPI} and~\eqref{eq:decayTimeB2HH} through the functions $\Omega_{\rm sig}(t,\vec{\xi},\vec{\eta})$ and $\overline{\Omega}_{\rm sig}(t,\vec{\xi},\vec{\eta})$. These functions are the same as already used in Ref.~\cite{LHCb-PAPER-2018-006} with the only difference being that they now depend on the decay time, as do the efficiencies of the SS taggers. This dependence is accommodated using separate efficiencies: one independent of the SS-tagger decision and one specific for the candidates tagged by the SS taggers. More details are reported in App.~\ref{sec:flavourTaggingAppendix}. 

The decay-time efficiency, $\varepsilon_{\rm sig}\left(t\right)$, is sculpted by the selection criteria presented in Sec.~\ref{sec:selection}. It is parameterised using an empirical function determined using the \BdToKpi calibration decay, whose time-dependent decay rate is independent of the flavour-tagging decision and described by an exponential distribution with  $\Gd=0.6588\pm0.0017\invps$~\cite{HFLAV18}. A sample of background-subtracted \BdToKpi candidates is obtained from the \Kpm\pimp sample in the invariant-mass window $5.23 < m(\Kpm\pimp) < 5.32~\gevcc$. The contributions of the combinatorial background, the only non-negligible background in this region, is subtracted by injecting, with negative weights, candidates from the sideband $m(\Kpm\pimp) > 5.6\gevcc$. As explained above, the procedure is repeated for the subsample with $\xi_{\rm SS}\neq 0$, in order to model the time dependence of the SS-tagging efficiency. For the \BdTopipi and \BsToKK decays, a small correction is applied to the efficiency in order to take into account the differences between signal and calibration modes. The correction for a given mode is a product of the efficiency determined from the \BdToKpi data and the ratio between the efficiencies of this mode and of the \BdToKpi decay, as determined from simulation. 

The final difference with respect to the model used in Ref.~\cite{LHCb-PAPER-2018-006} is that the decay-time resolution is no longer modelled on a per-candidate basis. This change is made since a correlation between the distributions of the decay-time and decay-time error is observed for the combinatorial background candidates. A full description of this correlation would imply a considerable complication of the fitting model that outweighs the small loss in statistical power that the use of an average decay-time resolution implies. A systematic uncertainty is established in order to cover for possible biases coming from using an average rather than per-candidate decay-time resolution.

The invariant-mass model for the combinatorial background components for each decays is an exponential function, with its slope depending on the decay time, in order to take into account a slight correlation between invariant mass and decay time observed in the high-mass sideband. The time dependence of the slope is studied using a two-dimensional unbinned maximum-likelihood fit to the invariant mass and decay time of the sample in the high-mass sideband above 5.6\gevcc, where only combinatorial background candidates contribute. The obtained time-dependent mass slope is used for the combinatorial background model in the entire invariant-mass window, going from 5.0 to 6.2\gevcc. The relative normalisation of each candidate in the sideband is scaled to reproduce that in the total invariant-mass window. A KDE method is applied to the weighted candidates and the output is used to model the decay-time shape of the combinatorial-background component. A dependence of the decay-time shape of combinatorial background candidates on the tagging assignment of the OS- and SS-taggers is also observed. Hence the time dependence of the mass slope is studied separately for the subsamples corresponding to the tagging decision $(|\xi_{\rm OS}|,\,|\xi_{\rm SS}|) = \left\{(1,1),(1,0),(0,1),(0,0)\right\}$. Different weights are determined for each subsample, and also the KDE method is applied separately to each of them. 
The weighting procedure is the same as employed for the background subtraction used to study the decay-time efficiency for \Bds decays. 
The functions taking into account the flavour-tagging assignment and mistag probabilities are the same used in Ref~\cite{LHCb-PAPER-2018-006}, but are generalised to consider all the possible combinations of $(|\xi_{\rm OS}|,\,|\xi_{\rm SS}|)$. Finally, in the case of the \Kpm\pimp samples, possible asymmetries in the flavour-tagging or reconstruction efficiencies for the two charge-conjugate final states are taken into account. 

The invariant-mass model of partially reconstructed \Bds decays is the same as that used in Ref.~\cite{LHCb-PAPER-2018-006}, comprising the sum of two Gaussian functions, which are defined using the same parameters as in the signal model and are convolved with ARGUS functions~\cite{Albrecht:1994aa}. For the \Kpm\pimp sample two three-body background components are used: one describing three-body \Bd and \Bp decays and another describing three-body \Bs decays. For the \pip\pim and $\Kp\!\Km$ samples a single ARGUS component is found to be sufficient to describe the invariant-mass shape in the low-mass region. The shape of the decay-time distribution is obtained by applying a KDE method to the candidates in the low-mass sideband below $5.2\gevcc$, after subtracting the combinatorial background contribution, as explained above. This is repeated separately for the candidates with $|\xi_{\rm SS}| = 0$ and $|\xi_{\rm SS}| \neq 0$, since a difference in the decay-time shape is observed in data for the two subsamples. The functions used to take into account the flavour-tagging information are the same as used for the combinatorial background model, but with independent parameters. Also for this component possible differences in flavour-tagging and reconstruction efficiencies between the \Kp\pim and \pip\Km final states are taken into account in the same way as used for the combinatorial background model.

\subsection{Per-candidate fit method}\label{sec:perEventMethod}

The per-candidate method relies on independent fits to the \pip\pim and \Kp\Km samples with all background components statistically subtracted using the \sFit technique~\cite{Pivk:2004ty,Xie:2009rka} with the \pip\pim and \Kp\Km  invariant mass as the discriminating variable. Hence only the decay-time distributions are modelled for the signal modes \BdTopipi and \BsToKK.

The invariant-mass distributions of the \BsToKK and \BdTopipi signal components are modelled with the sum of two Crystal Ball functions~\cite{Skwarnicki:1986xj} where the tail parameters are fixed to the values obtained from the simulation. The mean and width of the Gaussian core are allowed to vary in the fit for the \BsToKK and \BdTopipi signal modes, while these parameters are constrained for the \BdToKK and \BsTopipi signal components using the known mass difference between \Bd and \Bs and the ratio of resolutions obtained from simulations, respectively. 
The decay-time model for the signal components is the function described in Sect.~\ref{sec:signalModel}, multiplied on a per-candidate basis with the acceptance functions described below.

The invariant-mass distributions of the misidentified background candidates from other two-body \Bds decays are modelled with templates from simulations and their yields are constrained using efficiencies measured in data calibration samples. The three-body background components, which are the same as in the simultaneous method,  are modelled using an exponentially modified Gaussian PDF.

The decay-time resolution consists of a single Gaussian function with its width varying candidate by candidate,  depending on the decay-time error $\delta_t$ for each candidate and calibrated as presented in Sec.~\ref{sec:timeResolution}.  
The per-candidate acceptance function is determined with the swimming method~\cite{Bailey:1985zz,Rademacker:2005ay,Gligorov:2018lb,LHCb-PAPER-2011-014} by artificially changing the decay time of the \Bds meson and re-evaluating whether the candidate would have been accepted by the selection requirements that are known to bias the decay-time measurement. The decay time is changed by moving the position of every PV in the event along the direction of the \Bds momentum vector. For decay times for which the candidate is accepted the efficiency is 1, otherwise the efficiency is 0. By scanning a range of hypothetical decay times, a series of top-hat functions are constructed for each candidate as it changes from being rejected, to being accepted, finally to being rejected again.\footnote{A series of top hat functions are produced as each event can have more than one primary vertex.} The procedure is re-evaluated in steps of 50\mum along the \Bds momentum vector and, when the selection decision changes, the position at which this change occurs is determined with a finer granularity, giving an overall resolution of 0.5\mum on the decay-time efficiency. The effective lifetime measured on the fully simulated \BsToKK events, assuming an exponential decay-time model and using only the swimming-based efficiency for this simulation, is found to be 1.416~\ps. Compared to a generated effective lifetime of 1.394~\ps it exhibits a bias of $1.5\%$. 
This arises from effects that are not fully modelled in the swimming method and can result in an incorrect measurement of the parameter \ADGKK, for which high precision is expected. To correct for this, an additional decay-time efficiency weight is applied by comparing the decay-time efficiency extracted using the swimming method for the \BdToKpi data with the decay-time efficiency determined from the ratio of background-subtracted \BdToKpi events and the unbiased decay-time PDF. The unbiased decay-time PDF consists of an exponential function, whose decay time is fixed to the known \Bd lifetime, convolved with a Gaussian function to account for the intrinsic decay-time resolution. The width of the Gaussian is fixed to the effective decay-time resolution as detailed in Sec. \ref{sec:timeResolution}. The ratio of these efficiencies is modelled with an empirical function 
\begin{equation}
f(t) = p_0( 1 + \tanh[ p_1( t - p_2) ] ) + p_3t.
\label{eq:acceptance_ratio}	
\end{equation}
where $t$ is the decay time of the candidate and $p_{\{0,1,2,3\}}$ are free parameters measured in the fit to the ratio. Applying this weight to the swimming-based efficiency allows to correctly recover the effective lifetime of the simulated \BsToKK decays and the mean lifetime of \mbox{\BdToKpi} decays extracted from the \Kpm\pimp samples.

\section{Detection asymmetry between \boldmath{\Km\pip} and \boldmath{\Kp\pim} final states}\label{sec:asymmetrydet}

In order to extract  the \CP asymmetries \ACPBd~and \ACPBs from the asymmetries measured through the simultaneous fit, an estimation of the nuisance 
experimental detection asymmetry is required as indicated in Eq.~\ref{eq:decayTimeB2KPI}.
This asymmetry is a consequence of the different efficiency for selecting the \BdToKpi and \BsTopiK decays and their charge-conjugate final states.
To an excellent approximation, it can be expressed as the sum of two contributions
\begin{equation}\label{eq:rawAsymCorrection}
  A_{\rm D} = A^{\kaon\pion}_{\rm det} + A^{\kaon\pion}_{\rm PID},
\end{equation}
where $A^{\kaon\pion}_{\rm det}$ is the asymmetry between the selection efficiencies without the 
application of the PID requirements and $A^{\kaon\pion}_{\rm PID}$ is the asymmetry between the efficiencies of the PID requirements selecting the two final states. The convention used in the following to determine $A^{\kaon\pion}_{\rm det}$ and $A^{\kaon\pion}_{\rm PID}$ is such that a positive value of the asymmetry means a larger efficiency for the \Km\pip pair with respect to the \Kp\pim pair. As a consequence of this convention, the values reported below for the \Bd and \Bs asymmetries must be used with an inverted sign for the \BsTopiK decay.

The final-state detection asymmetry is determined using $\Dp \to \Km \pip \pip$ and $\Dp \to \Kzb \pip$ control modes, with the 
neutral kaon decaying to \pip\pim, following the strategy used in Ref.~\cite{LHCb-PAPER-2018-006}. Assuming negligible \CP 
violation in these Cabibbo-favoured \D-meson decays, the raw asymmetries between the measured yields of \Dp and \Dm decays can be written as
\begin{eqnarray}\label{eq:rawAsymKpipi}
A_{\rm RAW}^{\kaon\pion\pion} & = & A_{\rm P}^{\Dp}+A_{\rm det}^{\kaon\pion}+A_{\rm det}^{\pion}, \label{eq:rawAsymKpipi1} \\
A_{\rm RAW}^{\Kzb\pion} & = & A_{\rm P}^{\Dp}+A_{\rm det}^{\pion}-A_{\rm det}^{\Kz},\label{eq:rawAsymKpipi2}
\end{eqnarray}
where $A_{\rm P}^{\Dp}$ is the asymmetry between the production cross-sections of \Dp and \Dm mesons, and $A^{\pion}_{\rm det}$ 
($A^\Kz_{\rm det}$) is the asymmetry between the detection efficiencies of \pip (\Kz) and \pim (\Kzb) mesons. 
The difference between Eqs.~\eqref{eq:rawAsymKpipi1} and~\eqref{eq:rawAsymKpipi2} leads to
\begin{equation}\label{eq:adKPI}
A_{\rm det}^{\kaon\pion} = A_{\rm RAW}^{\kaon\pion\pion}-A_{\rm RAW}^{\Kzb\pion}-A_{\rm det}^{\Kz}.
\end{equation}
The asymmetry $A_{\rm det}^{\Kz}$ includes the effects from the kaon mixing and \CP violation, and was estimated to be $\left(0.054 \pm 0.014\right)\%$~\cite{LHCb-PAPER-2014-013}. 
The asymmetries $A_{\rm P}^{\Dp}$ and $A_{\rm det}^{\pion}$ can depend on the kinematics of the \Dp and \pip mesons. To obtain a
better cancellation of these nuisance asymmetries in Eq.~\eqref{eq:adKPI}, the momentum and the transverse momentum of the 
\Dp and \pip mesons from the $\Dp \to \Km \pip \pip$ sample are simultaneously weighted to match the corresponding 
distributions in the $\Dp \to \Kzb \pip$ sample. The
$A_{\rm det}^{\kaon\pion}$ is determined in intervals of the kaon momentum, to account for the kinematic-dependent variation of the interaction cross-sections of positive and negative kaons with the detector material. This binned asymmetry is averaged over the momentum distribution of the kaon in the \BdToKpi and \BsTopiK decays, giving no difference between the absolute values of the corrections for the two modes. 
\begin{figure}[h]
  \begin{center}
    \includegraphics[width=0.45\textwidth]{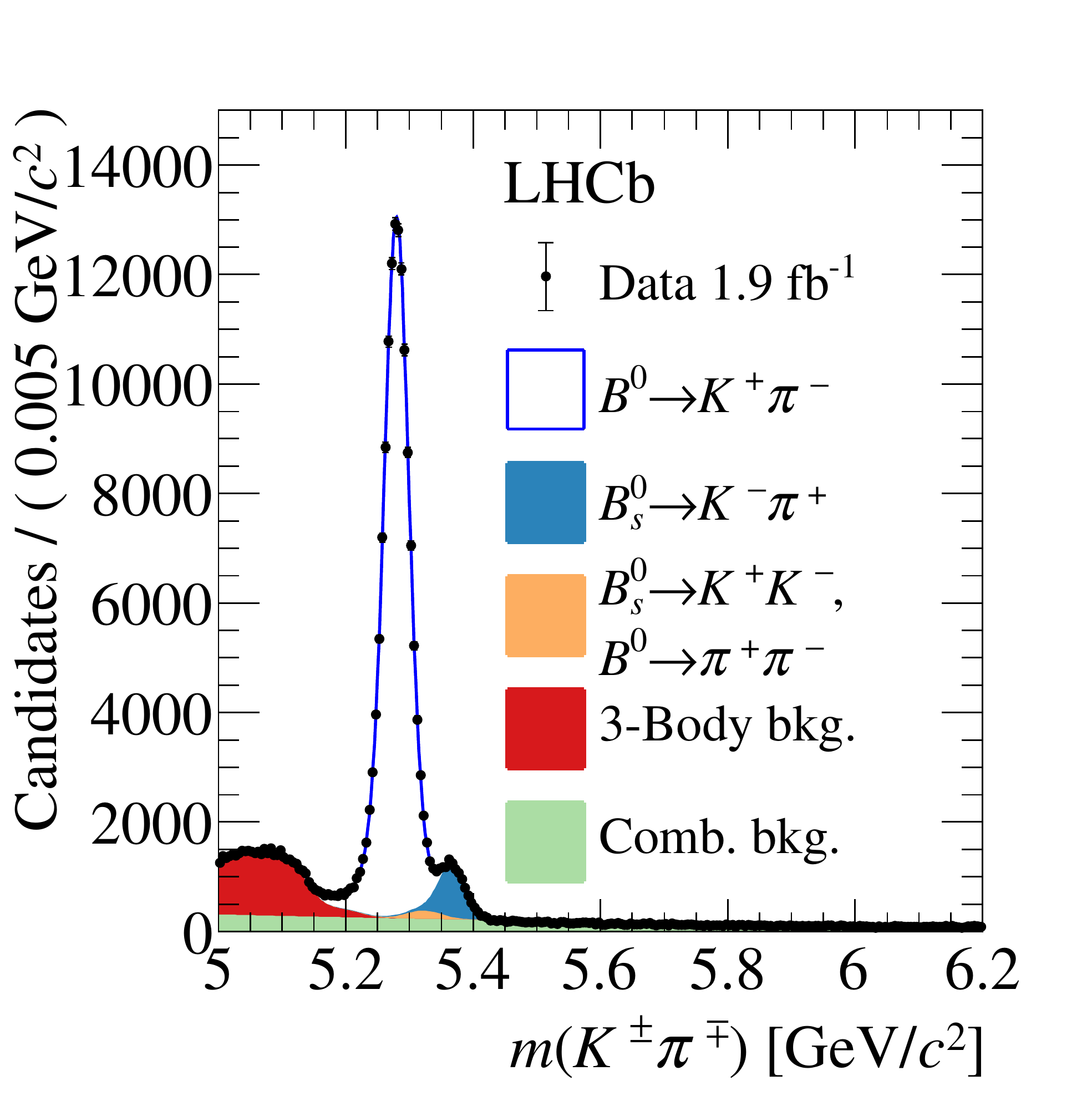}
    \includegraphics[width=0.45\textwidth]{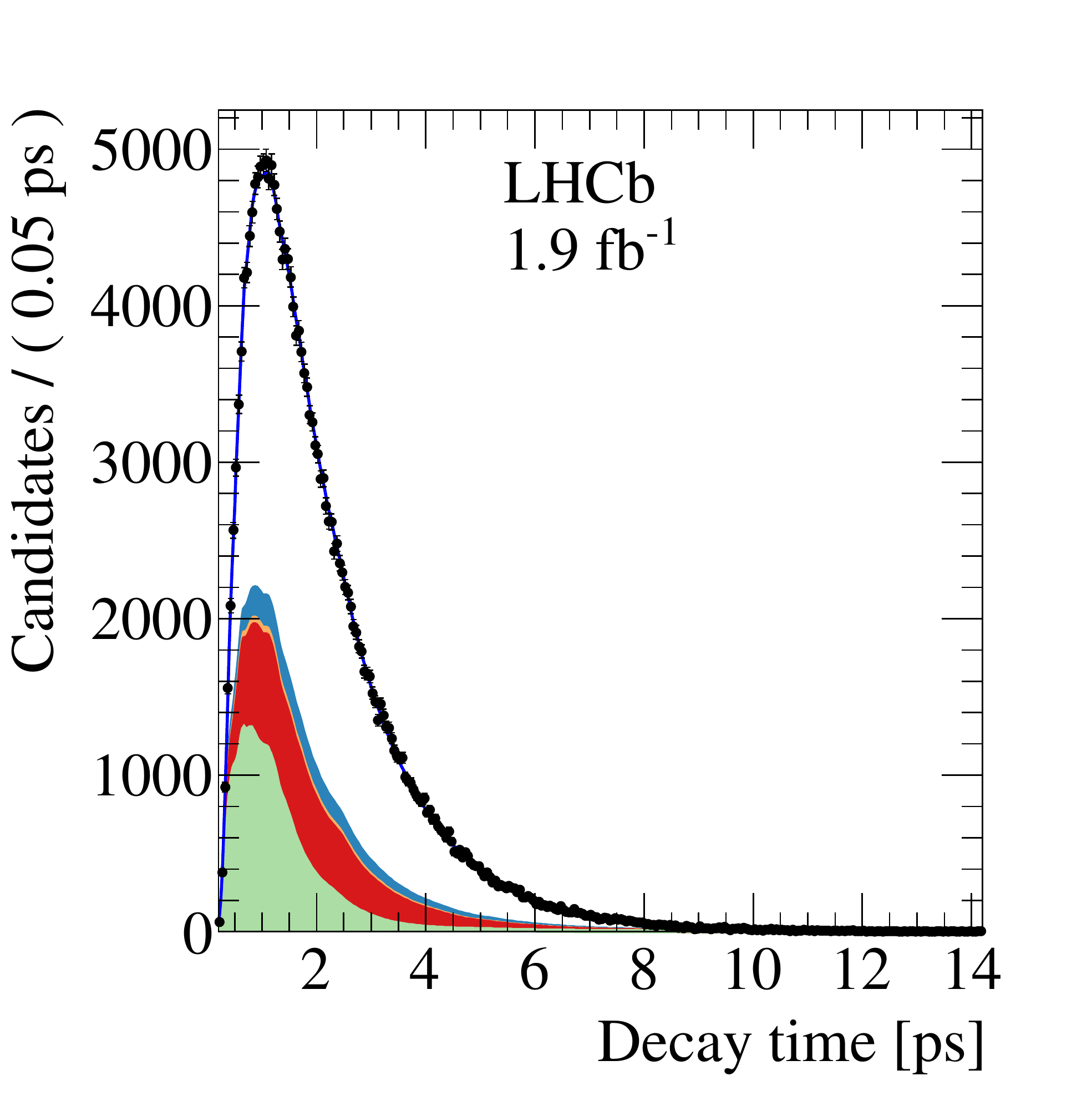}
    \includegraphics[width=0.45\textwidth]{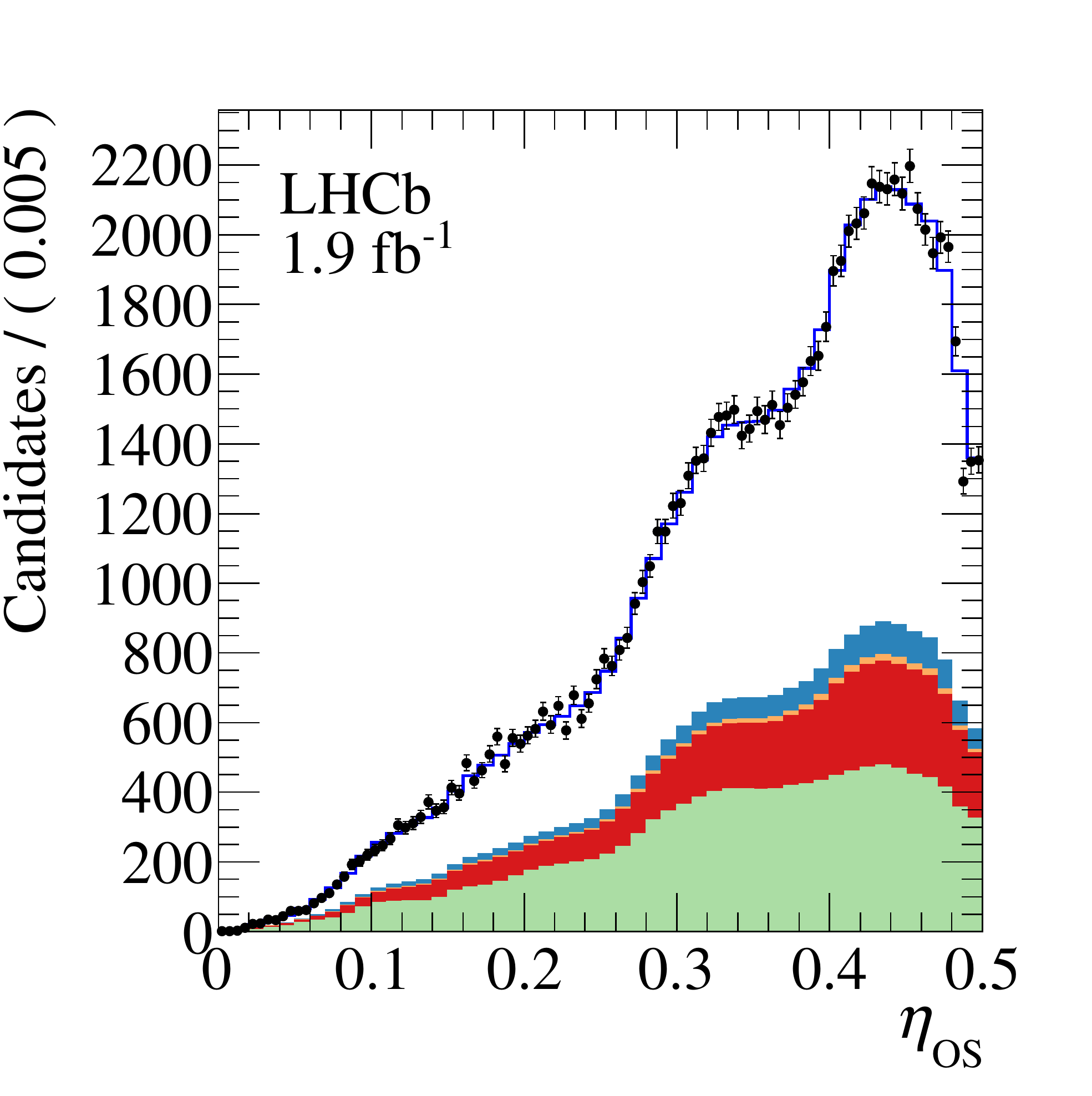}
    \includegraphics[width=0.45\textwidth]{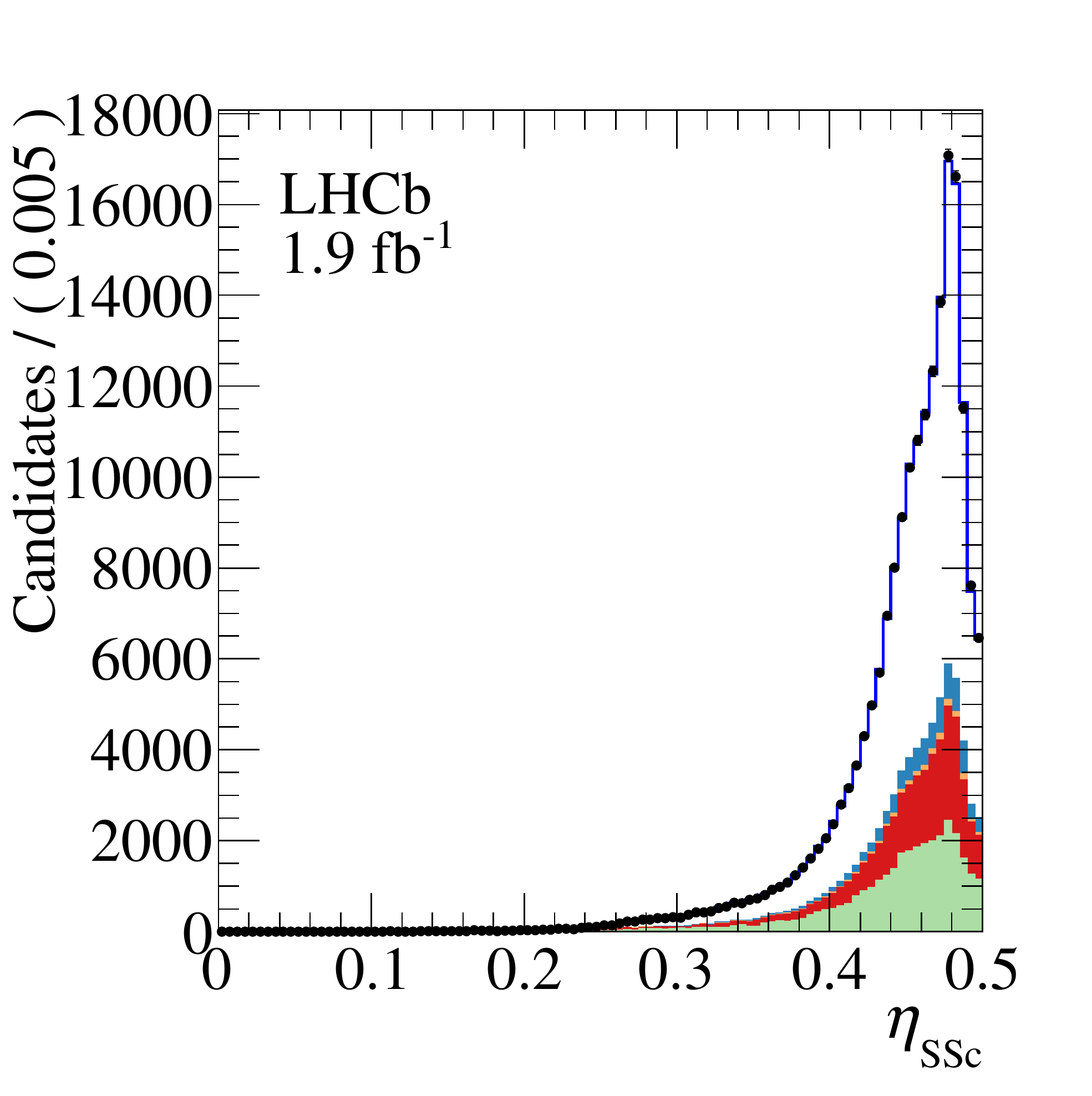}
\end{center}
  \vspace{-0.7cm}
  \caption{\small Distributions of (top left) \Kpm\pimp invariant mass, (top right) \Bds decay time, mistag fractions (bottom left) $\eta_{\rm OS}$ and (bottom right) $\eta_{\rm SSc}$ for \Kpm\pimp candidates. The result of the simultaneous fit is overlaid. The various components contributing to the fit model are drawn as stacked histograms.}
  \label{fig:plotsKPI}
\end{figure}
The final-state detection asymmetry values for the 2015 and 2016 data samples are
\begin{eqnarray}
	A_{\rm det}^{\kaon\pion}(2015) & = & \left(-0.96 \pm 0.32\right)\%, \label{eq:adkpiBd2KPI} \\
    A_{\rm det}^{\kaon\pion}(2016) & = & \left(-1.05 \pm 0.13\right)\%. \nonumber
\end{eqnarray}

The asymmetry between the PID efficiencies is computed in intervals of momentum, pseudorapidity and azimuthal angle of the two final-state particles, using the \decay{\Dstarp}{\Dz(\Km\pip)\pip} 
calibration samples, as discussed in Sec.~\ref{sec:selection}.
The computation is repeated using several binning schemes, and then the average and standard deviation of the PID asymmetries determined in each scheme are used as the central value and associated uncertainty for $A^{\kaon\pion}_{\rm PID}$, respectively.
\begin{figure}[h]
  \begin{center}
   \includegraphics[width=0.45\textwidth]{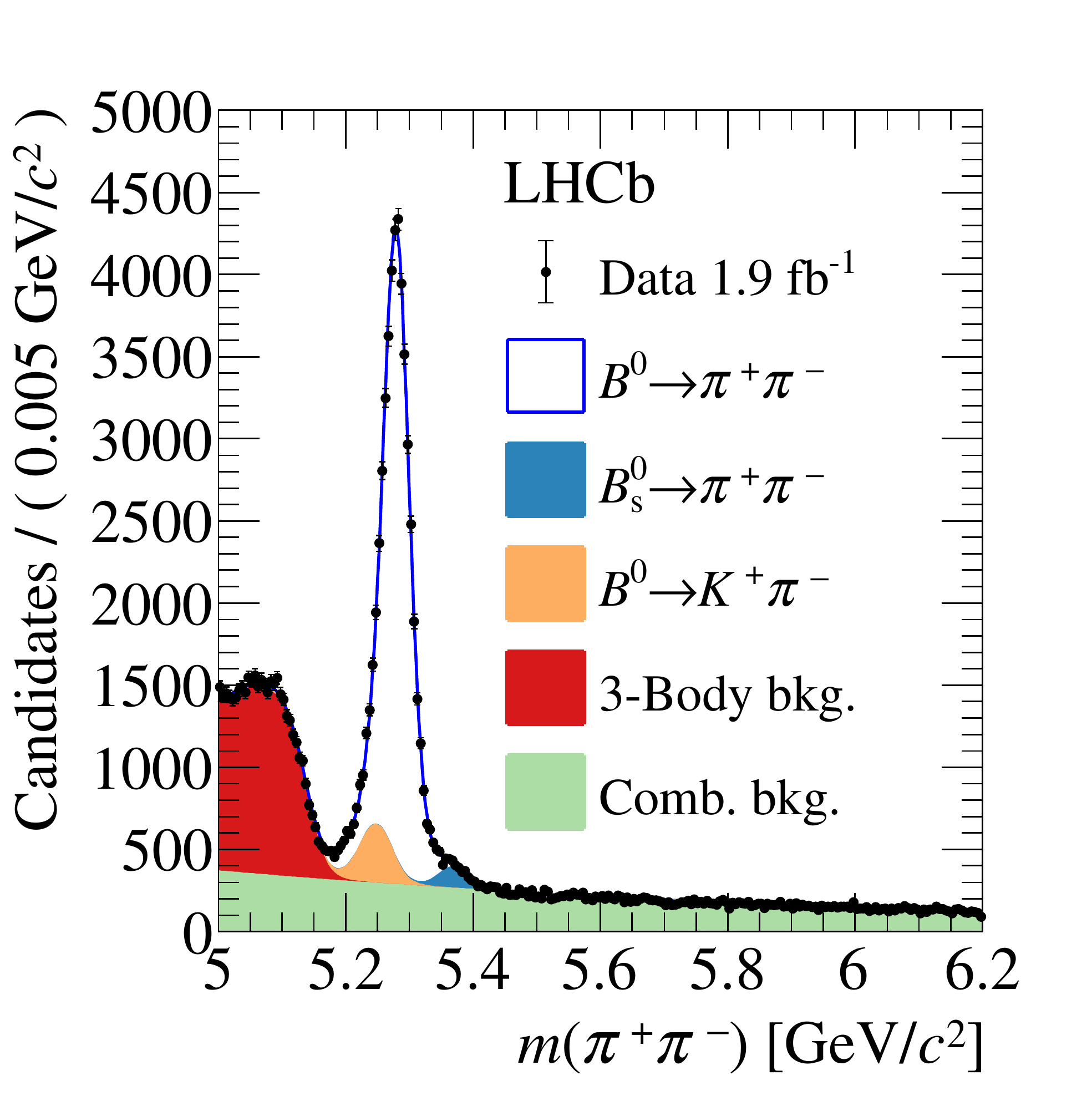}
   \includegraphics[width=0.45\textwidth]{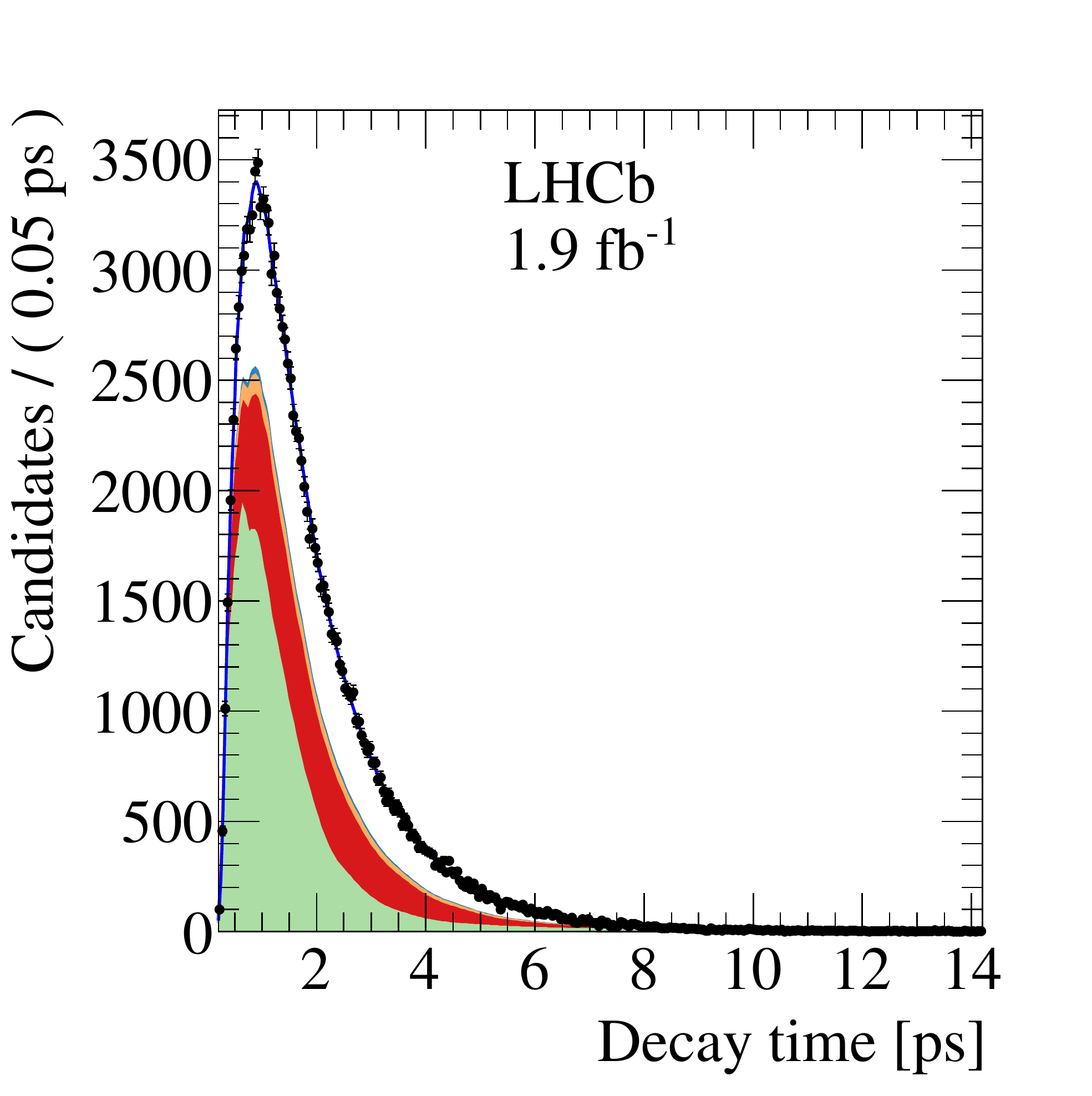}
   \includegraphics[width=0.45\textwidth]{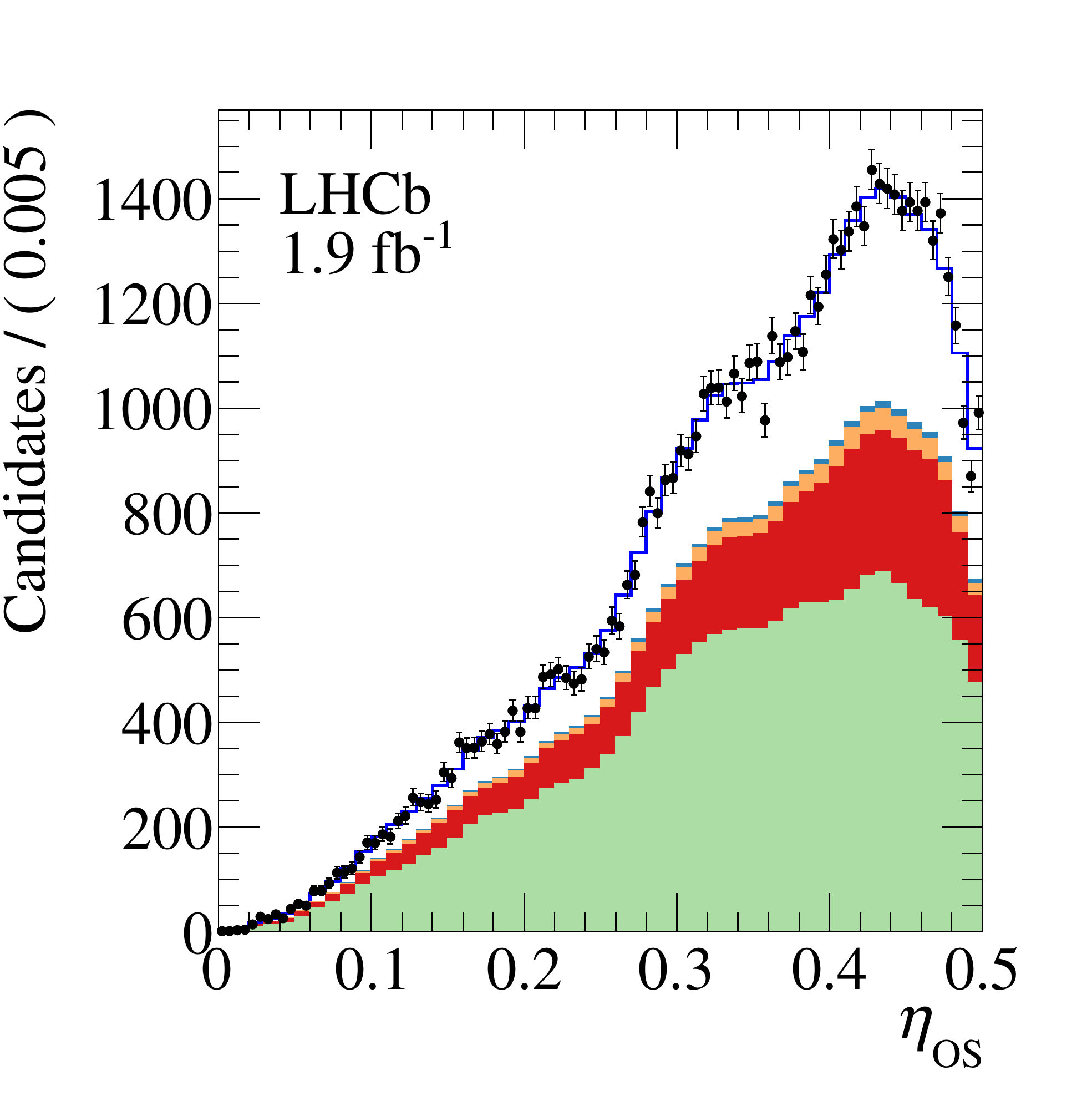}
   \includegraphics[width=0.45\textwidth]{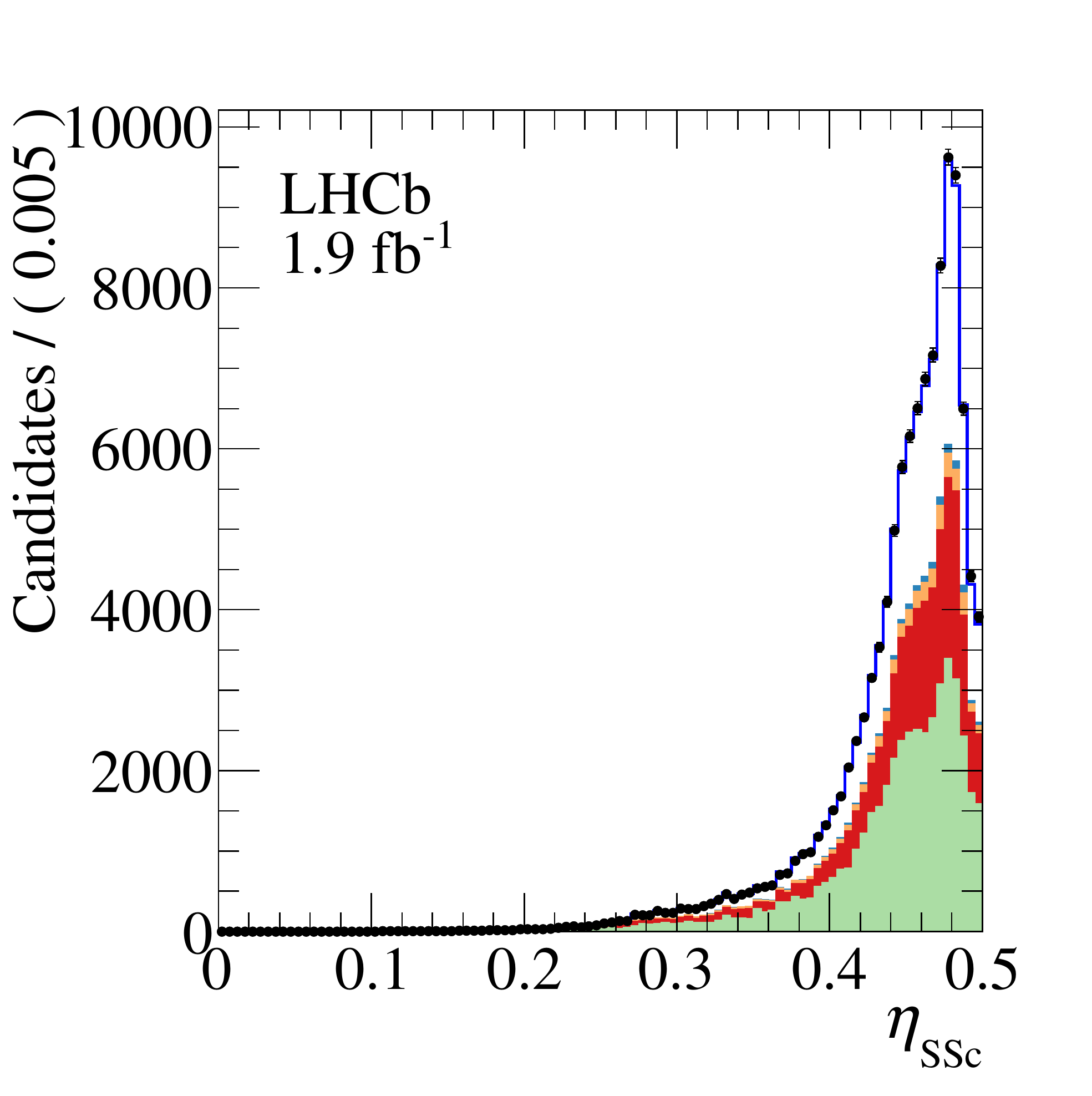}
 \end{center}
  \vspace{-0.7cm}
  \caption{\small Distributions of (top left) \pip\pim invariant mass, (top right) \Bds decay time, mistag fractions (bottom left) $\eta_{\rm OS}$ and (bottom) $\eta_{\rm SSc}$ for \pip\pim candidates. The result of the simultaneous fit is overlaid. The various components contributing to the fit model are drawn as stacked histograms.}
  \label{fig:plotsPIPI}
\end{figure}
The PID asymmetry is calculated taking into account the differences in the running conditions of the two years of data taking and the numerical results are:
\begin{eqnarray}
	A^{\kaon\pion}_{\rm PID}\left(2015\right) & = & \left(-1.2 \pm 0.7\right)\%, \label{eq:apidB2KPI} \\
	A^{\kaon\pion}_{\rm PID}\left(2016\right) & = & \left(0.5 \pm 0.3\right)\%.\nonumber
\end{eqnarray}

\section{Fit results}\label{sec:fitResult}

The results obtained from unbinned maximum likelihood fits to data of the models described in Secs.~\ref{sec:fitMethod} are presented in the following. Their comparison is also discussed.

\subsection{Simultaneous method}

\begin{figure}[h]
  \begin{center}
    \includegraphics[width=0.45\textwidth]{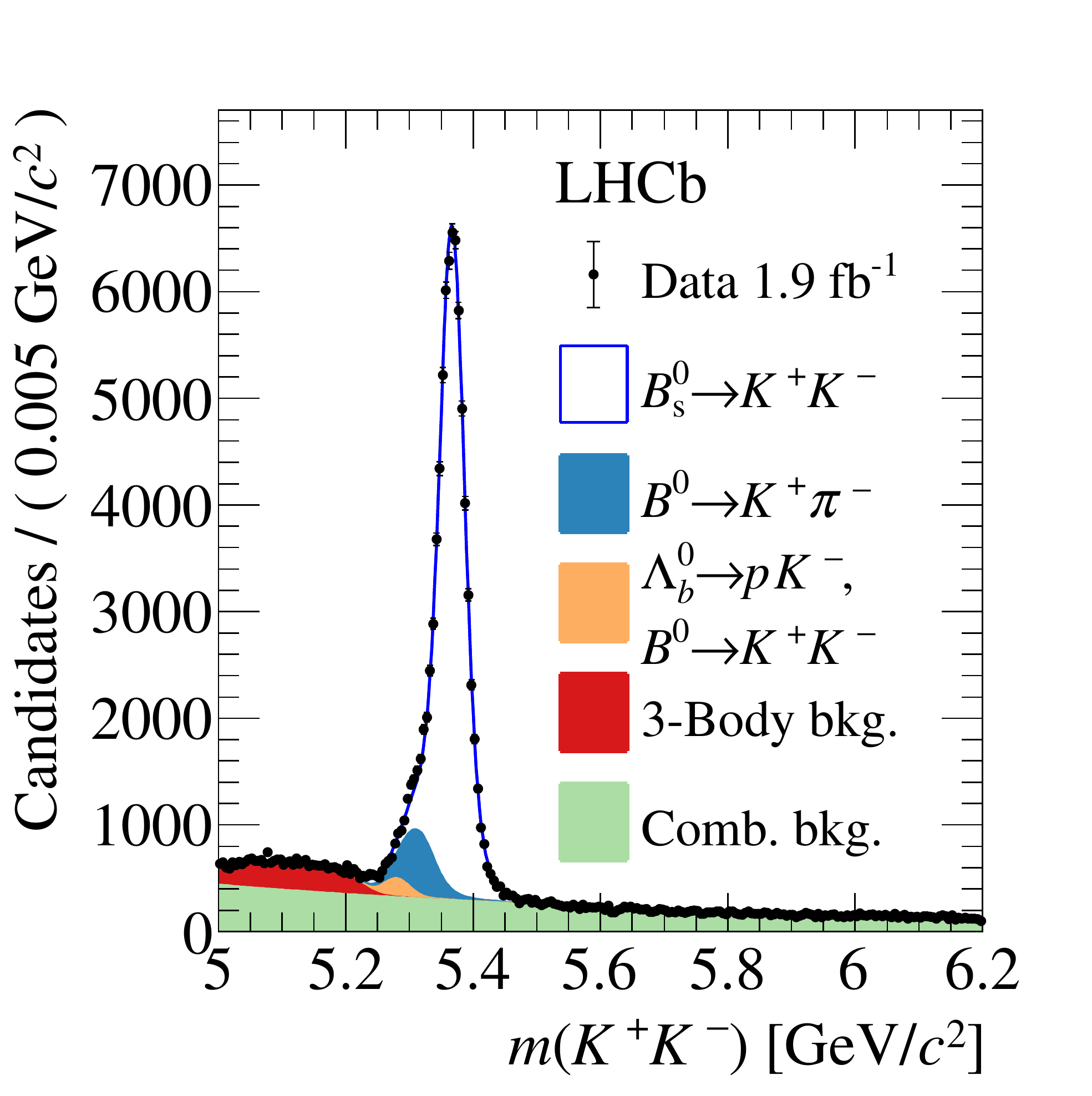}
    \includegraphics[width=0.45\textwidth]{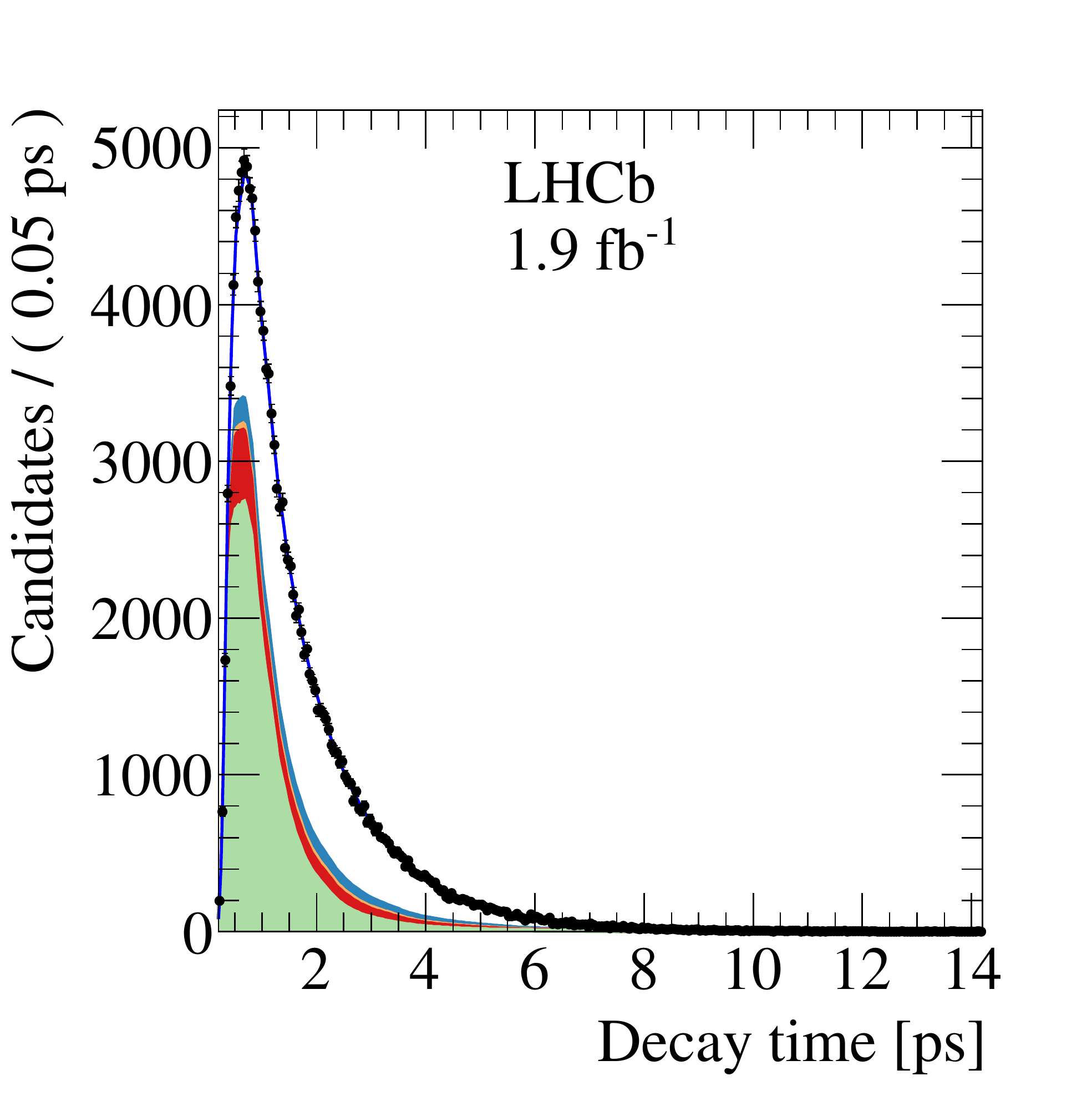}
    \includegraphics[width=0.45\textwidth]{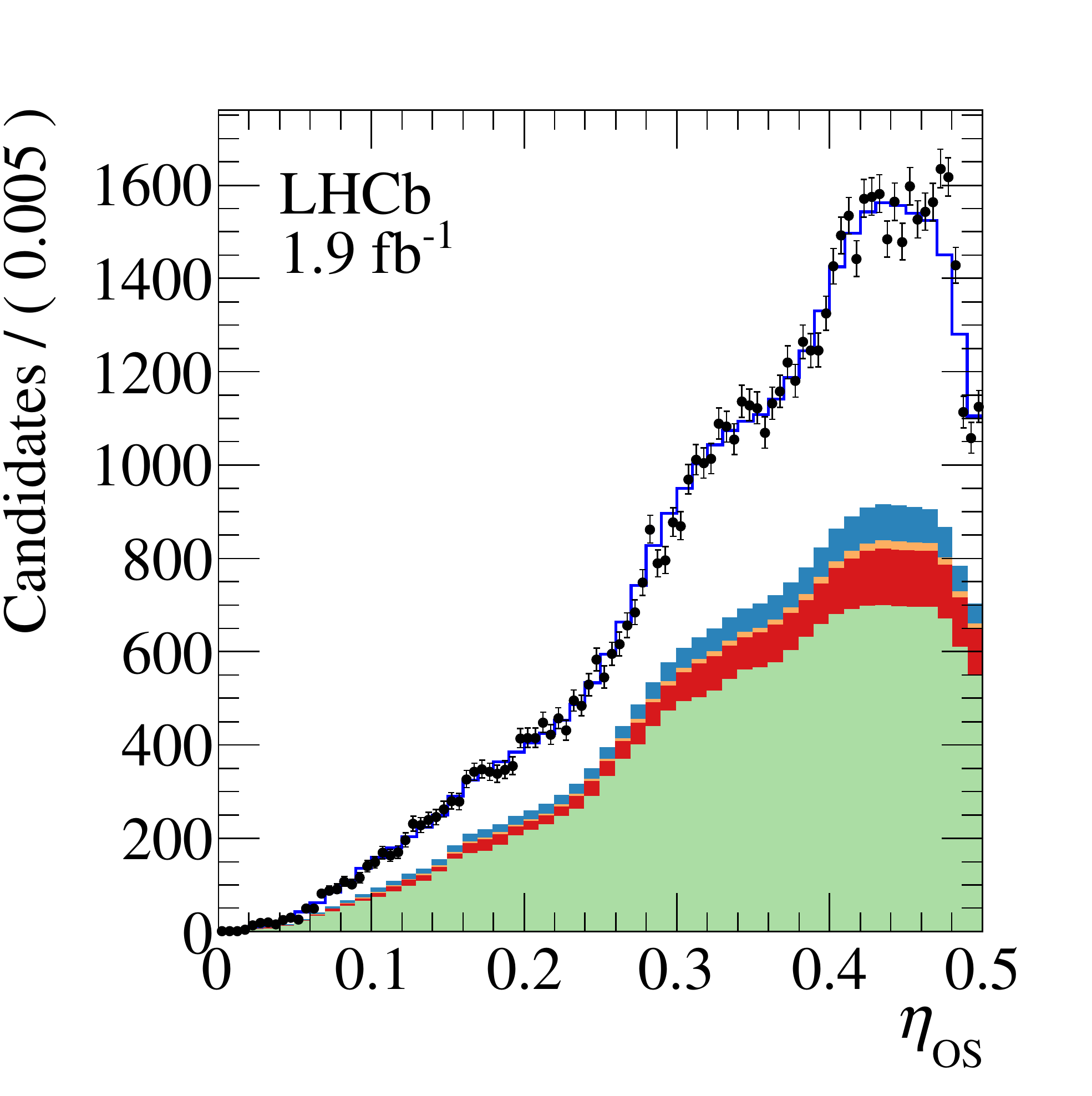}
    \includegraphics[width=0.45\textwidth]{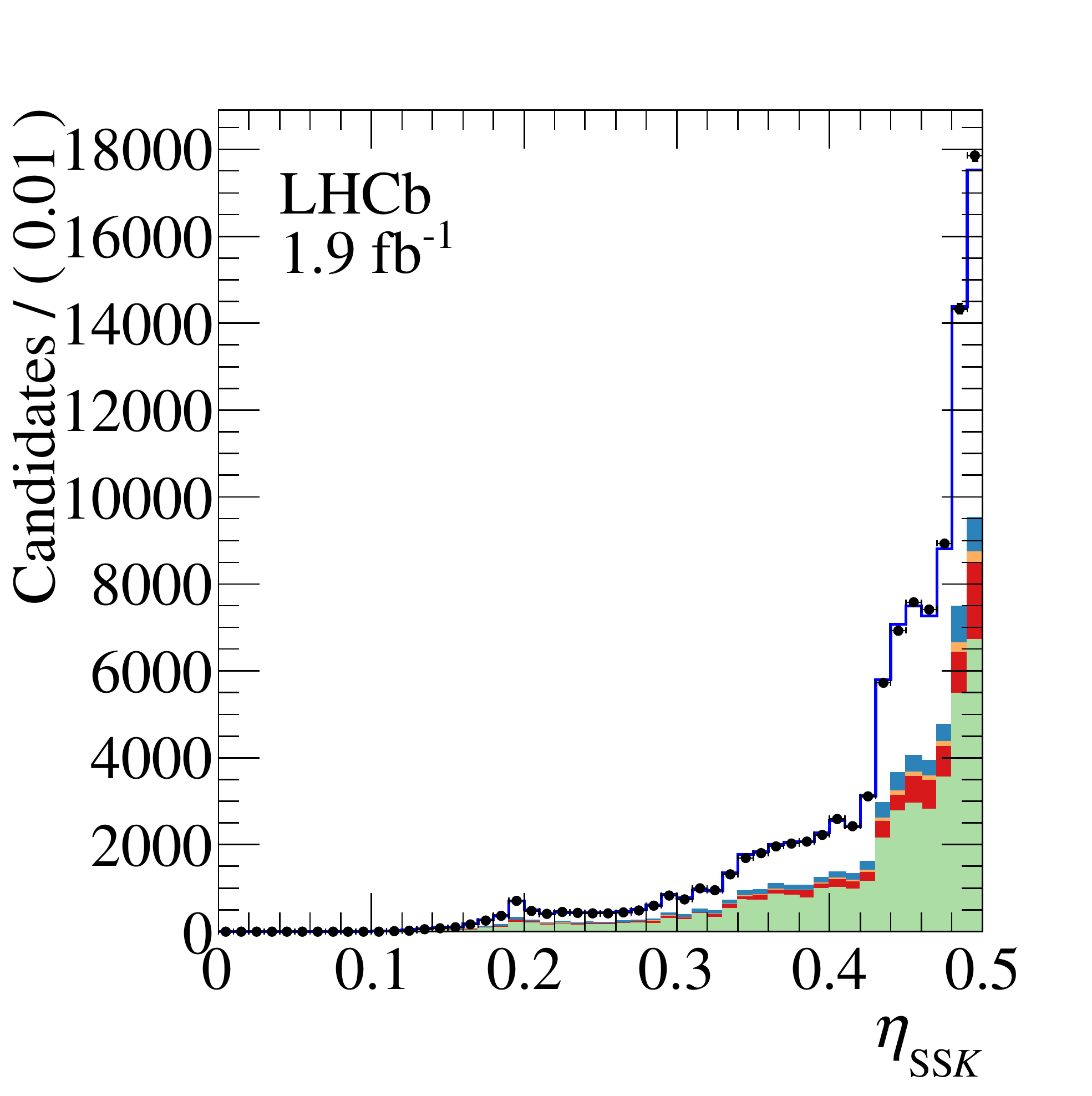}
 \end{center}
  \vspace{-0.7cm}
  \caption{\small Distributions of (top left) $\Kp\!\Km$ invariant mass, (top right) \Bds decay time, mistag fractions (bottom left) $\eta_{\rm OS}$ and (bottom right) $\eta_{{\rm SS}\kaon}$ for $\Kp\!\Km$ candidates. The result of the simultaneous fit is overlaid. The various components contributing to the fit model are drawn as stacked histograms.}
  \label{fig:plotsKK}
\end{figure}
 
The simultaneous fit to the final-state invariant mass, the \Bds decay time, and the tagging decisions and their associated mistag probabilities of the \pip\pim, $\Kp\!\Km$ and \Kpm\pimp samples determines the coefficients \Cpipi, \Spipi, \CKK, \SKK, \ADGKK~and the \CP asymmetries \ACPBd~and \ACPBs. The signal yields are $N(\BdTopipi) = 45\hspace{0.5mm}620 \pm 260$, $N(\BsToKK) = 70\hspace{0.5mm}310 \pm 320$, $N(\BdToKpi) = 140\hspace{0.5mm}340 \pm 420$ and $N(\BsTopiK) = 10\hspace{0.5mm}580 \pm 150$, where uncertainties are statistical only.
\begin{figure}[h]
  \begin{center}
	\includegraphics[width=0.45\textwidth]{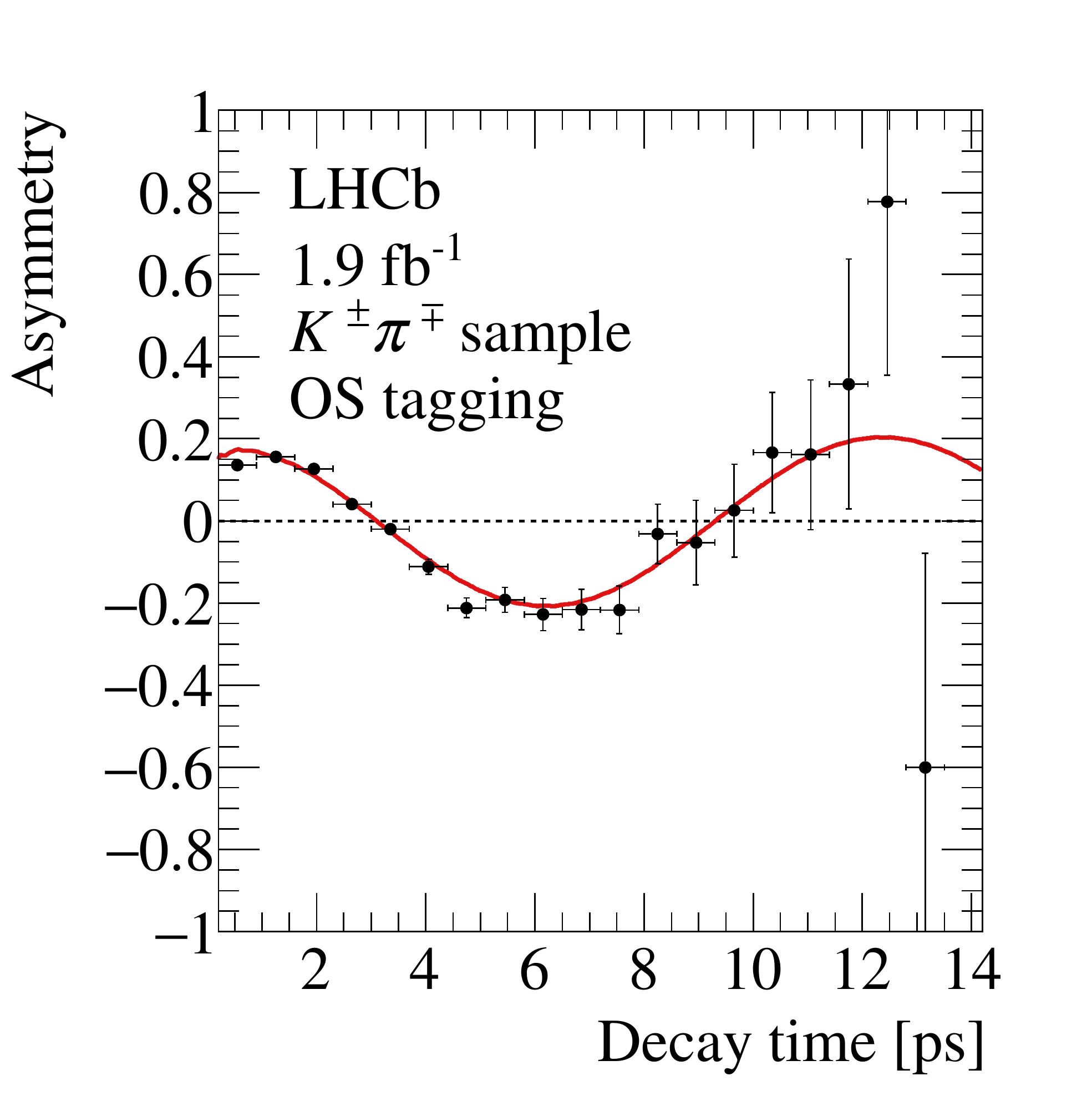}
	\includegraphics[width=0.45\textwidth]{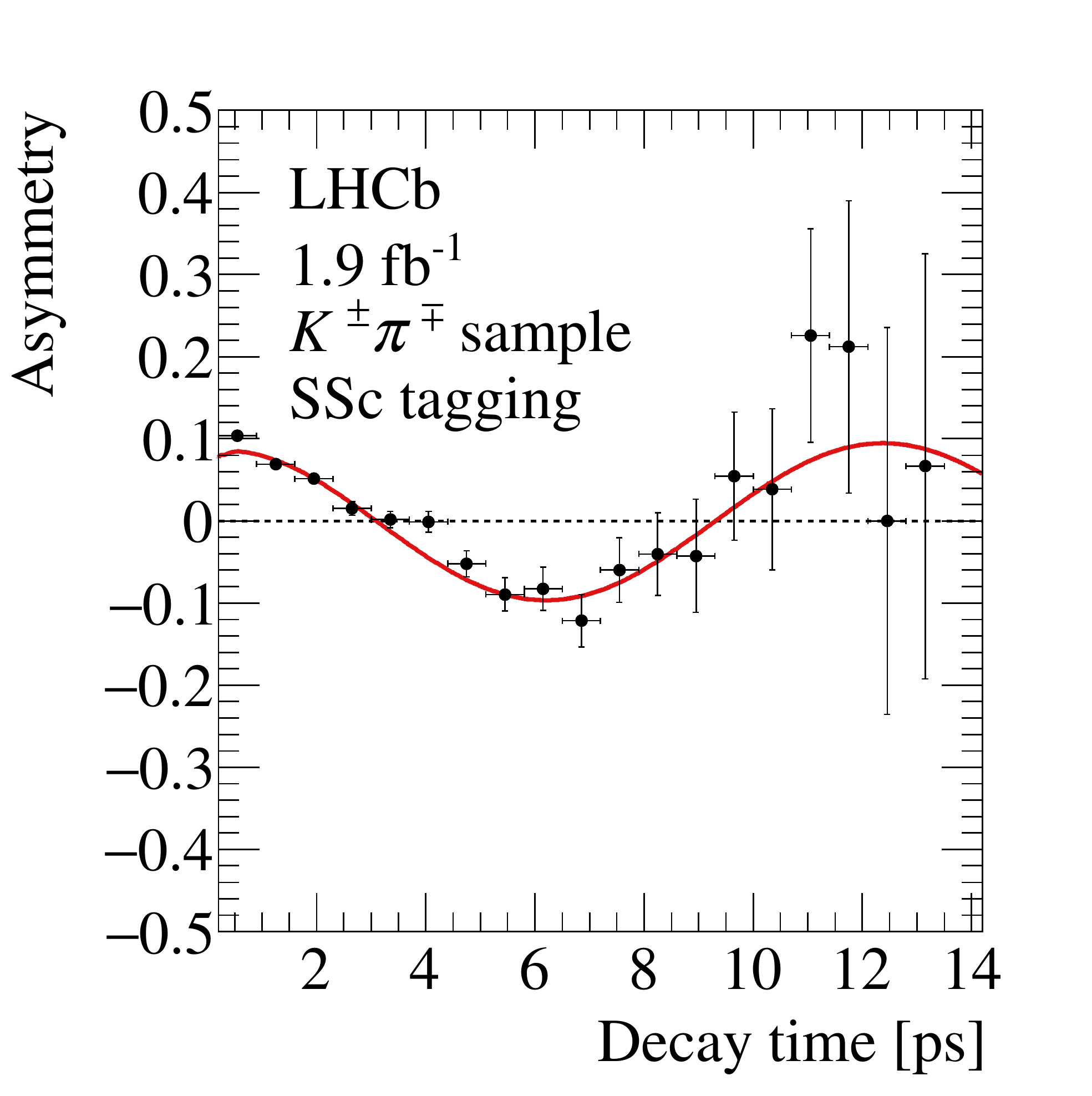}
 \end{center}
  \caption{\small Time-dependent asymmetries for \Kpm\pimp candidates with $5.20 < m(\Kpm\pimp) < 5.32\gevcc$: (left) using the OS-tagging decision and (right) the SSc-tagging decision. The result of the simultaneous fit is overlaid.}
  \label{fig:rawAsymmetryKPI}
\end{figure}
The distributions of the mass and decay time of the selected candidates are shown in Figs.~\ref{fig:plotsKPI}, \ref{fig:plotsPIPI} and~\ref{fig:plotsKK}, for the \Kpm\pimp, \pip\pim and \Kp\Km samples, respectively.
The time-dependent asymmetries, obtained separately by using the OS or the SS tagging decisions, for the \Bds candidates in the region $5.20 < m(\Kpm\pimp) < 5.32\gevcc$, dominated by the \BdToKpi decay, are shown in Fig.~\ref{fig:rawAsymmetryKPI}. 
\begin{figure}[h]
  \begin{center}
    \includegraphics[width=0.45\textwidth]{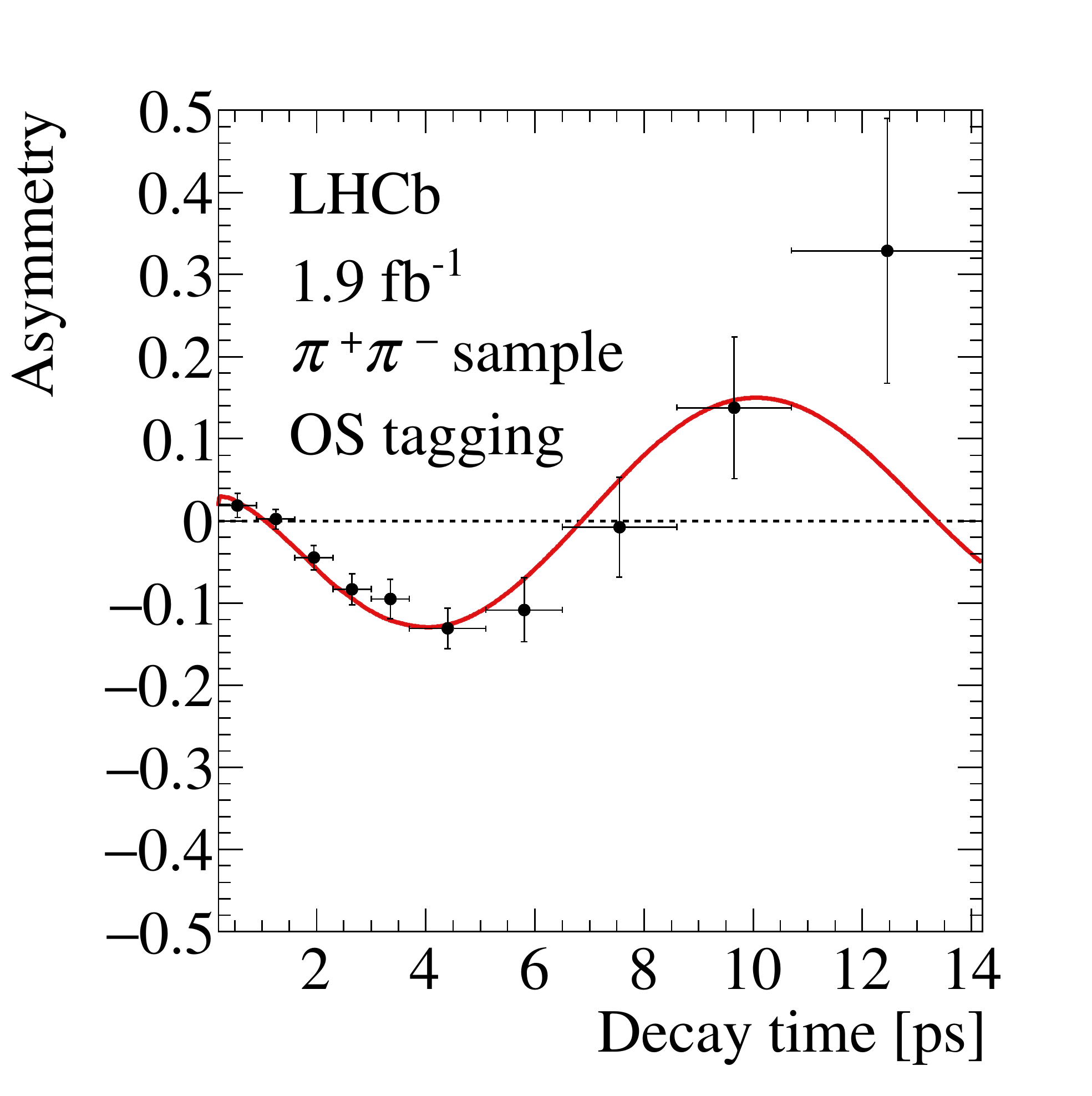}
    \includegraphics[width=0.45\textwidth]{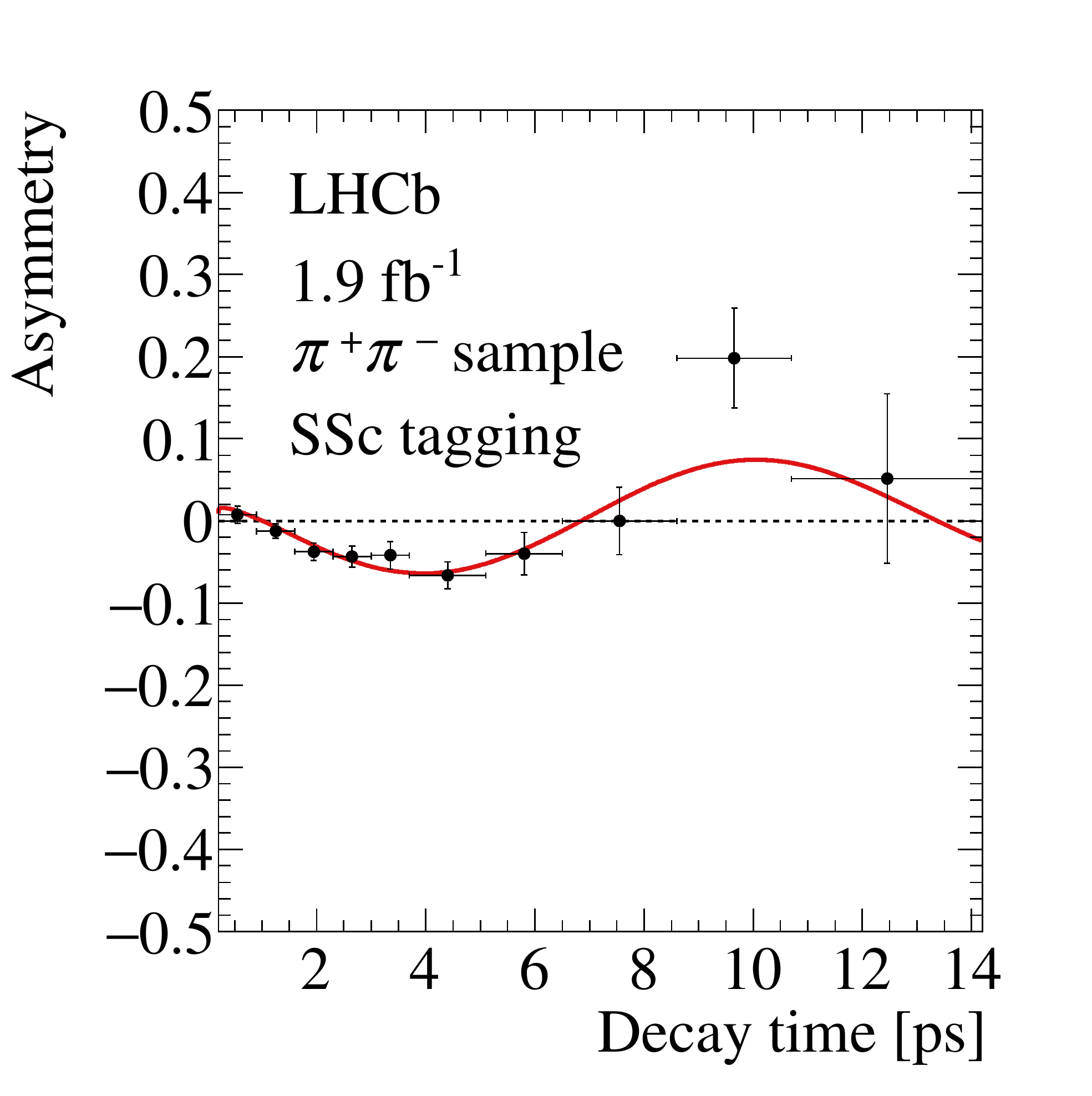}
    \includegraphics[width=0.45\textwidth]{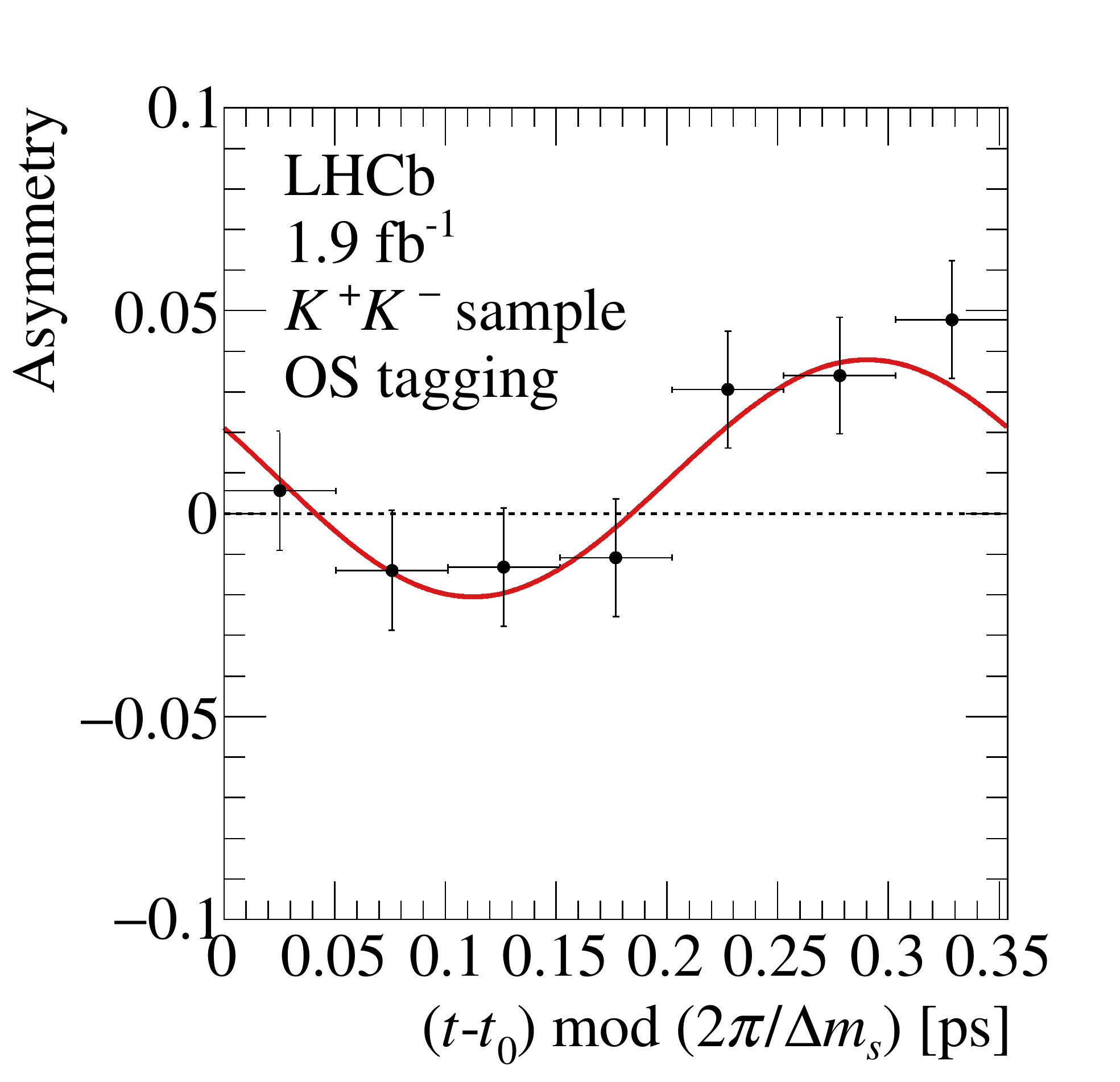}
    \includegraphics[width=0.45\textwidth]{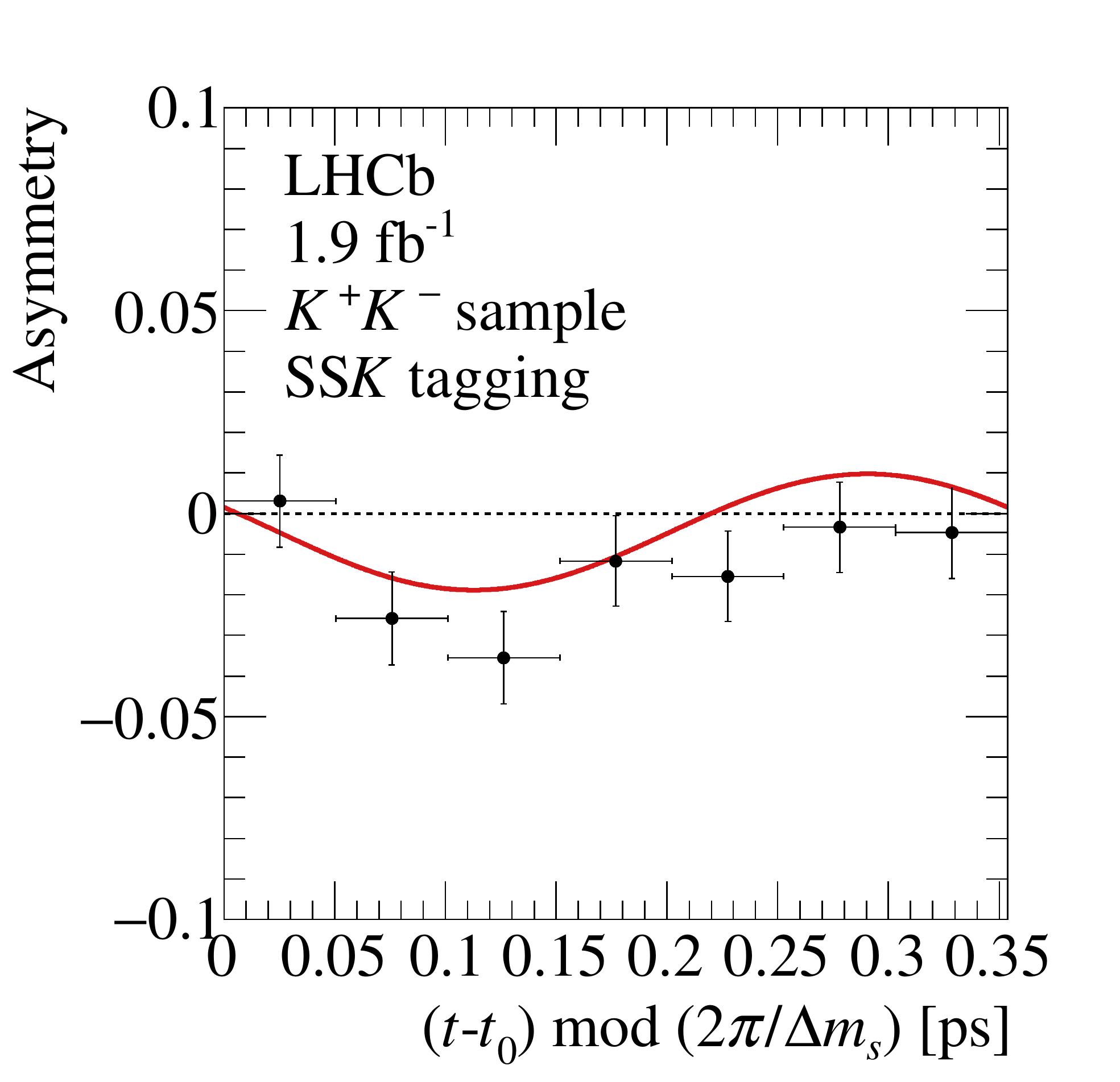}
 \end{center}
  \caption{\small Time-dependent asymmetries for (top) \pip\pim and (bottom) $\Kp\!\Km$ candidates with $5.20 < m(\pip\pim) < 5.35\gevcc$ and $5.30 < m(\Kp\Km) < 5.44\gevcc$, respectively: (left) using the OS-tagging decision and (right) using either the SSc-tagging decision (for the \pip\pim candidates) or the SS\kaon-tagging decision (for the $\Kp\!\Km$ candidates). The result of the simultaneous fit is overlaid. The asymmetry for the $\Kp\!\Km$ candidates is folded into one mixing period $2\pi/\dms$ and the parameter $t_0 = 0.2\ps$ corresponds to the minimum value of the decay-time used in the fit.}
  \label{fig:rawAsymCPEigenstate}
\end{figure}
The production asymmetries for the \Bd and \Bs mesons are determined to be $(-0.60 \pm 0.49)\%$ and $(-1.2 \pm 1.5)\%$, respectively, where uncertainties are statistical only. They are consistent with the expectations from Ref.~\cite{LHCb-PAPER-2016-062}. The time-dependent asymmetries for the \pip\pim candidates with $5.20 < m(\pip\pim) < 5.35\gevcc$, and for the $\Kp\!\Km$ candidates with $5.30 < m(\Kp\Km) < 5.45\gevcc$, dominated by the corresponding \BdTopipi and \BsToKK signal components, are shown in Fig.~\ref{fig:rawAsymCPEigenstate}, again separately for the OS and SS tagging decision. The effective tagging powers for the \BdTopipi and \BsToKK decays are $\left(4.5 \pm 0.2\right)\%$ and $\left(5.1 \pm 0.2\right)\%$, respectively. 
The results for the \CP-violating quantities are
\begin{eqnarray}
  \Cpipi  & = & -0.311\phantom{0} \pm 0.045, \nonumber\\
  \Spipi  & = & -0.706\phantom{0} \pm 0.042, \nonumber\\ 
  \ACPBd & = & -0.0824 \pm 0.0033,  \nonumber\\
  \ACPBs & = & \phantom{-}0.236\phantom{0} \pm 0.013, \label{eq:MiBo-Results}\\
  \CKK    & = & \phantom{-}0.164\phantom{0} \pm 0.034,\nonumber \\
  \SKK    & = & \phantom{-}0.123\phantom{0} \pm 0.034, \nonumber\\
  \ADGKK  & = & -0.833\phantom{0} \pm 0.054,\nonumber  
\end{eqnarray}
where the uncertainties are statistical, and the central values of \ACPBd~and \ACPBs~are corrected for the \Kp\pim detection and PID asymmetry.

\subsection{Per-candidate method}

\begin{figure}[htb]
  \begin{center}
    \includegraphics[width=0.45\textwidth]{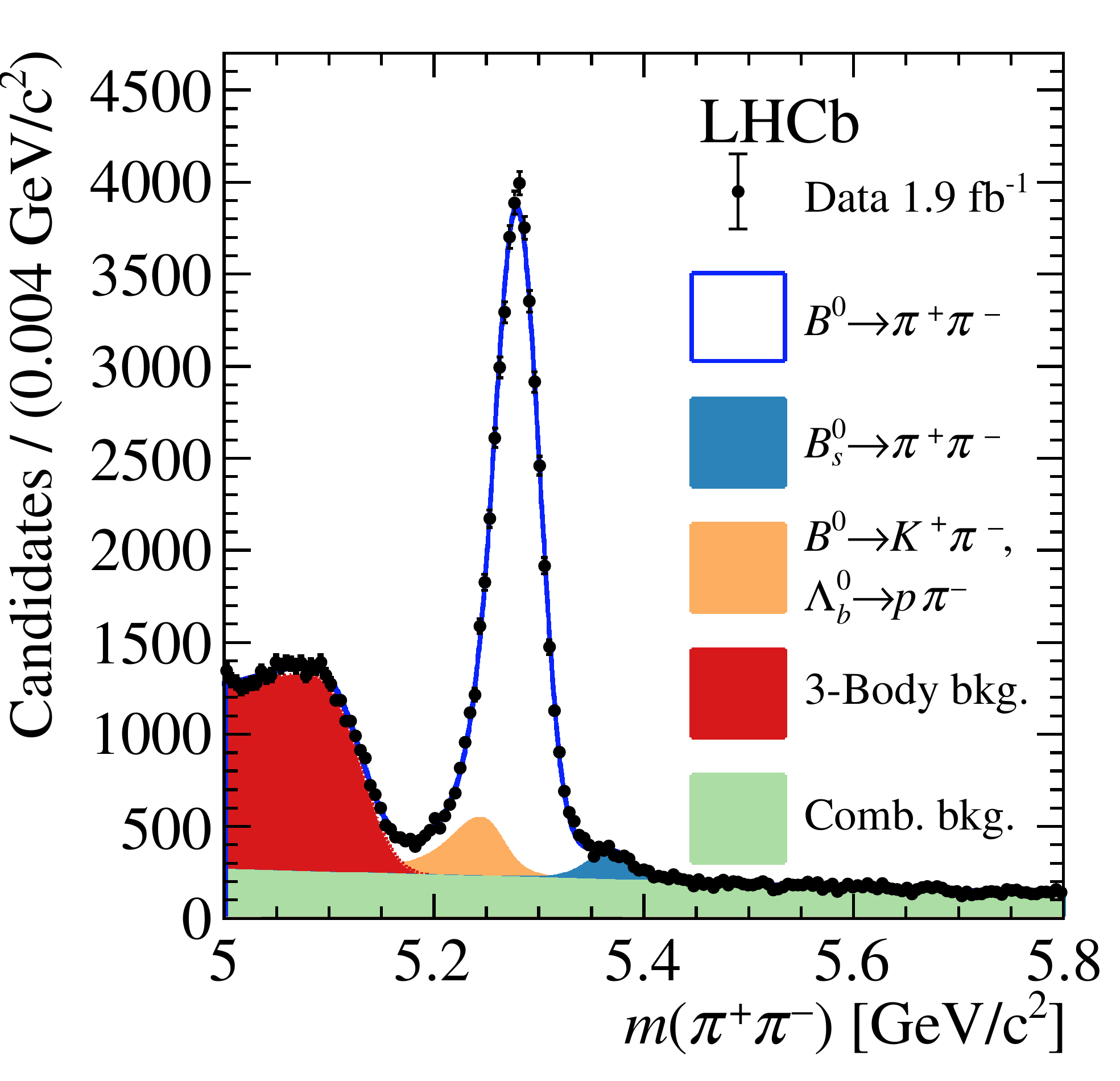}
    \includegraphics[width=0.45\textwidth]{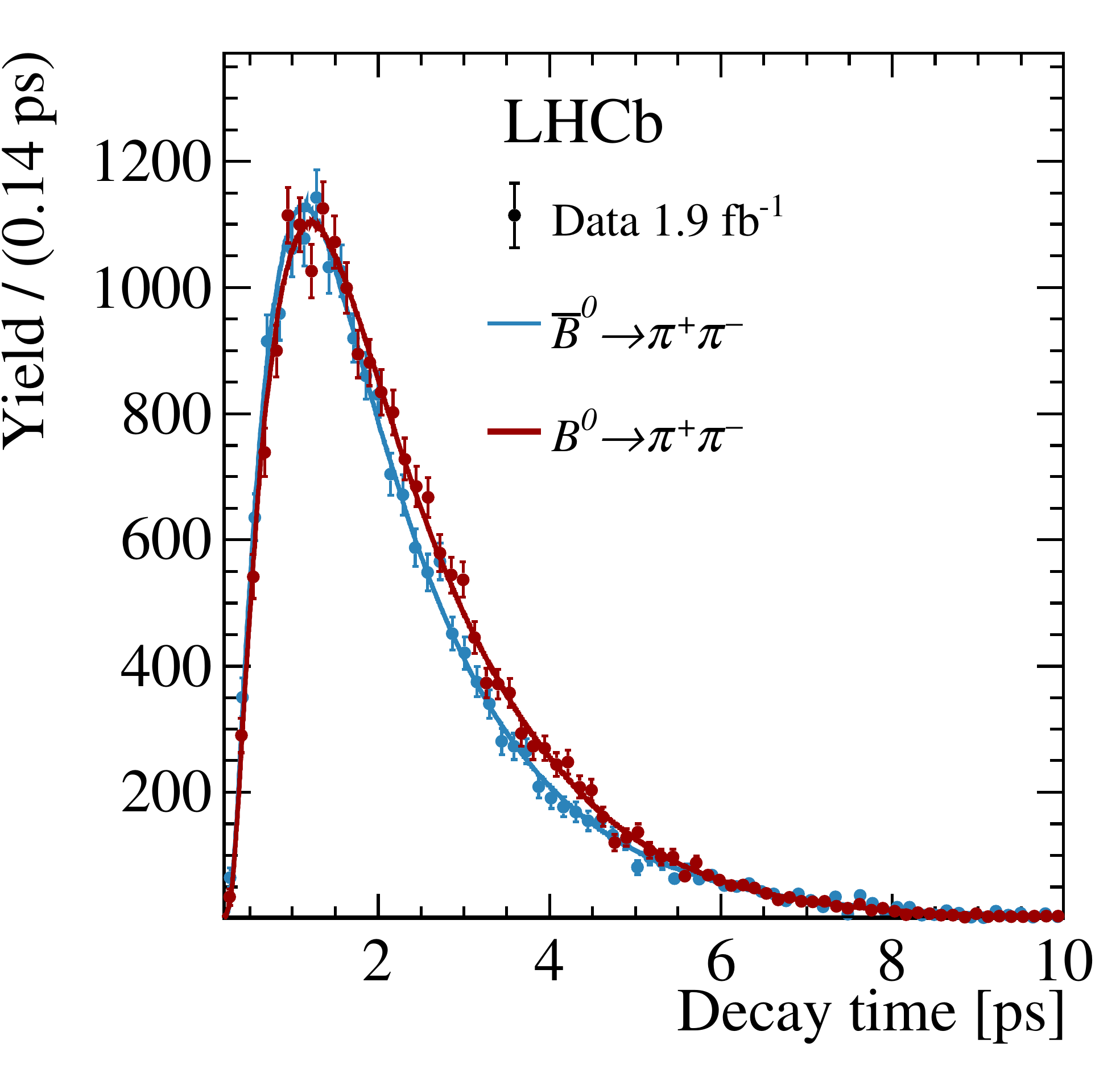}
    \includegraphics[width=0.45\textwidth]{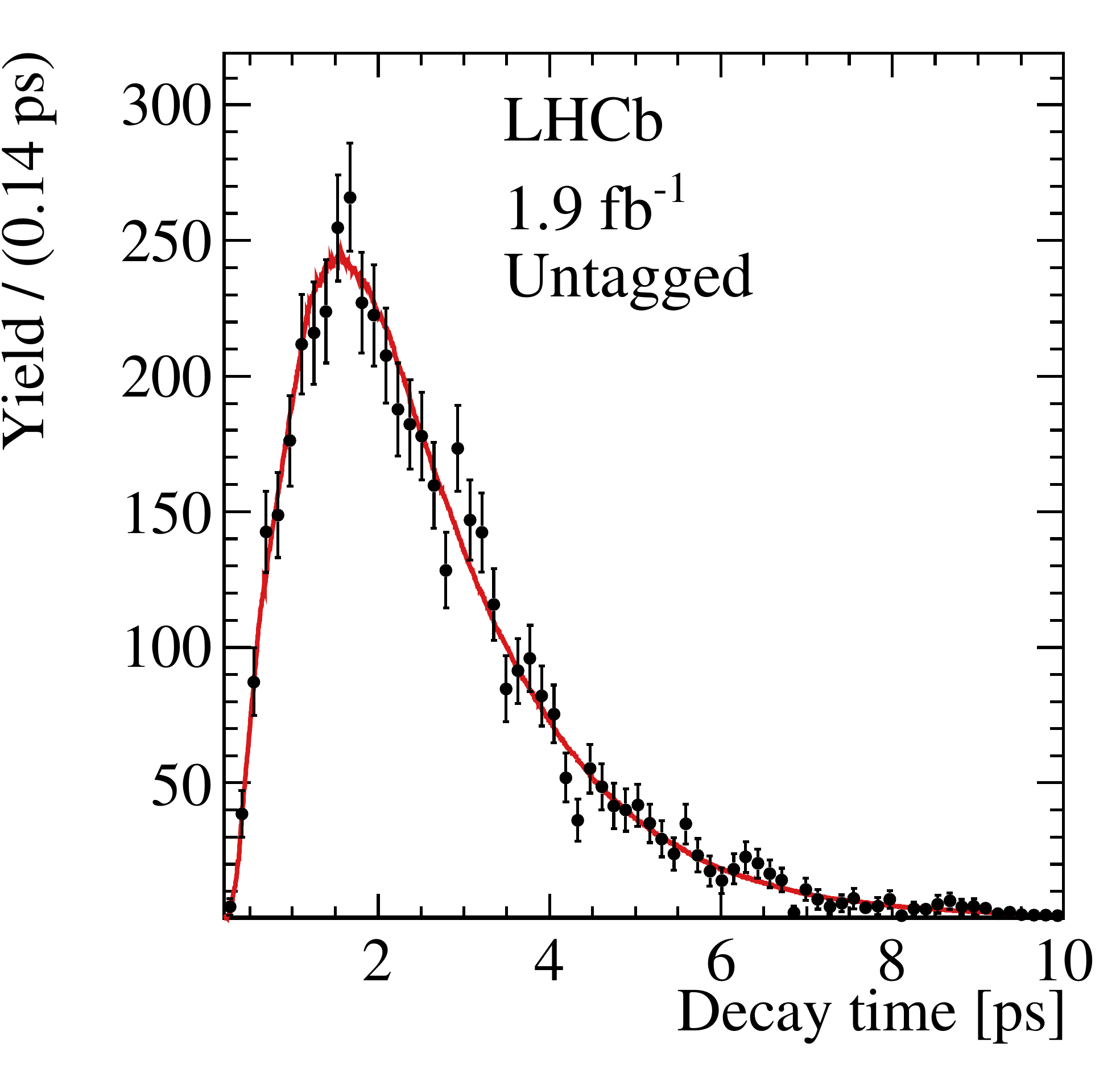}
    \includegraphics[width=0.45\textwidth]{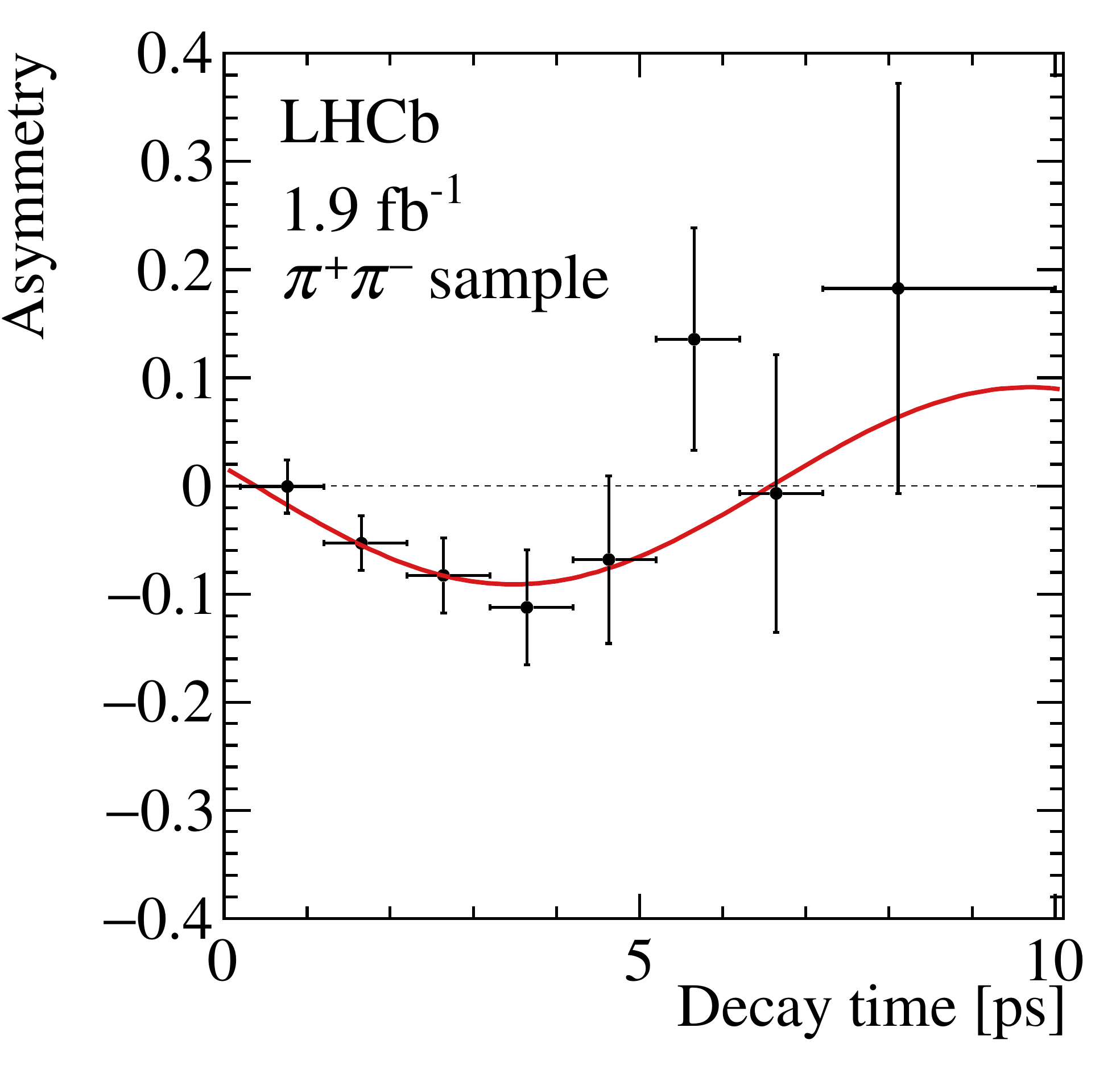}
 \end{center}
  \caption{\small Distributions of the (top left) \pip\pim invariant mass, (top right) decay time for tagged \Bd mesons, (bottom left) decay time for untagged \Bd mesons and (bottom right) asymmetry for the \BdTopipi decays. The individual components are shown for the invariant-mass spectrum while only tagged background-subtracted candidates are shown in the decay-time spectrum. The fit results to the different distributions are overlaid.  The various components contributing to the invariant mass model are drawn as stacked histograms.}
  \label{fig:plotsPIPI_gla}
\end{figure}

\begin{figure}[htb]
  \begin{center}
    \includegraphics[width=0.45\textwidth]{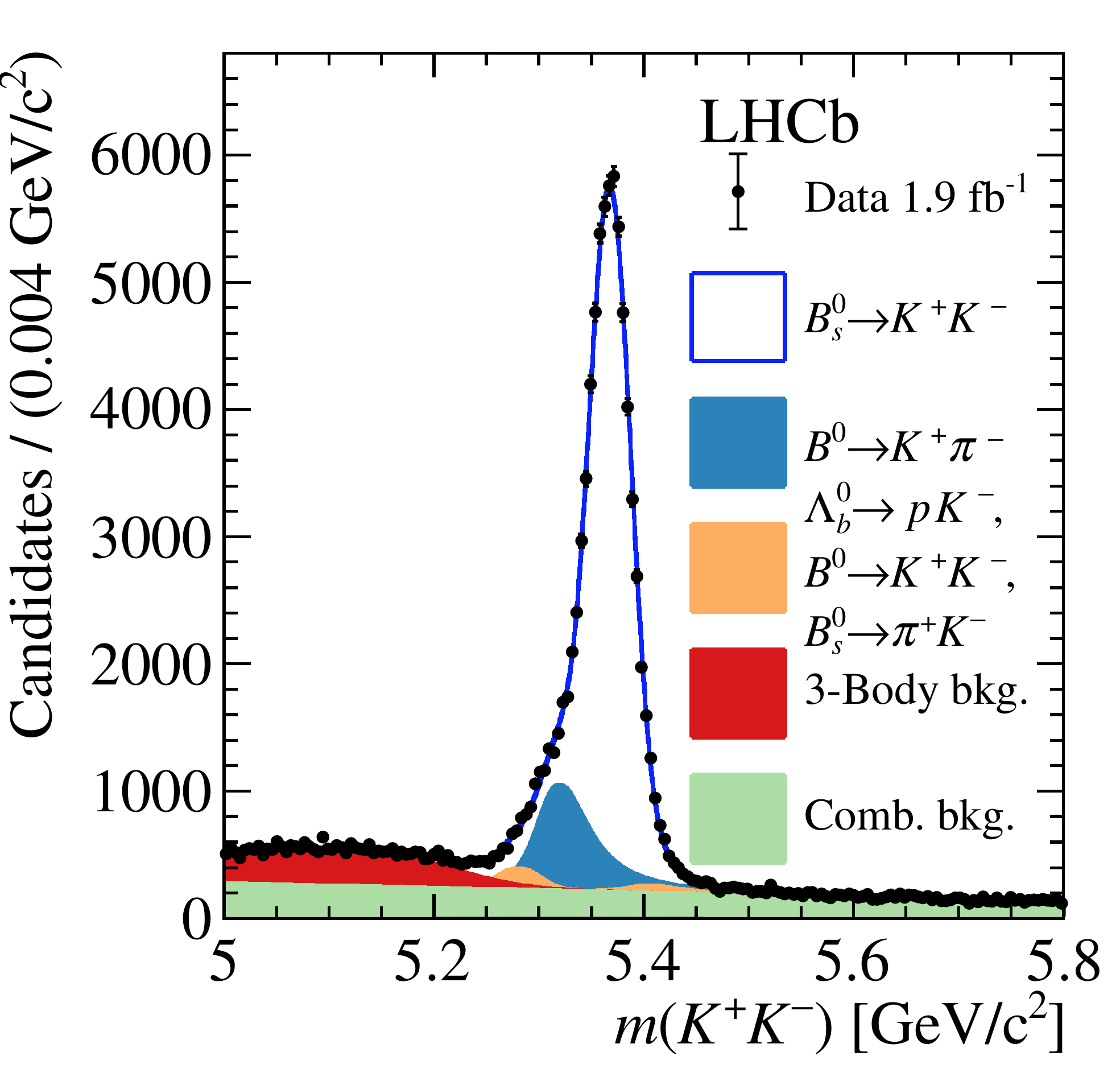}
    \includegraphics[width=0.45\textwidth]{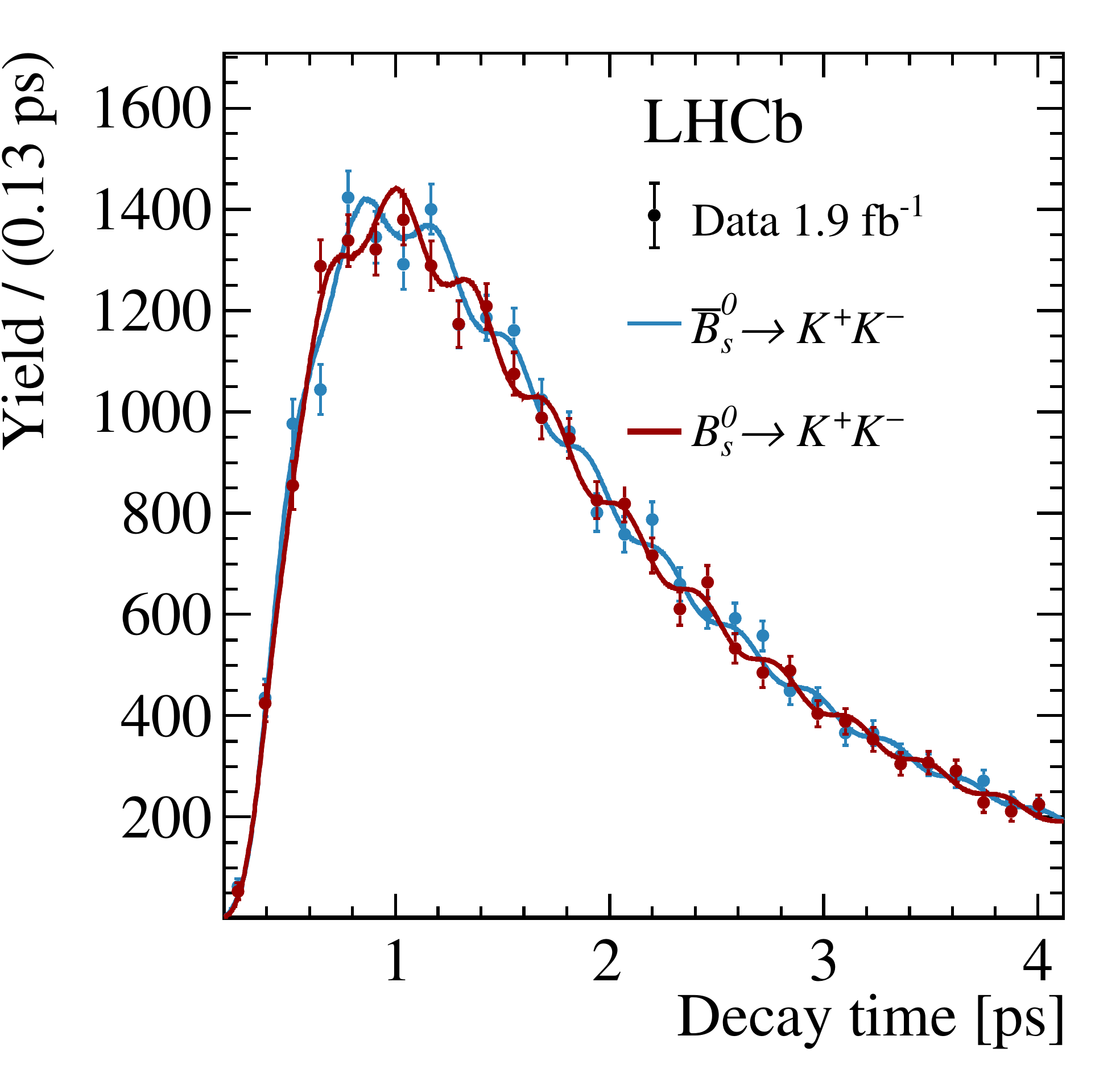}
    \includegraphics[width=0.45\textwidth]{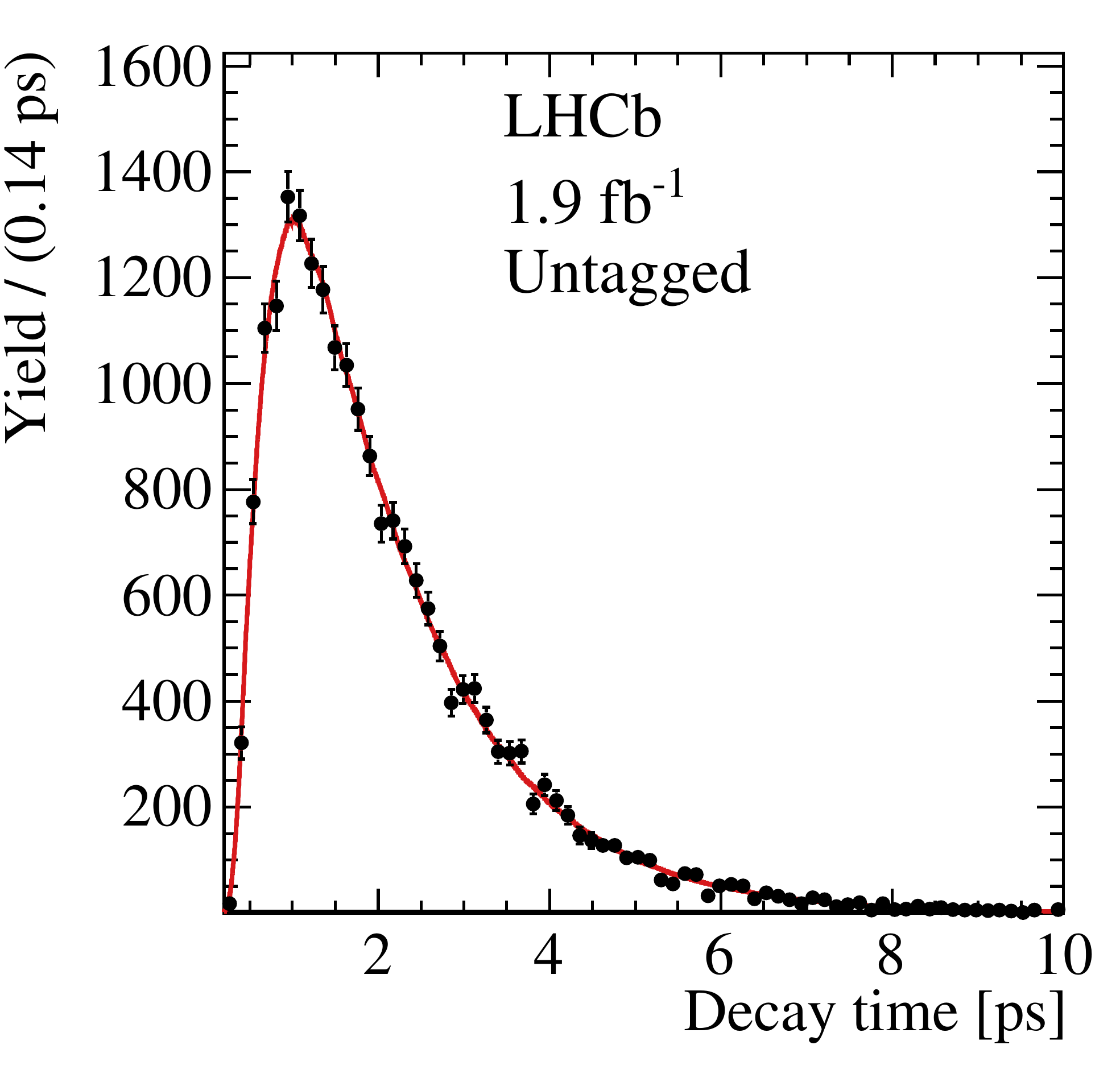}
    \includegraphics[width=0.45\textwidth]{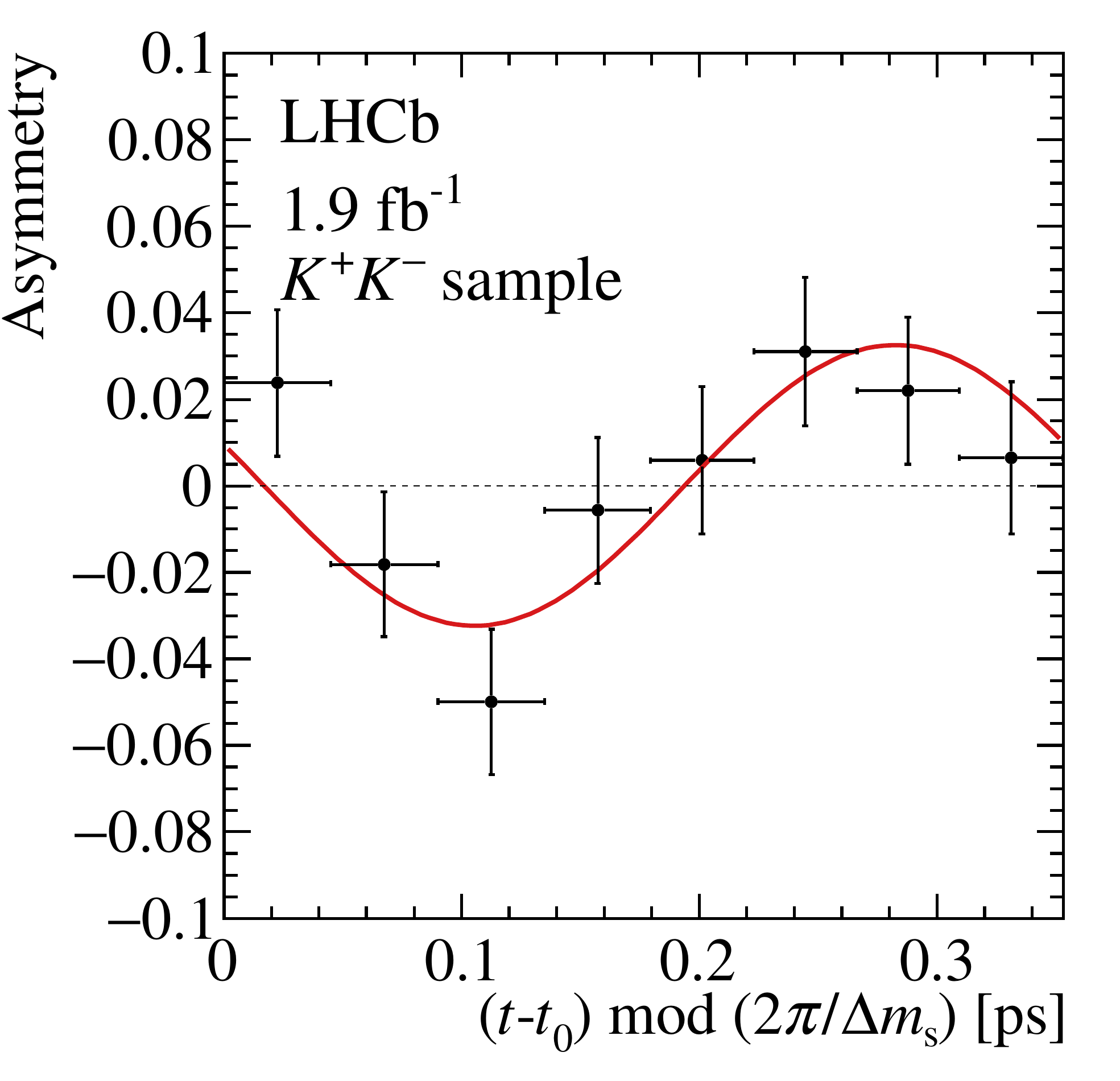}
 \end{center}
  \caption{\small Distributions of the \Kp\Km invariant mass (top left), decay time for tagged \Bs mesons (top right), decay time for untagged \Bs candidates (bottom left) and asymmetry (bottom right) for the \BsToKK decays. The individual components are shown for the invariant mass spectrum while only background-subtracted candidates are shown in the decay time spectrum. The fit results to the different distributions are overlaid. The various components contributing to the invariant mass model are drawn as stacked histograms. The asymmetry for the $\Kp\!\Km$ candidates is folded into one mixing period $2\pi/\dms$ and the parameter $t_0 = 0.2\ps$ corresponds to the minimum value of the decay-time used in the fit.}
  \label{fig:plotsKK_gla}
\end{figure}

The signal yields in the \BdTopipi and \BsToKK decays, used to determine the \CP-violating parameters with the per-candidate method, are in agreement with those of the simultaneous method. 
The parameters $\Delta m_{\dquark(\squark)}$, $\Gamma_{\dquark(\squark)}$, and $\Delta\Gamma_{\dquark(\squark)}$ are fixed to the values reported in Table~\ref{tab:lifetimeParameters}. The value of the production asymmetry is fixed to that measured by the simultaneous method. The fits to the \pip\pim and \Kp\Km invariant-mass spectra are shown in Figs. \ref{fig:plotsPIPI_gla} and \ref{fig:plotsKK_gla} along with the decay-time fits to the \Bds mesons having their flavours tagged.

The results for the \CP-violating parameters using the per-candidate method are
\begin{eqnarray*}
  \Cpipi & = & -0.338 \pm 0.048, \\
  \Spipi & = & -0.673 \pm 0.043, \\
  \CKK   & = & \phantom{-}0.173 \pm 0.042, \\
  \SKK   & = & \phantom{-}0.166 \pm 0.042, \\
  \ADGKK & = & -0.973 \pm 0.071,
\end{eqnarray*}
where the uncertainties are statistical only.


\subsection{Comparison}\label{sec:comparison}

\begin{figure}[htb]
  \begin{center}
    \includegraphics[width=0.45\textwidth]{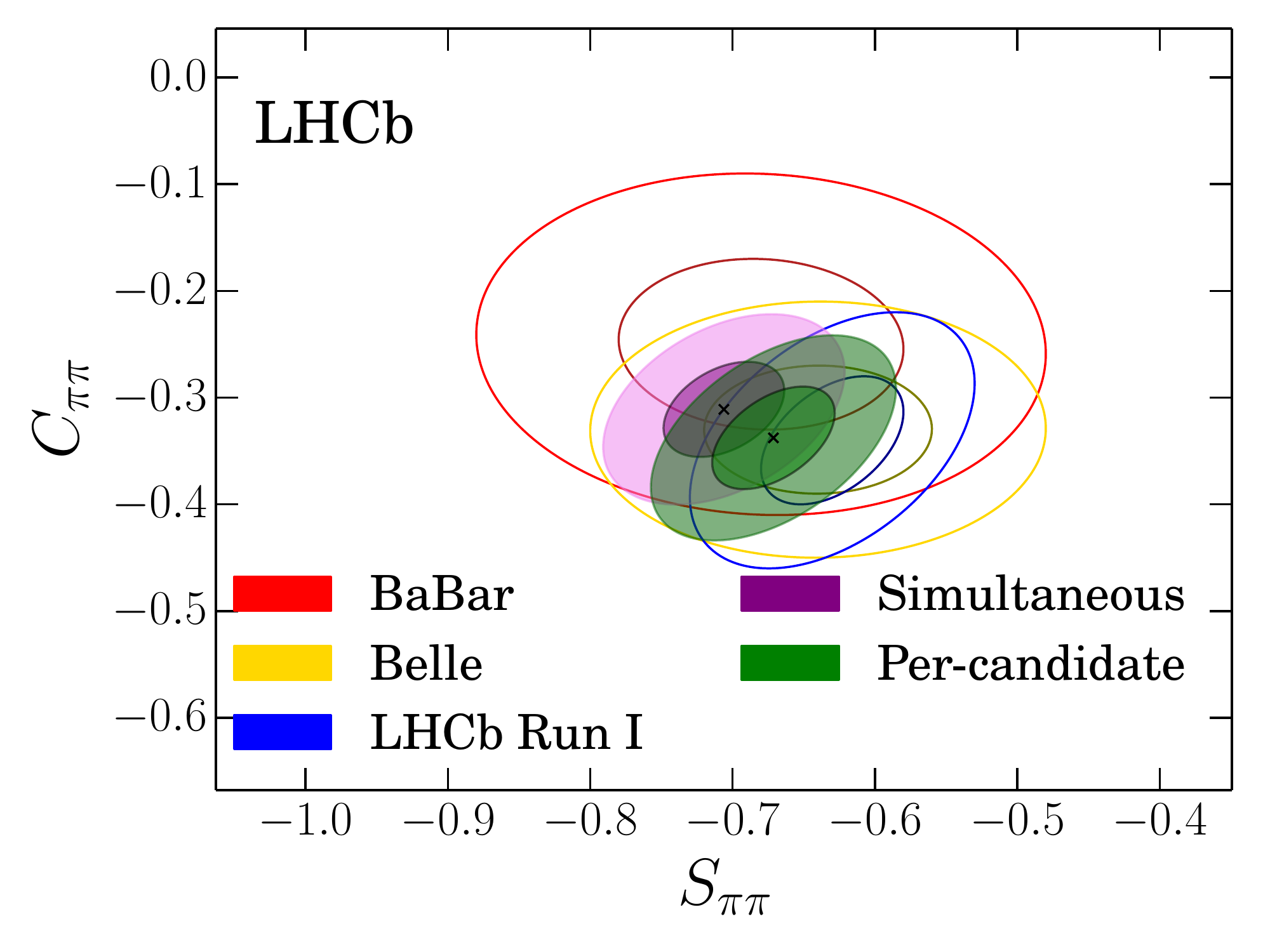}
    \includegraphics[width=0.45\textwidth]{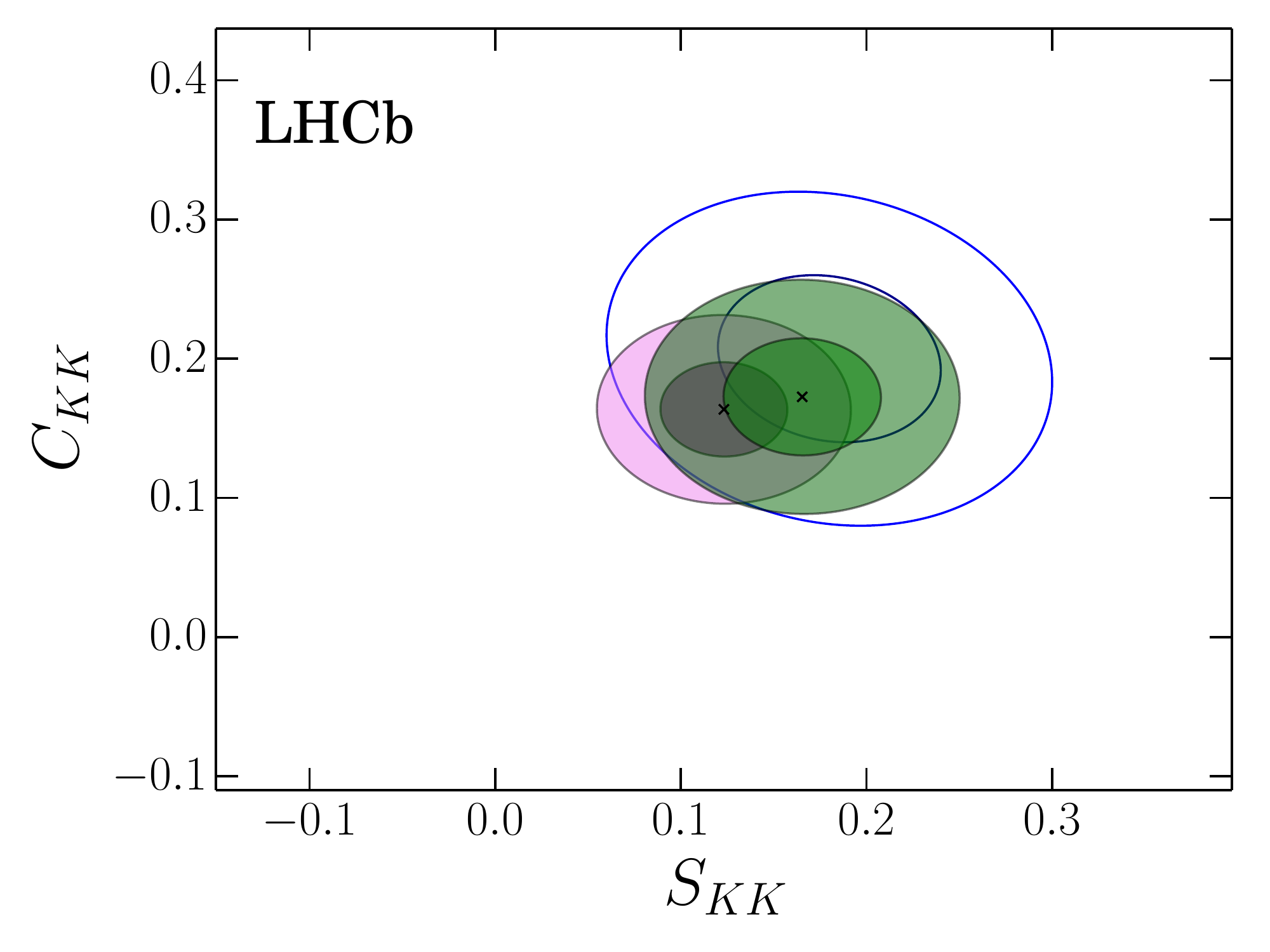}
 \end{center}
  \caption{\small Two-dimensional 68\% and 95\% confidence-level regions for the measured \CP-violating parameters of the \BdTopipi (left) and \BsToKK (right) decays from the two methods. The simultaneous method is shown in purple while the per-candidate method in green. Previous measurements of these parameters are also shown, with the LHCb Run~1 result in blue, the Belle result in yellow and the BaBar result in red. The confidence-level regions are calculated using only the statistical uncertainties of all the measurements. The correlation is found to be approximately 84\% for all \CP-violating parameters between the simultaneous and per-event methods.}
\label{fig:contours}
\end{figure}

To evaluate the compatibility of the results from the two methods, their statistical correlation is determined from 500 simulated pseudoexperiments. The correlation is found to be approximately 84\% for all \CP-violating parameters. This is used to determine the uncorrelated statistical uncertainty on the difference between the results of the two methods. The pseudoexperiments also confirm the smaller total uncertainty observed by the simultaneous method. A sizeable difference between the two results is observed for \ADGKK. This difference is reduced to approximately 1.5 standard deviations when taking into account the systematic uncertainties due to the determination of the decay-time efficiency (see Tab.~\ref{tab:systSummary}), which are completely uncorrelated between the two methods. Adding in quadrature the uncorrelated statistical and systematic uncertainties, the results are found to be compatible within one standard deviation. The resulting contour plots from measuring \Cpipi, \Spipi, \CKK and \SKK are given in Fig.~\ref{fig:contours}.

Given the large correlation between the two determinations, the values obtained from the simultaneous method are quoted as the \lhcb results. They are chosen due to the slightly smaller total uncertainty and the fact that the simultaneous method gives also the direct \CP asymmetries allowing for a complete combination with the results published in Ref.~\cite{LHCb-PAPER-2018-006}.

\section{Systematic uncertainties}\label{sec:systematics}

The systematic uncertainties are evaluated for both the simultaneous and the per-candidate methods, and the total systematic uncertainties for both results are given in Table~\ref{tab:systSummary}. A full description of the systematic uncertainties is only given for the simultaneous method since it is used as  the \lhcb result and for combination with the Run~1 measurement. Hence the description given in this section and the breakdown of the individual components in Table~\ref{tab:systSummary} refers to that method. The main differences in systematic uncertainties between the two methods are briefly discussed at the end of this section.

The systematic uncertainties on the \CP-violating parameters are determined following two approaches. In the first case the fit to data is repeated a large number of times, each time modifying the values of the input parameters. This approach is used to account for the knowledge of external inputs whose values are fixed in the fit. In the second case, pseudoexperiments are performed according to the default model and both the default model and modified models are used to fit the generated data. This strategy is used to account for the systematic uncertainties due to the assumptions on the fitting model. In both cases the difference between the default and alternative results for the \CP asymmetries is measured, and the mean and width of the obtained distribution is used to assign a systematic uncertainty. 

Three sources of systematic uncertainty are considered on the invariant-mass model. First, the systematic uncertainty due to a possibly imperfect description of the mass-resolution function, used for both 
signal and cross-feed background components, is determined by replacing the double Gaussian function with a single Gaussian model. Second, the systematic uncertainty associated to the combinatorial background model is assessed using an alternative model with no correlation between decay time and invariant mass. Finally, a systematic uncertainty associated with the model adopted for the three-body background components is determined by fitting a set of pseudoexperiments, after removing the candidates with an invariant mass below 5.2\gevcc and ignoring the components describing this background contributions in the model.

The PID efficiencies and misidentification probabilities govern the amount of cross-feed background components. A systematic uncertainty related to their calibration  is determined by repeating the fit to data changing those values according to their uncertainties estimated from the calibration samples.

The effect of ignoring the small fraction of \Bs candidates originating from decays of the \Bc meson is studied by injecting simulated \decay{\Bc}{\Bs\PX} decays (where \PX stands for any additional particle in the final state) into the pseudoexperiments, where the relative \Bc yield is  determined from Ref.~\cite{LHCb-PAPER-2013-044}. No systematic uncertainty is assigned for the \Bz\, \CP-violating parameters since the \decay{\Bc}{\Bz\PX} decay is Cabibbo suppressed.

Systematic uncertainties associated with the calibration of the OS and SSc flavour-tagging responses are determined using an alternative relation between $\eta_{\rm OS(SS)}$ and the calibrated mistag probability $\omega_{\rm OS(SS)}$. The linear relation connecting the two quantities in Eq.~\eqref{eq:taggingCalibrationParameters} is replaced with a second-order polynomial. A similar approach is also used for the SS\kaon tagger, but the values of the parameters of the alternative relations are first determined from the \BsToDspi sample and then used in the fit to data.
For the SS\kaon tagger an additional systematic uncertainty associated with the calibration of the 
flavour-tagging response is determined by varying the calibration parameters 
according to their uncertainties and correlations. 

Regarding the decay-time model, a systematic uncertainty associated with the uncertainties on the parameters reported in
Table~\ref{tab:lifetimeParameters} is determined by repeating the simultaneous fit using different fixed values, 
generated according to their uncertainties and correlations. 
The systematic effect due to the decay-time resolution can be decomposed into three contributions: one due to the calibration of the resolution width, another one due to the calibration of the bias in the determination of the decay time, and the last due to the usage of an average decay-time resolution instead of a per-candidate value. The first effect is estimated varying the value of the averaged decay-time resolution width according to a Gaussian distribution with mean equal to the default value, reported in Section~\ref{sec:timeResolution}, and with a width equal to the difference between the decay-time resolution for the fully simulated \JpsiTomumu and \BsToKK decays. The second effect is determined varying the mean of the decay-time resolution model according to a Gaussian 
centered at the default value and with the width of 2~\fs. The last contribution, due to the usage of an average decay-time resolution instead of a per-candidate value, is evaluated by fitting a set 
of pseudoexperiments with both the decay-time resolution models.

Three sources of systematic uncertainty related to the knowledge of the decay-time efficiency are identified. 
A systematic uncertainty on the chosen model is assessed by replacing the effective function with a cubic-spline
polynomial in an alternative model.
Second, a systematic uncertainty arising from the limited calibration-sample size
is computed by varying the parameters governing the decay-time acceptance according to their uncertainties and correlations.
An additional systematic uncertainty due to the imperfect description of the ratios between the decay-time efficiency of the various signal and calibration modes, 
determined from fully simulated samples, is estimated. In this case, the alternative model is created assuming that all the 
decay-time efficiencies  are equal to that of the \BdToKpi decay.

To determine the systematic uncertainty associated with the choice of the decay-time model for the cross-feed background component, an alternative model is created by disabling the oscillating component of the cross-feed background model. This means assuming
no $C\!P$ violation for both, the \BdToKpi component in the \pip\pim and $\Kp\Km$ samples, and for the 
\BdTopipi and \BsToKK components in the \Kpm\pimp sample.
Finally, the uncertainties associated with 
the detection and PID asymmetries reported in Eqs.~\eqref{eq:adkpiBd2KPI} and~\eqref{eq:apidB2KPI} are accounted for as systematic uncertainties on \ACPBd and \ACPBs.

The total systematic uncertainties are obtained as the quadratic sum of the individual contributions, and they are smaller than the corresponding statistical uncertainties for all the $C\!P$-violating parameters apart from \ADGKK. The dominating systematic uncertainty for this parameter is related to the knowledge of how the efficiency varies with the decay time, whose knowledge is limited by the size of the calibration sample of \BdToKpi decays.

Most of the sources of systematic uncertainties related to the per-candidate method are the same as those on the simultaneous method.
The systematic uncertainties are also similar in size. The main difference is a smaller uncertainty related to the decay-time acceptance in the per-candidate method. This uncertainty, which is uncorrelated between the two methods due to their different strategies, mainly affects \ADGKK and largely cancels in the other parameters. The second most important difference is due to  systematic uncertainties related to flavour tagging, where the uncertainties are larger in the per-candidate method, which arises from the different approaches of incorporating this information in the two fits. These differences in systematic uncertainties illustrate the strength of validating the result with two different methods.

\begin{table}[!htbp]
  \caption{\small Systematic uncertainties on the $C\!P$-violating parameters. The values given for each individual contribution to the systematic uncertainty are those for the simultaneous method. The total systematic uncertainties are given both for the simultaneous and the per-candidate methods. The dash indicates that the uncertainty is not applicable.}
  \begin{center}
    \resizebox{\columnwidth}{!}{
    \begin{tabular}{ll|ccccccc}
\hline
\multicolumn{2}{l|}{Source}             & $\Cpipi$ & \Spipi   & \ACPBd   & \ACPBs   & \CKK     & \SKK     & \ADGKK    \\
\hline
\multicolumn{2}{l|}{Time acceptance}    &          &          &          &          &          &          &           \\
 & Model                                & $0.005$ & $0.003$ & $0.0005$ & $0.001$ & $0.003$ & $0.003$ & $0.045$  \\
 & Calibration channel                  & $0.003$ & $0.001$ & $0.0003$ & $0.006$ & $0.001$ & $0.001$ & $0.047$  \\
 & Ratios between modes              & $0.004$ & $0.002$ & $0.0010$ & $0.000$ & $0.001$ & $0.001$ & $0.047$  \\
\multicolumn{2}{l|}{Time resolution}    &          &          &          &          &          &          &           \\
 & Width                                & $0.002$ & $0.003$ & $0.0001$ & $0.000$ & $0.0009$ & $0.010$ & $0.000$  \\
 & Bias                                 & $0.000$ & $0.000$ & $0.0000$ & $0.000$ & $0.004$ & $0.003$ & $0.000$  \\
 & Average                              & $0.000$ & $0.001$ & $0.0000$ & $0.000$ & $0.004$ & $0.004$ & $0.004$  \\
\multicolumn{2}{l|}{Input parameters}   & $0.003$ & $0.002$ & $0.0001$ & $0.000$ & $0.006$ & $0.007$ & $0.047$  \\
\multicolumn{2}{l|}{\Bs from \Bc}       & $-$      & $-$      & $-$      & $-$      & $0.004$ & $0.003$ & $0.004$  \\
\multicolumn{2}{l|}{Flavour tagging}    &          &          &          &          &          &          &           \\
 & Calibration model                    & $0.001$ & $0.001$ & $0.0000$ & $0.000$ & $0.004$ & $0.003$ & $0.001$  \\
 & \SSK calibration                     & $-$      & $-$      & $-$      & $-$      & $0.003$ & $0.004$ & $0.000$  \\
\multicolumn{2}{l|}{PDF modeling}       &          &          &          &          &          &          &           \\
 & Signal mass                          & $0.007$ & $0.008$ & $0.0004$ & $0.007$ & $0.002$ & $0.002$ & $0.006$  \\
 & Cross-feed bkg.                      & $0.008$ & $0.004$ & $0.0001$ & $0.000$ & $0.001$ & $0.000$ & $0.002$  \\
 & Combinatorial bkg                    & $0.006$ & $0.003$ & $0.0001$ & $0.002$ & $0.001$ & $0.001$ & $0.006$  \\
 & 3-body bkg.                          & $0.004$ & $0.006$ & $0.0005$ & $0.004$ & $0.001$ & $0.001$ & $0.011$       \\
\multicolumn{2}{l|}{PID in fit model}   & $0.002$ & $0.003$ & $0.0002$ & $0.002$ & $0.000$ & $0.001$ & $0.001$  \\
\multicolumn{2}{l|}{PID asymmetry}      & $-$      & $-$      & $0.0028$ & $0.003$ & $-$      & $-$      & $-$       \\
\multicolumn{2}{l|}{Det. asymmetry}     & $-$      & $-$      & $0.0012$ & $0.001$ & $-$      & $-$      & $-$       \\
\hline
\multicolumn{2}{l|}{Total (simultaneous)} & $0.015$ & $0.013$  & $0.0033$ & $0.011$ & $0.014$ & $0.015$ & $0.094$   \\
\hline
\multicolumn{2}{l|}{Total (per-candidate)} & $0.018$ & $0.016$  & $-$     & $-$     & $0.021$ & $0.012$ & $0.067$   \\
\hline
   \end{tabular}
    }
  \end{center}
  \label{tab:systSummary}
\end{table}
 
\section{Results}\label{sec:finalResults}

The final results for the time-dependent \CP violation in \BdTopipi and \BsToKK decays, and of the \CP asymmetries in \BdToKpi and \BsTopiK decays are
\begin{eqnarray*}\label{eq:finalResultsRun2}
  \Cpipi  & = & -0.311\phantom{0} \pm 0.045\phantom{0} \pm 0.015, \\
  \Spipi  & = & -0.706\phantom{0} \pm 0.042\phantom{0} \pm 0.013, \\ 
  \ACPBd & = & -0.0824 \pm 0.0033  \pm 0.0033, \\
  \ACPBs & = & \phantom{-}0.236\phantom{0} \pm 0.013\phantom{0} \pm 0.011, \\
  \CKK    & = & \phantom{-}0.164\phantom{0} \pm 0.034\phantom{0} \pm 0.014, \\
  \SKK    & = & \phantom{-}0.123\phantom{0} \pm 0.034\phantom{0} \pm 0.015, \\
  \ADGKK  & = & -0.83\phantom{00} \pm 0.05\phantom{00} \pm 0.09, 
\end{eqnarray*}
where the first uncertainties are statistical and the second systematic. The corresponding statistical and systematic correlation matrices are reported in Tables~\ref{tab:correlationStat} and~\ref{tab:correlationSyst}, respectively. The results are 
compatible with the previous \lhcb measurement in Ref.~\cite{LHCb-PAPER-2018-006}.
\begin{table}[bt]
  \caption{\small Correlations of statistical uncertainties among the \CP-violating parameters.}
  \begin{center}
    \begin{tabular}{l|ccccccc}
            & \Cpipi  & \Spipi & \ACPBd   & \ACPBs & \CKK   & \SKK  & \ADGKK\rule[-1.3ex]{0pt}{0pt}            \\
     \hline
      \Cpipi  & $1$      &         &          &         &          &         &     \\
      \Spipi  & $\phantom{-}0.394$ & $1$      &         &          &         &        &     \\
      \ACPBd  & $-0.035$           & $\phantom{-}0.011$ & $1$      &         &        &           &    \\
      \ACPBs  & $\phantom{-}0.000$ & $\phantom{-}0.000$ & $\phantom{-}0.052$ & $1$    &           &       &    \\
      \CKK    & $-0.008$           & $-0.029$           & $\phantom{-}0.002$ & $\phantom{-}0.001$ & $1$   &            &    \\
      \SKK    & $-0.008$           & $\phantom{-}0.005$ & $-0.006$           & $\phantom{-}0.001$ & $-0.010$           & $1$     &    \\
      \ADGKK  & $\phantom{-}0.000$ & $\phantom{-}0.000$ & $\phantom{-}0.001$ & $\phantom{-}0.000$ & $\phantom{-}0.025$ & $\phantom{-}0.023$ & $1$ \\
    \end{tabular}
  \end{center}
  \label{tab:correlationStat}
\end{table}
\begin{table}[bt]
  \caption{\small Correlations of systematic uncertainties among the \CP-violating parameters.}
  \begin{center}
    \begin{tabular}{l|ccccccc}
            & \Cpipi  & \Spipi & \ACPBd   & \ACPBs & \CKK   & \SKK  & \ADGKK\rule[-1.3ex]{0pt}{0pt}            \\
    \hline
     \Cpipi & $1$  &             &      &             &          &          &          \\
     \Spipi & $\phantom{-}0.306$ & $1$  &             &          &          &         &          \\
     \ACPBd & $-0.044$           & $-0.024$           & $1$                 &         &          &          &          \\
     \ACPBs & $\phantom{-}0.075$ & $\phantom{-}0.010$ & $-0.238$            & $1$     &          &          &          \\
     \CKK   & $-0.050$           & $-0.022$           & $\phantom{-}0.028$  & $-0.009$           & $1$ &    &          \\
     \SKK   & $\phantom{-}0.053$ & $\phantom{-}0.045$ & $-0.025$            & $\phantom{-}0.011$ & $\phantom{-}0.197$ & $1$ &          \\
     \ADGKK & $-0.117$           & $-0.090$           & $\phantom{-}0.050$  & $-0.006$           & $\phantom{-}0.082$ & $\phantom{-}0.018$ &  $1$\\
    \end{tabular}
  \end{center}
  \label{tab:correlationSyst}
\end{table}

A combination is performed between the results in this paper and those based on the Run~1 data sample reported in Ref.~\cite{LHCb-PAPER-2018-006}. Since the values of \Gs and \DGs used as input to the fit have changed with respect to Ref.~\cite{LHCb-PAPER-2018-006}, the Run~1 analysis is updated to account for the new values. The main variation is observed for the central value of \ADGKK that changes from $-0.79 \pm 0.07$ to $-0.97 \pm 0.07$. The large variation of \ADGKK is expected, given its correlation of 0.91 with \Gs, and the significant change in the value of \Gs from $0.6654\pm0.0022\invps$ to $0.6563\pm0.0021\invps$~\cite{LHCb-PAPER-2019-013}. The only other variation is for \CKK, moving from $0.20 \pm 0.06$ to $0.19 \pm 0.06$. The compatibility between the updated Run~1 result and the numbers reported in Eq.~\eqref{eq:MiBo-Results} is computed by means of $\chi^2$ test statistic, finding the two sets of results in agreement with a $p$-value of 0.68.

The full statistical and systematic covariance matrices of the two results are taken into account in the combination. The only relevant correlation between the two results is related to the values of the input parameters in Table~\ref{tab:lifetimeParameters}, hence the corresponding systematic uncertainties are removed from the covariance matrices of the two results, before combining them. The systematic uncertainty due to these input parameters is included again by summing the corresponding covariance matrix to the covariance matrix of the combination. The results of the combination are:
\begin{eqnarray*}\label{eq:finalResultsCombination1}
  \Cpipi  & = & -0.320\phantom{1} \pm 0.038, \\
  \Spipi  & = & -0.672\phantom{1} \pm 0.034, \\
  \ACPBd & = & -0.0831 \pm 0.0034, \\
  \ACPBs & = & \phantom{-}0.225\phantom{1}  \pm 0.012, \\
  \CKK    & = & \phantom{-}0.172\phantom{1}  \pm 0.031, \\
  \SKK    & = & \phantom{-}0.139\phantom{1}  \pm 0.032, \\
  \ADGKK  & = & -0.897\phantom{1} \pm  0.087\label{eq:finalResultsCombination2}
\end{eqnarray*}
and their correlation matrix is reported in Table~\ref{tab:correlationMatrixFinal}.
\begin{table}[bt]
  \caption{\small Correlation matrix for the \CP violation parameters obtained from the combination with Run-1 results.}
  \begin{center}
    \begin{tabular}{l|ccccccc}
            & \Cpipi  & \Spipi & \ACPBd   & \ACPBs & \CKK   & \SKK  & \ADGKK\rule[-1.3ex]{0pt}{0pt}            \\
     \hline
  \Cpipi & $1$  &             &      &             &         &          &          \\
  \Spipi & $\phantom{-}0.405$ & $1$  &             &         &          &         &          \\
  \ACPBd & $-0.019$           & $\phantom{-}0.001$ & $1$     &          &         &          &          \\
  \ACPBs & $\phantom{-}0.014$ & $-0.002$           & $-0.063$           & $1$     &          &         &          \\
  \CKK   & $-0.009$           & $-0.032$           & $\phantom{-}0.008$ & $\phantom{-}0.000$ & $1$     &          &  \\
  \SKK   & $-0.004$           & $\phantom{-}0.004$ & $-0.007$           & $\phantom{-}0.002$ & $\phantom{-}0.007$ & $1$                & \\
  \ADGKK & $-0.019$           & $-0.014$           & $\phantom{-}0.019$ & $-0.003$           & $\phantom{-}0.027$ & $\phantom{-}0.043$ &  $1$ \\  
    \end{tabular}
  \end{center}
  \label{tab:correlationMatrixFinal}
\end{table}

\section{Concluding remarks}\label{sec:conclusions}

The time-dependent \CP asymmetries of \BdTopipi and \BsToKK decays and the time-integrated \CP asymmetries in \BdToKpi and \BsTopiK decays are measured using a data sample of \proton\proton collisions corresponding to an integrated luminosity of 1.9\invfb, collected with the \lhcb detector at a centre-of-mass energy of 13\tev. The measurements are compatible with previous \lhcb determinations of the same quantities obtained with \mbox{Run 1} data~\cite{LHCb-PAPER-2018-006} and are combined with them. The measurements of \Cpipi, \Spipi, \ACPBd and \ACPBs are in good agreement with previous results from other experiments~\cite{Lees:2012mma,Adachi:2013mae,Duh:2012ie,Aaltonen:2014vra} and are the most precise from a single experiment to date.

A $\chi^2$ test statistic is used to determine the significance for $(\CKK,\,\SKK,\,\ADGKK)$ to differ from $(0,\,0,\,-1)$ and for $(\CKK,\,\SKK)$ to differ from $(0,\,0)$. The significance for the combined \lhcb results is found to be of $6.5$ and $6.7$ standard deviations, respectively. This constitutes the first observation of time-dependent \CP violation in decays of the \Bs meson. 

The unitary relation among \CKK, \SKK and \ADGKK is tested, giving $\sqrt{\left(\CKK\right)^2+\left(\SKK\right)^2+\left(\ADGKK\right)^2} = 0.93 \pm 0.08$. This is compatible with unity within one standard deviation.

According to the test of the SM proposed in Ref.~\cite{LIPKIN2005126}, the following sum must be satisfied
\begin{equation}\label{eq:lipkinTest}
\Delta \equiv \frac{\ACPBd}{\ACPBs}+\frac{\mathcal{B}\left(\BsTopiK\right)}{\mathcal{B}\left(\BdToKpi\right)}\frac{\Gs}{\Gd} = 0,
\end{equation}
where $\mathcal{B}\left(\BdToKpi\right)$ and $\mathcal{B}\left(\BsTopiK\right)$ are \CP-averaged branching fractions. The \lhcb measurements of the relative fragmentation-fraction ratio between \Bs and \Bz mesons $f_\squark/f_\dquark = 0.259 \pm 0.015$~\cite{fsfd}, $f_\squark/f_\dquark\times \mathcal{B}\left(\BsTopiK\right)/\mathcal{B}\left(\BdToKpi\right)$~\cite{LHCb-PAPER-2012-002} and $\Gs/\Gd$~\cite{LHCb-PAPER-2019-013} are used in this test along with the measurements of \ACPBd and \ACPBs. The value \mbox{$\Delta = -0.085 \pm 0.025 \pm 0.035$} is obtained, where the first uncertainty is from the measurements of the \CP asymmetries and the second is from the other inputs in Eq.~\eqref{eq:lipkinTest}. With the present experimental precision, $\Delta$ is in agreement with zero within two standard deviations.

Owing to the measurements reported in this paper, improved constraints on the CKM angles and \Bs mixing phase can be obtained, as outlined in Refs.~\cite{Fleischer:2010ib,Ciuchini:2012gd,LHCb-PAPER-2014-045}. 
The comparison of these precises determinations, based on decays receiving sizeable loop-level contributions, with those provided by the study of the decays dominated by tree-level amplitudes, will constitute a stringent test of the SM hypothesis.

\section*{Acknowledgements}
%
%
\noindent We express our gratitude to our colleagues in the CERN accelerator departments for the excellent performance of the LHC. We thank the technical and administrative staff at the LHCb institutes. We acknowledge support from CERN and from the national agencies: CAPES, CNPq, FAPERJ and FINEP (Brazil); MOST and NSFC (China); CNRS/IN2P3 (France); BMBF, DFG and MPG (Germany); INFN (Italy); NWO (Netherlands); MNiSW and NCN (Poland); MEN/IFA (Romania); MSHE (Russia); MICINN (Spain); SNSF and SER (Switzerland); NASU (Ukraine); STFC (United Kingdom); DOE NP and NSF (USA). We acknowledge the computing resources that are provided by CERN, IN2P3 (France), KIT and DESY (Germany), INFN (Italy), SURF (Netherlands), PIC (Spain), GridPP (United Kingdom), RRCKI and Yandex LLC (Russia), CSCS (Switzerland), IFIN-HH (Romania), CBPF (Brazil), PL-GRID (Poland) and OSC (USA). We are indebted to the communities behind the multiple open-source software packages on which we depend. Individual groups or members have received support from AvH Foundation (Germany); EPLANET, Marie Sk\l{}odowska-Curie Actions and ERC (European Union); A*MIDEX, ANR, Labex P2IO and OCEVU, and R\'{e}gion Auvergne-Rh\^{o}ne-Alpes (France); Key Research Program of Frontier Sciences of CAS, CAS PIFI, Thousand Talents Program, and Sci. \& Tech. Program of Guangzhou (China); RFBR, RSF and Yandex LLC (Russia); GVA, XuntaGal and GENCAT (Spain); the Royal Society and the Leverhulme Trust (United Kingdom).



\clearpage
{\noindent\normalfont\bfseries\Large Appendix}
\appendix
\section{Additional information on flavour-tagging}\label{sec:flavourTaggingAppendix}

\subsection{Formalism}

The functions $\Omega_{\rm sig}(t,\vec{\xi},\,\vec{\eta})$ and $\overline{\Omega}_{\rm sig}(t,\vec{\xi},\,\vec{\eta})$ in Eqs.~\eqref{eq:decayTimeB2KPI} and~\eqref{eq:decayTimeB2HH} are 
\begin{eqnarray}\label{eq:omegaPdfSig}
  \Omega_{\rm sig}(t,\xi_{\rm OS},\xi_{\rm SS},\eta_{\rm OS},\eta_{\rm SS}) = \Omega_{\rm OS}^{\rm sig}(\xi_{\rm OS},\eta_{\rm OS})\Omega^{\rm SS}_{\rm sig}(t,\xi_{\rm SS},\eta_{\rm SS}), \\
  \overline{\Omega}_{\rm sig}(t,\xi_{\rm OS},\xi_{\rm SS},\eta_{\rm OS},\eta_{\rm SS}) = \overline{\Omega}_{\rm OS}^{\rm sig}(\xi_{\rm OS},\eta_{\rm OS})\overline{\Omega}^{\rm SS}_{\rm sig}(t,\xi_{\rm SS},\eta_{\rm SS}), \nonumber
\end{eqnarray}
where $\eta$ is the mistag probability computed by the flavour-tagging algorithms and discussed in Sec.~\ref{sec:flavourTagging}, $\Omega^{\rm sig}_{\rm OS}$ and $\overline{\Omega}^{\rm sig}_{\rm OS}$ are the same functions used in Ref.~\cite{LHCb-PAPER-2018-006}, \ie
\begin{eqnarray}\label{eq:tagginProbsOS}
  \begin{split}
    \Omega^{\rm sig}_{\rm OS}(\xi_{\rm OS},\,\eta_{\rm OS}) = & \left\{\delta_{\xi_{\rm OS},\,+1}\,\varepsilon^{\rm sig}_{\rm OS}\,\left[1-\omega^{\rm sig}_{\rm OS}(\eta_{\rm OS})\right]+\delta_{\xi_{\rm OS},\,-1}\,\varepsilon^{\rm sig}_{\rm OS}\,\omega^{\rm sig}_{\rm OS}(\eta_{\rm OS})\right\}\,h^{\rm sig}_{\rm OS}(\eta_{\rm OS})\,+ \\
                               & \delta_{\xi_{\rm OS},\,0}\,(1-\varepsilon^{\rm sig}_{\rm OS})\,U(\eta_{\rm OS}),
  \end{split} \\
  \begin{split}
    \overline{\Omega}^{\rm sig}_{\rm OS}(\xi_{\rm OS},\,\eta_{\rm OS}) = & \left\{\delta_{\xi_{\rm OS},\,-1}\,\bar{\varepsilon}^{\rm sig}_{\rm OS}\,\left[1-\overline{\Omega}^{\rm sig}_{\rm OS}(\eta_{\rm OS})\right]+\delta_{\xi_{\rm OS},\,+1}\,\bar{\varepsilon}^{\rm sig}_{\rm OS}\,\overline{\Omega}^{\rm sig}_{\rm OS}(\eta_{\rm OS})\right\}h^{\rm sig}_{\rm OS}(\eta_{\rm OS})\,+ \\
                                     & \delta_{\xi_{\rm OS},\,0}\,(1-\bar{\varepsilon}^{\rm sig}_{\rm OS})\,U(\eta_{\rm OS}), 
  \end{split} \nonumber
\end{eqnarray}
while $\Omega^{\rm sig}_{\rm SS}$ and $\overline{\Omega}^{\rm sig}_{\rm SS}$ are
\begin{eqnarray}\label{eq:tagginProbsSS}
  \begin{split}
    \Omega^{\rm sig}_{\rm SS}(t,\xi_{\rm SS},\,\eta_{\rm SS}) = & \left\{\delta_{\xi_{\rm SS},\,+1}\,\varepsilon^{\rm sig}_{\rm SS}(t)\,\left[1-\omega^{\rm sig}_{\rm SS}(\eta_{\rm SS})\right]+\delta_{\xi_{\rm SS},\,-1}\,\varepsilon^{\rm sig}_{\rm SS}(t)\,\omega^{\rm sig}_{\rm SS}(\eta_{\rm SS})\right\}h^{\rm sig}_{\rm SS}(\eta_{\rm SS})\,+ \\
                               & \delta_{\xi_{\rm SS},\,0}\,\left[\varepsilon(t)-\varepsilon^{\rm sig}_{\rm SS}(t)\right]\,U(\eta_{\rm SS}),
  \end{split} \\
  \begin{split}
    \overline{\Omega}^{\rm sig}_{\rm SS}(\xi_{\rm SS},\,\eta_{\rm SS}) = & \left\{\delta_{\xi_{\rm SS},\,-1}\,\bar{\varepsilon}^{\rm sig}_{\rm SS}(t)\,\left[1-\overline{\Omega}^{\rm sig}_{\rm SS}(\eta_{\rm SS})\right]+\delta_{\xi_{\rm SS},\,1}\,\bar{\varepsilon}^{\rm sig}_{\rm SS}(t)\,\overline{\Omega}^{\rm sig}_{\rm SS}(\eta_{\rm SS})\right\}h^{\rm sig}_{\rm SS}(\eta_{\rm SS})\,+ \\
                                     & \delta_{\xi_{\rm SS},\,0}\,\left[\varepsilon(t)-\bar{\varepsilon}^{\rm sig}_{\rm SS}(t)\right]\,U(\eta_{\rm SS}). \nonumber
  \end{split} 
\end{eqnarray}
Here, $\varepsilon^{\rm sig}_{\rm tag}$ ($\bar{\varepsilon}^{\rm sig}_{\rm tag}$) is the probability that the flavour of a \Bds (\Bdsb) meson is tagged, which in the case of the SS tagger depends on the decay time\footnote{From now on, in order to simplify the notation, the dependency of SS-tagger efficiency on the decay-time is omitted.}; $\varepsilon(t)$ is the decay-time efficiency independent from the decision of the SS-tagger, such that $\varepsilon(t)-\varepsilon^{\rm sig}_{\rm SS}(t)$ is the decay-time efficiency for candidates that have $\xi_{\rm SS} = 0$; $\omega^{\rm sig}_{\rm tag}(\eta_{\rm tag})$ and $\overline{\Omega}^{\rm sig}_{\rm tag}(\eta_{\rm tag})$ are the calibrated mistag probabilities as a function of $\eta_{\rm tag}$ for \Bds and \Bdsb mesons; $h^{\rm tag}_{\rm sig}(\eta_{\rm tag})$ is the PDF describing the distribution of $\eta_{\rm tag}$ for tagged candidates, and $U(\eta_{\rm tag})$ is a uniform distribution of $\eta_{\rm tag}$. It is empirically observed that, to a good approximation, $\eta_{\rm tag}$ and $\omega_{\rm tag}$ are related by a linear function, \ie
\begin{eqnarray}\label{eq:taggingCalibrationParameters}
\omega^{\rm sig}_{\rm tag}(\eta_{\rm tag}) & = & p^{\rm tag}_0\,+\,p^{\rm tag}_1\,(\eta_{\rm tag}\,-\,\hat{\eta}_{\rm tag}), \\
\overline{\omega}^{\rm sig}_{\rm tag}(\eta_{\rm tag}) & = & \bar{p}^{\rm tag}_0\,+\,\bar{p}^{\rm tag}_1\,(\eta_{\rm tag}\,-\,\hat{\eta}_{\rm tag}), \nonumber
\end{eqnarray}
where $\hat{\eta}_{\rm tag}$ is a fixed value, chosen to be equal to the mean value of the $\eta_{\rm tag}$ distribution to minimise the correlation among the parameters. To reduce the correlation among $\varepsilon^{\rm sig}_{\rm tag}$ and $\bar{\varepsilon}^{\rm sig}_{\rm tag}$, and $p^{\rm tag}_0$, $\bar{p}^{\rm tag}_0$, $p^{\rm tag}_1$, and $\bar{p}^{\rm tag}_1$, these variables are conveniently parameterised as
\begin{eqnarray}\label{eq:taggingCalibrationReparameterisation}
  \varepsilon^{\rm sig} & = & \hat{\varepsilon}^{\rm sig}_{\rm tag}(1+\Delta\varepsilon^{\rm sig}_{\rm tag}), \\
  \bar{\varepsilon}^{\rm sig} & = & \hat{\varepsilon}^{\rm sig}_{\rm tag}(1-\Delta\varepsilon^{\rm sig}_{\rm tag}), \nonumber \\
  p^{\rm tag}_0 & = & \hat{p}^{\rm tag}_0(1+\Delta p^{\rm tag}_0), \nonumber \\
  \bar{p}^{\rm tag}_0 & = & \hat{p}^{\rm tag}_0(1-\Delta p^{\rm tag}_0), \nonumber \\
  p^{\rm tag}_1 & = & \hat{p}^{\rm tag}_1(1+\Delta p^{\rm tag}_1), \nonumber \\
  \bar{p}^{\rm tag}_1 & = & \hat{p}^{\rm tag}_1(1-\Delta p^{\rm tag}_1), \nonumber
\end{eqnarray}
where $\hat{p}^{\rm tag}_{0,1}$ and $\Delta p^{\rm tag}_{0,1}$ are the average and the asymmetry between $p^{\rm tag}_{0,1}$ and $\bar{p}^{\rm tag}_{0,1}$, and $\hat{\varepsilon}^{\rm sig}_{\rm tag}$ and $\Delta\varepsilon^{\rm sig}_{\rm tag}$ are the average and the asymmetry between $\varepsilon^{\rm sig}_{\rm tag}$ and $\bar{\varepsilon}^{\rm sig}_{\rm tag}$. The dependence on the decay-time is considered only for the averaged efficiency $\hat{\varepsilon}^{\rm sig}_{\rm SS}$ and not for the asymmetry $\Delta\varepsilon^{\rm sig}_{\rm tag}$. The strategy used to determine the decay-time efficiencies $\varepsilon(t)$ and $\hat{\varepsilon}^{\rm sig}_{\rm SS}(t)$ is reported in Sec.~\ref{sec:simFitMethod}. The description of $h^{\rm SS}_{\rm sig}(\eta)$ for the SS taggers is presented in Secs.~\ref{sec:ssdCombination} and~\ref{sec:sskCalibration}, respectively.

The PDF $\Omega_{\rm comb}(\vec{\xi},\vec{\eta})$ for the combinatorial background is empirically parameterised by
\begin{equation}\label{eq:omegaPdfBkg}
  \begin{split}
    \Omega_{\rm comb}(\vec{\xi},\vec{\eta}) = 
    & \left[\delta_{\xi_{\rm SS},\,+1}\varepsilon^{\rm SS}_{\rm comb}\,+
    \,\delta_{\xi_{\rm SS},\,-1}\bar{\varepsilon}^{\rm SS}_{\rm comb}
    \right]h^{\rm SS}_{\rm comb}(\eta_{\rm SS})\Omega^{\rm OS,1}_{\rm comb}(\xi_{\rm OS},\eta_{\rm OS})+ \\
    & \delta_{\xi_{\rm SS},\,0}\,(1-\varepsilon^{\rm SS}_{\rm comb}-\bar{\varepsilon}^{\rm SS}_{\rm comb})\,U(\eta_{\rm SS})\Omega^{\rm OS,0}_{\rm comb}(\xi_{\rm OS},\eta_{\rm OS}),
  \end{split}
\end{equation}
where the functions $\Omega^{{\rm OS},j}_{\rm comb}(\xi_{\rm OS},\eta_{\rm OS})$ ($j=0,1$) are the PDF for the $\xi_{\rm OS}$ and $\eta_{\rm OS}$, defined as
\begin{equation}\label{eq:omegaBkgOS}
  \begin{split}
    \Omega^{{\rm OS},j}_{\rm comb}(\xi_{\rm OS},\,\eta_{\rm OS}) = 
    & \left[\delta_{\xi_{\rm OS},\,1}\varepsilon^{{\rm OS},j}_{\rm comb}\,+
    \,\delta_{\xi_{\rm OS},\,-1}\bar{\varepsilon}^{{\rm OS},j}_{\rm comb}\right]\,h^{\rm OS}_{\rm comb}(\eta_{\rm OS})\,+ \\
    & \delta_{\xi_{\rm OS},\,0}\,(1-\varepsilon^{{\rm OS},j}_{\rm comb}-\bar{\varepsilon}^{{\rm OS},j}_{\rm comb})\,U(\eta_{\rm OS}).
  \end{split}
\end{equation}
The variables $\varepsilon^{\rm tag}_{\rm comb}$ and $\bar{\varepsilon}^{\rm tag}_{\rm comb}$ are the probabilities to tag a combinatorial background candidate as \Bds or \Bdsb and $h^{\rm tag}_{\rm comb}(\eta_{\rm tag})$ is the distribution of $\eta_{\rm tag}$. The distribution is described using histograms taken from the right-hand sideband with invariant-mass range between 5.6\gevcc and 6.2\gevcc. 
The tagging efficiencies are parameterised by
\begin{eqnarray}\label{eq:effTagBkg}
  \varepsilon^{\rm tag}_{\rm comb} & = & \frac{\hat{\varepsilon}^{\rm tag}_{\rm comb}}{2}(1+\Delta\varepsilon^{\rm tag}_{\rm comb}), \\
  \bar{\varepsilon}^{\rm tag}_{\rm comb} & = & \frac{\hat{\varepsilon}^{\rm tag}_{\rm comb}}{2}(1-\Delta\varepsilon^{\rm tag}_{\rm comb}), \nonumber
\end{eqnarray}
such that the fit to data determines the average probability to tag combinatorial background as \Bds or \Bdsb, $\hat{\varepsilon}^{\rm tag}_{\rm comb}$, and the asymmetry between the two probabilities, $\Delta\varepsilon^{\rm tag}_{\rm comb}$. For the OS tagger, the distinction labelled by the index $j=0,1$ is used to differentiate the OS-tagger probability between cases that have $\xi_{\rm SS}=0$ ($j=0$) and $\xi_{\rm SS}\neq0$ ($j=1$). In the case of the \Kp\pim and \Km\pip samples,  Eq.~\eqref{eq:omegaPdfBkg} is modified in order to include the dependence on the final-state tag $\psi$
\begin{equation}\label{eq:omegaPdfBkgKPI}
    \Omega_{\rm comb}(\psi,\vec{\xi},\vec{\eta}) = 
    \frac{(1-\psi A_{\rm raw}^{\rm comb})(1-\psi\xi_{\rm OS} A_{\rm OS}^{\rm comb})(1-\psi\xi_{\rm SS} A_{\rm SS}^{\rm comb})}
  {\sum_{\psi=-1,1}{(1-\psi A_{\rm raw}^{\rm comb})(1-\psi\xi_{\rm OS} A_{\rm OS}^{\rm comb})(1-\psi\xi_{\rm SS} A_{\rm SS}^{\rm comb})}}\times \Omega_{\rm comb}(\vec{\xi},\vec{\eta}),
\end{equation}
where $A_{\rm raw}^{\rm comb}$ is the total asymmetry between the combinatorial-background yields in the \Kp\pim and \Km\pip samples, $A_{\rm OS}^{\rm comb}$ and $A_{\rm SS}^{\rm comb}$ are additional parameters that take into account the possibility that the flavour-tagging probabilities $\varepsilon^{\rm tag}_{\rm comb}$ and $\bar{\varepsilon}^{\rm tag}_{\rm comb}$ may depend on the final state.

The PDF $\Omega_{\rm 3-body}(\vec{\xi},\vec{\eta})$ for the partially reconstructed \B decays are empirically parametersied as
\begin{equation}\label{eq:omegaPdfPhys}
  \Omega_{\rm 3-body}(\vec{\xi},\vec{\eta}) = \Omega^{\rm OS}_{\rm 3-body}(\vec{\xi},\vec{\eta})\Omega^{\rm SS}_{\rm 3-body}(\vec{\xi},\vec{\eta}),
\end{equation}
where $\Omega^{\rm OS}_{\rm 3-body}(\vec{\xi},\vec{\eta})$ and $\Omega^{\rm SS}_{\rm 3-body}(\vec{\xi},\vec{\eta})$ are
\begin{equation}\label{eq:omegaTagPhys}
    \begin{split}
        \Omega^{\rm tag}_{\rm 3-body}(\xi_{\rm tag},\,\eta_{\rm tag}) = & \delta_{\xi_{\rm tag},\,+1}\varepsilon^{\rm tag}_{\rm 3-body}\,h^{\rm tag}_{\rm 3-body}(\eta_{\rm tag})\,+\,\delta_{\xi_{\rm tag},\,-1}\bar{\varepsilon}^{\rm tag}_{\rm 3-body}\,h^{\rm tag}_{\rm 3-body}(\eta_{\rm tag})\,+ \\
        & \delta_{\xi_{\rm tag},\,0}\,(1-\varepsilon^{\rm tag}_{\rm 3-body}-\bar{\varepsilon}^{\rm tag}_{\rm 3-body})\,U(\eta_{\rm tag}),
    \end{split}
\end{equation}
where $\varepsilon^{\rm tag}_{\rm 3-body}$ and $\bar{\varepsilon}^{\rm tag}_{\rm 3-body}$ are the probabilities to tag a background candidate as \Bds or \Bdsb, and $h^{\rm tag}_{\rm 3-body}(\eta_{\rm tag})$ is the distribution of $\eta_{\rm tag}$. As before, the tagging efficiencies are parameterised as a function of the total efficiency ($\hat{\varepsilon}^{\rm tag}_{\rm 3-body}$) and asymmetry ($\Delta\varepsilon^{\rm tag}_{\rm 3-body}$)
\begin{eqnarray}
    \varepsilon^{\rm tag}_{\rm 3-body} & = & \frac{\hat{\varepsilon}^{\rm tag}_{\rm 3-body}}{2}(1+\Delta\varepsilon^{\rm tag}_{\rm 3-body}), \\
    \bar{\varepsilon}^{\rm tag}_{\rm 3-body} & = & \frac{\hat{\varepsilon}^{\rm tag}_{\rm 3-body}}{2}(1-\Delta\varepsilon^{\rm tag}_{\rm 3-body}).\nonumber
\end{eqnarray}
The PDF $h_{\rm 3-body}^{\rm tag}(\eta_{\rm tag})$ is determined as a histogram from the low-mass sideband, where the residual contamination of combinatorial-background candidates is subtracted by injecting candidates with negative weights. In the case of the \Kp\pim and \pip\Km samples Eq.~\eqref{eq:omegaTagPhys} is modified in order to include the dependence on the final-state tag $\psi$. Analogously to Eq.~\eqref{eq:omegaPdfBkgKPI}, the parmeterisation is
\begin{equation}\label{eq:omegaPdfPhysKPI}
\begin{split}
    \Omega_{\rm 3-body}(\psi,\vec{\xi},\vec{\eta}) = & 
    \frac{(1-\psi A_{\rm raw}^{\rm 3-body})(1-\psi\xi_{\rm OS} A_{\rm OS}^{\rm 3-body})(1-\psi\xi_{\rm SS} A_{\rm SS}^{\rm 3-body})}
  {\sum_{\psi=-1,1}{(1-\psi A_{\rm raw}^{\rm 3-body})(1-\psi\xi_{\rm OS} A_{\rm OS}^{\rm 3-body})(1-\psi\xi_{\rm SS} A_{\rm SS}^{\rm 3-body})}} \\
  & \times \Omega_{\rm 3-body}(\vec{\xi},\vec{\eta}),
\end{split}
\end{equation}
where $A_{\rm raw}^{\rm 3-body}$, $A_{\rm OS}^{\rm 3-body}$ and $A_{\rm SS}^{\rm 3-body}$ have the same meaning of the corresponding quantities as for the combinatorial-background component.

The PDFs in Eqs.~\eqref{eq:omegaPdfSig}, \eqref{eq:omegaPdfBkg} and~\eqref{eq:omegaPdfPhys} are valid if $\eta_{\rm OS}$ and $\eta_{\rm SS}$ are uncorrelated. This assumption is verified by means of background-subtracted~\cite{Pivk:2004ty} signal candidates, and of candidates from the high- and low-mass sidebands for the combinatorial and three-body backgrounds components, respectively. 

\subsection{Combination of the single SS and OS taggers}\label{sec:ssdCombination}

The calibration parameters governing the relations in Eqs.~\eqref{eq:decayTimeB2KPI} are determined separately for the individual SS and OS taggers by means of a binomial regression to the tagged decay-time distribution of
background-subtracted \BdToDpi decays.
Then the extracted calibration parameters of the SS\pion and SS\proton taggers are used to combine the two taggers 
into a unique one (SSc) with decision $\xi_{\rm SSc}$ and mistag probability $\eta_{\rm SSc}$. 
The assumption of a linear relation between $\eta_{\rm tag}$ and $\omega_{\rm tag}$ for each tagger is validated 
splitting the sample in bins of $\eta_{{\rm SS}(\pion, \proton)}$, estimating the average mistag fraction 
in each bin by means of the binomial regression. 
Similarly, the various OS taggers are combined together into a unique OS tagger with decision $\xi_{\rm OS}$ and 
mistag probability $\eta_{\rm OS}$, and the same linearity check is performed. 

The PDFs $h^{\rm SS}_{\rm sig}(\eta_{\rm SSc})$ describing the $\eta_{\rm SSc}$ distributions for the signal \Bd mesons 
are determined using background-subtracted distributions of \BdToDpi decays. 
As the pion and proton kinematics are correlated with those of the \Bd meson, the performance of the SS\pion and 
SS\proton taggers also depend on the latter. The differences between the \Bd-meson kinematics and other relevant 
distributions in \BdToDpi and \BdTopipi decays, due to the different topologies and selection requirements, are 
taken into account by means of a weighting procedure to the \BdToDpi sample. 
It is empirically observed that the \B meson transverse momentum and the number of hits in the SPD detector distributions need to be equalised.

\subsection{Calibration of the SS\boldmath{\kaon} tagger}\label{sec:sskCalibration}

The natural control mode to calibrate the response of the SS\kaon tagger would be the \BsTopiK decay.
However, since the signal yield of this decay is approximately 8\% of that of the \BdToKpi decay and 20\% of that 
of the \BsToKK decay, it would not be possible obtaining a reliable calibration. Furthermore, the calibration 
parameters of the SS\kaon tagger would be affected by large uncertainties, limiting the precision on \CKK~and \SKK. 
Therefore, the calibration is performed with a large sample of \BsToDspi decays. Analogously to the SS\pion and 
SS\proton case, the SS\kaon-calibration parameters are determined using a binomial regression to the tagged 
decay-time distribution of the \BsToDspi decays. Also in this case the regression is performed using the 
flavour-tagging information on a per-candidate basis, determining the calibration parameters directly, and a 
check of the linear relation between $\eta_{{\rm SS}\kaon}$ and $\omega_{{\rm SS}\kaon}$ is performed. As described for the \BdToDpi sample in the previous section, a weighting procedure is applied to the \BsToDspi 
sample in order to equalise the signal distribution of the \B meson transverse momentum and the number of hits in 
the SPD detector. 
The PDF $h^{{\rm SS}\kaon}_{\rm sig}(\eta_{{\rm SS}\kaon})$ for \BsToKK decays is determined using a 
background-subtracted histogram of the same weighted sample of \BsToDspi decays used for the calibration.


\clearpage

\addcontentsline{toc}{section}{References}
\bibliographystyle{LHCb}
\bibliography{main,standard,LHCb-PAPER,LHCb-CONF,LHCb-DP,LHCb-TDR}

\newpage
\centerline
{\large\bf LHCb collaboration}
\begin
{flushleft}
\small
R.~Aaij$^{31}$,
C.~Abell{\'a}n~Beteta$^{49}$,
T.~Ackernley$^{59}$,
B.~Adeva$^{45}$,
M.~Adinolfi$^{53}$,
H.~Afsharnia$^{9}$,
C.A.~Aidala$^{84}$,
S.~Aiola$^{25}$,
Z.~Ajaltouni$^{9}$,
S.~Akar$^{64}$
% ...

{\footnotesize \it

% ...
$ ^{9}$Universit{\'e} Clermont Auvergne, CNRS/IN2P3, LPC, Clermont-Ferrand, France\\
% ...
$ ^{25}$INFN Sezione di Milano, Milano, Italy\\
% ...
$ ^{31}$Nikhef National Institute for Subatomic Physics, Amsterdam, Netherlands\\
% ...
$ ^{45}$Instituto Galego de F{\'\i}sica de Altas Enerx{\'\i}as (IGFAE), Universidade de Santiago de Compostela, Santiago de Compostela, Spain\\
% ...
$ ^{49}$Physik-Institut, Universit{\"a}t Z{\"u}rich, Z{\"u}rich, Switzerland\\
% ...
$ ^{53}$H.H. Wills Physics Laboratory, University of Bristol, Bristol, United Kingdom\\
% ...
$ ^{59}$Oliver Lodge Laboratory, University of Liverpool, Liverpool, United Kingdom\\
% ...
$ ^{64}$University of Cincinnati, Cincinnati, OH, United States\\
% ...
$ ^{67}$Syracuse University, Syracuse, NY, United States\\
% ...
$ ^{84}$University of Michigan, Ann Arbor, United States, associated to $^{67}$\\
\bigskip
% ...
\medskip
}
\end{flushleft}

\end{document}